\documentclass[twocolumn]{aastex63}
\usepackage{hyperref}
\usepackage{amsmath}
\usepackage{amssymb}
\usepackage{graphicx}
\usepackage{color}
\usepackage[utf8]{inputenc}
\DeclareUnicodeCharacter{2212}{-}

\newcommand{\kms}{km\,s$^{-1}$}

\newcommand{\logg}{$\log g$}

\newcommand{\rproc}{$r$-process}

\begin{document}
\shorttitle{R-process-rich stellar streams in the Milky Way}

\title{R-process-rich stellar streams in the Milky Way} \footnote{This paper includes data gathered with the 6.5 meter
  Magellan Telescopes located at Las Campanas Observatory, Chile.}

\author{Maude Gull}
\affiliation{Department of Astronomy, University of California Berkeley, Berkeley, CA 94720, USA}
\affiliation{Department of Physics \& Kavli Institute for Astrophysics and Space Research, Massachusetts Institute of Technology, Cambridge, MA 02139, USA}

\author{Anna Frebel}
\affiliation{Department of Physics \& Kavli Institute for Astrophysics and Space Research, Massachusetts Institute of Technology, Cambridge, MA 02139, USA}

\author{Karina Hinojosa}
\affiliation{Department of Physics \& Kavli Institute for Astrophysics and Space Research, Massachusetts Institute of Technology, Cambridge, MA 02139, USA}

\author{Ian U.\ Roederer}
\affiliation{Department of Astronomy, University of Michigan, Ann Arbor, MI 48109, USA}
\affiliation{Joint Institute for Nuclear Astrophysics -- Center for the
Evolution of the Elements (JINA-CEE), USA}

\author[0000-0002-4863-8842]{Alexander~P.~Ji}
\affiliation{Observatories of the Carnegie Institution for Science, 813 Santa Barbara St., Pasadena, CA 91101, USA}

\author{Kaley Brauer}
\affiliation{Department of Physics \& Kavli Institute for Astrophysics and Space Research, Massachusetts Institute of Technology, Cambridge, MA 02139, USA}

\begin{abstract}

We present high-resolution Magellan/MIKE spectra of 22 bright ($9<V<13.5$) metal-poor stars ($-3.18<\mbox{[Fe/H]}<-1.37$) in three different stellar streams, the Helmi debris stream, the Helmi trail stream, and the $\omega$~Centauri progenitor stream. We augment our Helmi debris sample with results for ten stars by \citet{Roederer10}, for a total of 32 stars.
Detailed chemical abundances of light elements as well as heavy neutron-capture elements have been determined for our 22 stars. All three streams contain carbon-enhanced stars. For 13 stars, neutron-capture element lines were detectable and they all show signatures in agreement with the scaled solar $r$-process pattern, albeit with a large spread of $-0.5<\mbox{[Eu/Fe]}<+1.3$. Eight of these stars show an additional small $s$-process contribution superposed onto their $r$-process pattern. This could be discerned because of the relatively high $S/N$ of the spectra given that the stars are close by in the halo. 
Our results suggest that the progenitors of these streams experienced one or more $r$-process events, such as a neutron star merger or another prolific $r$-process source, early on that widely enriched these host systems before their accretion by the Milky Way. The small $s$-process contribution suggests the presence of AGB stars and associated local (inhomogeneous) enrichment as part of the ongoing chemical evolution by low mass stars.
Stars in stellar streams may thus be a promising avenue for studying the detailed history of large dwarf galaxies and their role in halo assembly with easily accessible targets for high-quality spectra of many stars.

\end{abstract}


\section{Introduction} \label{s:intro}

The halo of the Milky Way provides a unique window into the chemical and dynamical evolution of the universe through the study of its dwarf galaxies, globular clusters, and stellar streams. Dwarf galaxies and clusters constitute the essential building blocks for the hierarchical growth of galaxies, while stellar streams arise from accretion and disruption of satellite dwarf galaxies and globular clusters in the Galaxy's potential.  
Streams may also form through the on-going disruption of a satellite with the host parent galaxy remaining partly intact, such as the tidal debris stream arising from the Sagittarius (Sgr) dwarf spheroidal galaxy  (\citealt{majewski03,penarrubia05}). Furthermore, there are other pathways for stream formation,  including high-eccentricity flyby encounters  \citep{younger08} and flares or warps in the galactic disk (e.g., \citealt{momany06}). 

Most stellar streams have been identified through stellar overdensities, common kinematic signatures of their members, or simply discrepancies in angular momentum and velocity space between stream and background stars (e.g., \citealt{Helmi06, grillmair08, malhan18, shipp18}).
It is usually challenging to distinguish stream stars from background, in-situ halo stars based on distinct chemical abundance signatures, given their typically similar compositions.
However, once kinematically identified, an in-depth analysis of the chemical composition of the stars in a stream allows for a broad characterization of the progenitor dwarf galaxy (e.g., \citealt{casey13, ji20}).

Here we present a detailed chemical abundance analysis of 22 stars, identified by \citet{beers17}, associated with debris stripped from the putative parent dwarf galaxy of $\omega$~Centauri (NGC 5139, $\omega$~Cen) (\citealt{dinescu02,klement09,majewski12}) or the debris and trail streams of unknown origin, as discovered by \citet{Helmi99b}.

Increasing the sample of chemically studied stream stars is important to advance our understanding of this putative, yet important progenitor. \citet{Kepley07} estimated that the progenitor was accreted 6-9\,Gyr ago based on a smaller stellar sample, and taking advantage of the bimodality of the $v_{z}$ distribution. Meanwhile, the N-body simulations of \citet{koppelman19}, based on the kinematics of 523 potential members, suggest an accretion 5-8\,Gyr ago. Furthermore, they postulate an age of 11-13\,Gyr for the progenitor. The progenitor was likely a massive dwarf galaxy with a stellar mass of $\sim$10$^8$\,M$_{\odot}$. 
\citet{naidu20} also estimate a mass of $\approx$0.5-1.0 $\times 10^{8}$\,M$_{\odot}$ for the Helmi streams progenitor, based on the stellar metallicity distribution and the dwarf galaxy mass-metallicity relation \citep{Kirby13b}.
\citet{koppelman19} strengthen the original \citet{Helmi99a} claim that the progenitor of these streams may have been responsible for $\sim$10 to 14\% of the stars currently present in the Galactic halo.
While \citeauthor{naidu20}\ agree with these authors about the stellar mass of the progenitor, their analysis of the in-situ halo leads to a lower estimate of the fractional stellar halo contribution from the progenitor, $\sim$~1\%.

The globular cluster $\omega$~Centauri (NGC 5139, $\omega$~Cen) is an extremely unique object given its structure, kinematic signature, and stellar content. It is the most massive Galactic globular cluster ($\approx 4 \times 10^{6}$\,M$_{\odot}$; \citealt{dsouza13}), and it displays several stellar sub-populations (e.g., \citealt{gratton11,bellini17}). Origin scenarios of this cluster center around it being the stripped core of a dwarf galaxy, or that it is the result of a merger of several less massive clusters into a dwarf galaxy that was then accreted by the Milky Way (\citealt{norris97, bekki03, deboer15}). Most recent studies support the idea that $\omega$~Cen is the stripped core or the most massive globular cluster of the progenitor that also produced either Sequoia \citep{myeong19} or Gaia-Enceladus \citep{massari19}. Such a scenario then supports $\omega$~Cen being a major contributor of local retrograde stars \citep{majewski12}, in the same way as  N-body simulations \citep{Ibata19} and chemical tagging \citep{simpson20} have shown that the recently discovered ``Fimbulthul'' structure is the likely the main trailing tidal stream of $\omega$~Cen. 

Previously, based on a common signature of retrograde orbits, the $\omega$~Cen stream had been postulated to be a debris stream of the dwarf galaxy progenitor of the globular cluster $\omega$~Cen \citep{dinescu02, beers17}. It remains to be seen if this association can be confirmed. Here, we analyse ten stars, originally classified by \citet{beers17} as putative members of the $\omega$~Cen debris stream, to help shed light on the nature of the progenitor. In \citet{bekki03}, the initial stellar mass of the  $\omega$~Cen progenitor was estimated to be $\sim10^7$\,M$_\odot$. The system was then accreted by the Milky Way around $\sim$10\,Gyr ago. These estimates agree with the initial stellar mass estimates ($\sim10^7$\,M$_\odot$) and an infall time around $\sim$9\,Gyr ago for the progenitor of Sequoia. While an accretion time of $\sim$10\,Gyr ago for Gaia-Enceladus agrees with the putative progenitor accretion time, the putative progenitor of Gaia-Enceladus mass estimate is much larger, $\sim6\times10^8$\,M$_\odot$.

Signatures of the build up of the halo from accreted systems are encoded in the chemical composition of halo stars. One approach to uncovering this information is to use known stream stars to establish key elemental characteristics of their progenitor. Another one is to use metal-poor halo stars showing enhancements in heavy rapid neutron-capture ($r$-)process elements, as many or even most of these stars may have originated in small dwarf galaxies that experienced an $r$-process event, such as Reticulum\,II (\citealt{Ji16a,Ji16b,Roederer16b,Roederer18}), thus tracing accreted systems. Indeed, this idea has been modeled \citep{brauer19} to further explore the relationship between  low-metallicity halo stars and accreted dwarf galaxies over cosmic time. In this paper, we are able to combine both approaches by studying stream stars, most of which show moderate enhancements of $r$-process elements. Being able to investigate a disrupted galaxy in the form of a stream thus offers unique clues into the origin scenario(s) of $r$-process enhanced halo stars along with new insights about their progenitor systems.

\begin{deluxetable*}{lrrccrrcrrr}
\tablewidth{0pt}
\tabletypesize{\scriptsize}
\tablecaption{\label{Tab:obs} Observing Details }
\tablehead{                                                     
\colhead{Star} & 
\colhead{$\alpha$}&
\colhead{$\delta$}&
\colhead{UT date}&
\colhead{slit}&
\colhead{$t_{\rm {exp}}$}&
\colhead{$V$} &
\colhead{$B-V$} &
\colhead{$S/N$} &
\colhead{$S/N$} &
\colhead{$v_{hel}$}   \\
\colhead{}&
\colhead{(J2000)}&
\colhead{(J2000)}&
\colhead{}&
\colhead{arcsec}&
\colhead{min}&
\colhead{mag}&
\colhead{mag}&
\colhead{4500\,{\AA}}&
\colhead{6000\,{\AA}}&
\colhead{\kms}
}

\startdata
\multicolumn{11}{c}{Helmi trail stream}\\
\hline
BM235       & 17 52 37.6 &$-$69 01 44.0  & 2019 05 01    &0.7 & 10 & 9.48 & 1.06 & 200 & 250 & $+$138.4 \\ 
HE~0033$-$2141 & 00 35 42.0 &$-$21 24 56.8  & 2014 06 24    &0.7 & 8& 12.29 & 0.72 & 70& 85 &$-$185.1 \\
HE~0050$-$0918 & 00 52 41.9 &$-$09 02 24.2  & 2016 10 14    &0.7 & 10& 11.06 & 0.71 & 130& 175& $-$238.0 \\
HE~1210$-$2729 & 12 13 06.9 &$-$27 45 45.6  & 2013 01 08    &0.7 &  7& 12.54 & 0.86 & 75 & 100 &$-$82.7\\
HE~2234$-$4757 & 22 37 20.5 &$-$47 41 44.1  & 2014 06 23 	 &0.7 & 10& 12.39 & 0.92 & 90& 125& $-$102.8 \\ 
\hline
\multicolumn{11}{c}{Helmi debris stream}\\
\hline
BM028 & 02 47 37.5 &$-$36 06 25.2  & 2016 10 13       &0.7 &   5& 9.94 & 0.46 & 180& 220&$+$308.5\\ 
BM209 & 14 36 49.2 &$-$29 07 07.6 & 2019 02 11 & 0.7 & 4 & 8.02 & 0.64 & 30 & 70& $-$78.1 \\ 
BM308 & 22 37 08.2 &$-$40 30 39.6  & 2016 10 13       &0.7 &   5& 9.11 & 0.97 & 200& 300&$-$364.9\\ 
HE~0012$-$5643 & 10 56 17.9 &$-$30 57 49.3  & 2014 06 22       &0.7 & 10& 12.29 & 0.46 & 95& 100&$-$265.5\\
HE~0017$-$3646 & 00 20 26.3 &$-$36 30 18.3  & 2015 10 02       &0.7 & 15& 13.02 & 0.54 & 80& 100&$+$347.6\\
HE~0048$-$1109 & 00 51 26.5 &$-$10 53 18.4 & 2014 06 24 &0.7 & 3 &10.83 & 0.49 & 100& 145&$+$50.5\\
HE~0324$-$0122 & 03 27 02.4 &$+$01 32 29.3  & 2016 10 14       &0.7 & 15& 12.13& 0.72 & 75& 100&$+$184.1\\
\hline
\multicolumn{11}{c}{$\omega$~Cen progenitor stream}\\
\hline
BM056 &05 10 49.6  & $−$ 37 49 03.0  & 2016 04 16       &0.7 & 30& 9.50 & 0.86 & 155& 320&$-$33.4\\
BM121 & 09 53 38.4 &$-$22 50 11.4  & 2019 02 10       &0.7 & 10& 9.39 & 1.16 & 350 & 500 &$+$151.6\\
HE~0007$-$1752 & 00 10 17.8 &$-$17 35 37.3  & 2016 10 14       &0.7 & 20& 11.54 & 0.65 & 150& 200&$+$191.2\\
HE~0039$-$0216 & 00 41 53.7 &$-$02 00 34.7  & 2014 06 24       &0.7 & 10& 13.35 & 0.37 & 55& 65&$+$211.9\\
HE~0429$-$4620 & 04 30 49.0 &$-$46 13 55.7  & 2016 10 14       &0.7 & 10& 13.10 & 0.62 & 35& 40&$+$138.9\\
HE~1120$-$0153 & 11 22 42.6 &$-$02 09 37.9  & 2013 01 08        &0.7 & 15& 11.68 & 0.44 & 140& 180&$+$288.2\\
HE~1401$-$0010 & 14 04 03.4 &$-$00 24 30.2  & 2010 08 07       &1.0 & 25& 13.51 & 0.41 & 70& 83&$+$387.7\\
HE~2138$-$0314 & 21 40 41.6 &$-$03 01 19.1  & 2010 06 06       &0.7 & 5& 13.23 & 0.57 & 62& 90&$-$373.0\\
HE~2315$-$4306 & 23 18 18.7 &$-$42 50 22.7  & 2016 08 29   &1.0 & 11& 11.28 & 0.65 & 70&105&$+$201.7\\
HE~2319$-$5228 & 23 21 57.4 &$-$52 11 45.9  & 2014 06 23       &0.7 & 15& 13.25 & 0.90 & 75& 100&$+$293.7\\
HE~2322$-$6125 & 23 25 34.6  &$-$61 09 10.0 & 2016 04 16       &0.7 & 30& 12.47 & 0.63 & 38& 45&$+$328.8\\
\enddata

\end{deluxetable*}

\section{Sample selections and Observations}

\begin{figure}[!ht]
  \includegraphics[clip=false,width=8.4cm
   ]{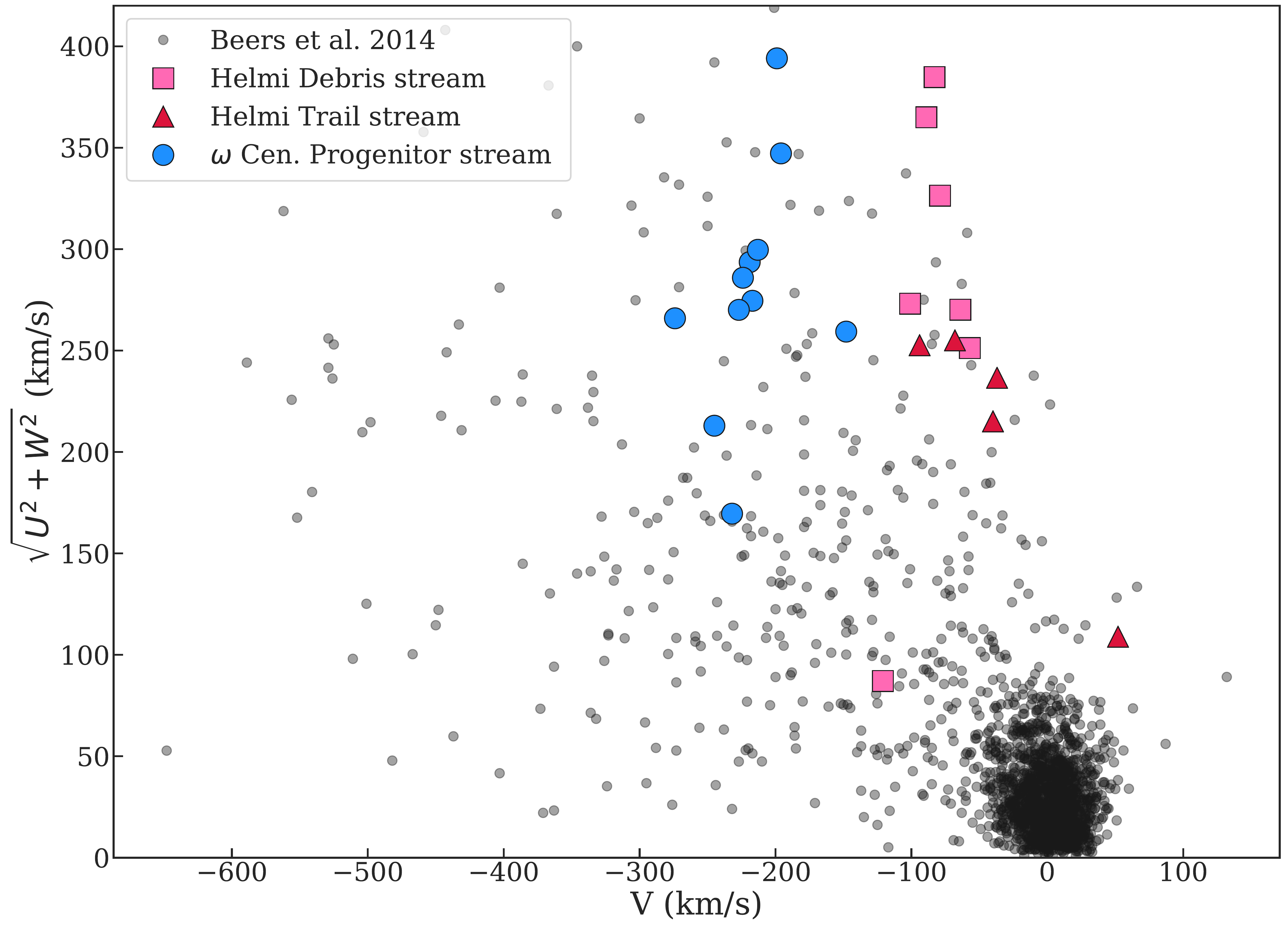} 
   \figcaption{\label{toomre} Toomre diagram of the \citet{beers14} sample with stars in the three streams marked.} 
\end{figure}

\citet{beers17} re-analysed a combination of the bright (9.1 $<$ V $<$ 13.5) metal-poor halo stars by \citet{Frebel06a} and ``weak-metal'' candidates \citep{beers14}, originally analyzed by \citet{bidelman1973}, by performing a comprehensive chemodynamical investigation. They combined a Toomre diagram analysis, constructed from using derived orbital parameters of all stars, with the Lindblad diagram that compares energy to angular momentum space to obtain information about the kinematics of this sample. We reproduce the Toomre diagram that shows the streams' signature in Figure~\ref{toomre}.
In this way we selected 25 candidates associated with the Helmi debris and trail stellar stream, and the $\omega$~Cen progenitor stream. These stars are the focus of this high-resolution spectroscopy study.

A diverse history of stream member discoveries exists for the Helmi stream stars.
\citet{Helmi99a} found 13 members of the now so-called debris stream. 
\citet{Roederer10} performed a detailed abundance analysis of 12 of those 13 members. Additionally, \citet{chiba00} detected a secondary stream associated with the Helmi debris stream, and they postulated a membership of 9 stars belonging to the Helmi trail stream. The Helmi trail stream distinguishes itself from the Helmi debris stream kinematically \citep{yuan20}. The Helmi trail stars display positive $v_z$ (vertical velocity) motions, slightly higher energy, larger radial motions, and are more diffuse without clear features when assessing kinematic diagrams, such as a Toomre diagram (See Fig.\ref{toomre}). On the other hand, the Helmi debris stars manifest themselves in a well-defined stream, with prominent negative $v_z$ motion \citep{myeong19}.

\citet{beers17} then re-identified 2 of the original 13 Helmi debris stream stars, and they identified 8 additional ones. They also confirmed one of the suggested nine stars suggested to be part of the Helmi trail stream, and they found an additional four stars with the same kinematic profile. By 2018, a total of $\sim$33 members (20 Helmi Debris stream, 13 Helmi Trail stream) were known. 

\citet{koppelman19} carried out a more extensive kinematic analysis using data from the second data release (DR2) of the Gaia mission and several large spectroscopic surveys. They
identified 40 potential members within 1\,kpc of the Sun, re-identifying 8 of the original 13 stars. All 13 members could be recovered by increasing the search radius to 2.5\,kpc. In total, \citeauthor{koppelman19}\ find 523 potential members based on their energy and angular momenta. That study also confirmed two different streams, and it found that a ratio of 1:2 was the best descriptor between trail and debris streams, suggesting $\sim$350 stars to be of the debris and $\sim$175 of the trail stream.

The history of discovery of field stars associated with $\omega$~Cen is more limited.
\citet{majewski12} identified 35 candidates,
\citet{beers17} identified 11 candidates, and \citet{ibata19a} identified 340 candidate members of the Fimbulthul stream.
Neither \citeauthor{majewski12}\ nor \citeauthor{ibata19a} published complete lists of candidates, except for the 12~stars for which \citeauthor{majewski12}\ obtained followup spectroscopy, so we only observed the 11 candidates identified by \citeauthor{beers17}.
These stars are spaced across the sky and are not members of the Fimbulthul stream.

We observed 23 stars from \citet{beers17}. We did not observe one star in the \citet{beers17} list of Helmi debris stream stars (HE~2215$-$3842), and we exclude one star (HE~1120$-$0153) in the $\omega$~Cen progenitor stream from further analysis given its status as a double-lined spectroscopic binary. Of the 22~stars analyzed further, 5 are Helmi trail stars, 7 are Helmi debris stars, and 10 are $\omega$~Cen progenitor stream stars. We note that two of the seven debris stars had already been analyzed by \citet{Roederer10}, but we carried out our own analysis based on our spectra.

Spectroscopic observations were taken over the span of nine years at the Magellan-Clay telescope at Las Campanas Observatory, as part of our continuous follow up of bright metal-poor stars \citep{Frebel06b}, many of which turned out to be part of these streams investigated here.
Using the Magellan Inamori Kyocera Echelle (MIKE) spectrograph \citep{Bernstein03}, high-resolution spectra covering $\sim$3500\,{\AA} to $\sim9000$\,{\AA} were obtained using the $0\farcs7$ slit for all but three stars, for which a $1\farcs0$ slit was employed. The $0\farcs7$ slit yields a nominal spectral resolving power ranging from $R \sim28,000$ in the red and $R\sim35,000$ in the blue wavelength regime. The $1\farcs0$ slit yields a resolving power of $R\sim22,000$ in the red and $R\sim28,000$ in the blue wavelength regime. The red and blue arms are split by a dichroic at $\sim$5000~{\AA}.

\begin{figure*}[!ht]
 \begin{center}
  \includegraphics[clip=false,width=17.7cm ,
  ]{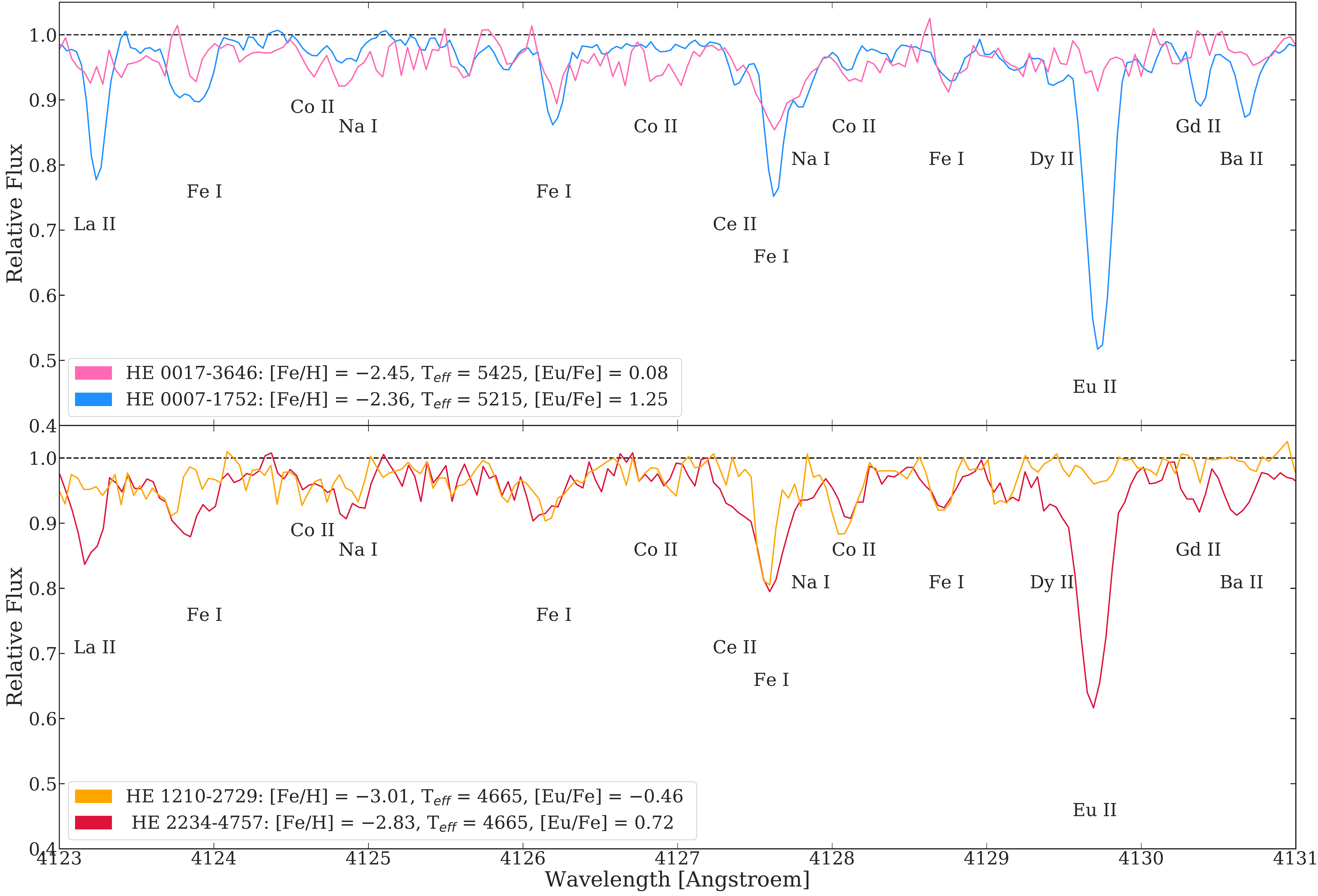} \figcaption{\label{spectra} Spectral comparison around the Eu\,II line at 4129\,{\AA} of various stars (top: HE~0017$-$3646: Helmi debris stream, HE~0007$-$1752: $\omega$ Cen prog. stream. Bottom: both Helmi trail stream). The different line strengths of stars with varying neutron-capture element abundances are easily apparent.   }
 \end{center}
\end{figure*}

Exposure times varied from 5 to 30\,min. Data reduction was carried out with the CarPy pipeline \citep{Kelson03}. The resulting signal-to-noise (S/N) ratios per pixel range from 30 to 350 at $\lambda \sim 4500$\,\AA$ $ and 40 to 500 at $\lambda \sim 6000$\,\AA. Further details on the observations, stellar magnitudes, and resulting heliocentric-corrected radial velocity measurements are listed in Table~\ref{Tab:obs}. 
Radial velocities were obtained from cross-correlating the Ca triplet region with a template metal-poor star of similar stellar parameters. 
Typical uncertainties in radial velocity measurements obtained with MIKE are $\sim1$-2\,\kms. We find good agreement overall with previous literature values. 

Whenever possible we compare the radial velocity measurements of our stars with the Gaia DR2 results \citep{gaia18}, and if those are not available we consider the medium-resolution measurements of \citet{beers17}. We find that six stars have significant radial velocity differences that suggest binarity: HE~2234$-$4757, HE~0012$-$5643, BM056, HE~1120$-$0153 (a double-lined spectroscopic binary), HE~1401$-$0010 and HE~2319$-$5228. For CS~29513-032, no radial velocity measurement is available in Gaia DR2, but we recently obtained a  high-resolution measurements in addition to those given in \citet{Roederer10}. The velocity remains unchanged from previous measurements, $-215.3 +/- 0.7$\,\kms (2021 January 12), $-215.8$\,\kms (2009 Juli 25) and 
$-215.4$\,\kms (2009 October 26). This does not exclude the star to be in a binary system but somewhat strengthens the case of a single star despite the significant enhancement in $r+s$ elements.

\section{Spectral Analysis and Stellar parameters}\label{sec3}

For each star in the sample, we shifted the spectrum to rest wavelengths and proceeded to measure the equivalent widths of various absorption lines by fitting Gaussian profiles to them. Our linelist was originally presented in \citep{roederer08} that includes atomic data sources but updated over time to include Fe $\log gf$ values from \citet{obrian91}, \citet{kurucz_lines}, \citet{mel09}, \citet{den14}, and \citet{ruf14}.
Neutron-capture lines were added also, based on data from \citet{Hill02,hill17}.

\subsection{Measurements and stellar parameter determination}

The equivalent widths measurements are presented in Table~\ref{ew_tab_stub}. For blended spectral features (e.g., CH, NH, neutron-capture absorption features), lines with hyperfine structure, as well as upper limits, we performed spectrum synthesis.

Assuming local thermodynamic equilibrium (LTE), we compute abundances from our equivalent width measurements or from spectrum synthesis for blended features or features affected by hyperfine splitting (HFS) structure or isotope shifts (IS).
These elements include C, Sc, V, Mn, Co, Cu , Ba, La, Pr, Nd, Sm, Eu, Tb, Ho, Yb, Ir, and Pb. Tools for creating linelists for spectrum synthesis are publicly available\footnote{https://github.com/vmplacco/linemake (originating from C. Sneden, priv. comm.)}$^{,}$\footnote{http://kurucz.harvard.edu/linelists.html} and contain atomic data from \citet{Sneden09, sneden14, Sneden16}.

\tabcolsep=0.05cm

\begin{deluxetable*}{lrrrrrrrrrrrrrrrrrrrrrrrrr} 
\tiny
\tablecaption{Equivalent width measurements}
\tablewidth{0pc}
\tablehead{
\colhead{Elm.} & \colhead{$\lambda$} & \colhead{$\chi$} & \colhead{$\log$ gf} & 
\colhead{1} & \colhead{2} & \colhead{3} & 
\colhead{4} & \colhead{5} & \colhead{6} & 
\colhead{7} & \colhead{8} & \colhead{9} & 
\colhead{10} & \colhead{11} & \colhead{12} & 
\colhead{13} & \colhead{14} & \colhead{15} & 
\colhead{16} & \colhead{17} & \colhead{18} & 
\colhead{19} & \colhead{20} & \colhead{21} & 
\colhead{22}}
\startdata
O I & 6300.3 & 0.0 &$-$9.82 &\ldots&\ldots&\ldots&\ldots&\ldots&\ldots&\ldots&\ldots&\ldots&\ldots&\ldots&\ldots& 12.0 & 12.4 &\ldots&\ldots&\ldots&\ldots&\ldots&\ldots&\ldots&\ldots\\
O I & 6363.78 & 0.02 &$-$10.3 & 8.2 &\ldots&\ldots&\ldots&\ldots&\ldots&\ldots&\ldots&\ldots&\ldots&\ldots&\ldots&\ldots&\ldots&\ldots&\ldots&\ldots&\ldots&\ldots&\ldots&\ldots&\ldots\\
Na I & 5682.63 & 2.1 &$-$0.7 & 17.9 &\ldots&\ldots&\ldots&\ldots&\ldots& 7.2 &\ldots&\ldots&\ldots&\ldots&\ldots&\ldots&\ldots&\ldots&\ldots&\ldots&\ldots&\ldots&\ldots& 30.9 &\ldots\\
Na I & 5688.2 & 2.1 &$-$0.45 & 36.0 &\ldots&\ldots&\ldots&\ldots&\ldots&\ldots&\ldots&\ldots& 8.3 &\ldots&\ldots& 18.2 &\ldots&\ldots&\ldots&\ldots&\ldots&\ldots&\ldots& 34.6 &\ldots\\
Na I & 5895.92 & 0.0 &$-$0.19 &\ldots& 128.2 & 242.6 & 151.8 & 150.6 & 202.0 & 137.8 & 169.2 &\ldots& 136.6 & 77.3 & 165.8 & 192.0 & 193.9 & 94.4 & 53.7 & 125.0 & 51.8 & 91.0 & 147.0 &\ldots& 143.7 \\
Na I & 5889.95 & 0.0 & 0.11 & 261.8 & 144.1 & 259.3 & 178.3 & 178.8 & 276.3 & 160.1 & 191.6 & 28.4 & 168.9 & 100.4 & 190.5 & 228.8 & 223.4 & 100.3 & 67.7 & 163.5 & 70.6 & 112.5 & 164.5 &\ldots& 159.5 \\
Mg I & 3838.29 & 2.72 & 0.49 &\ldots&\ldots&\ldots&\ldots& 245.2 &\ldots&\ldots&\ldots&\ldots&\ldots&\ldots&\ldots&\ldots&\ldots&\ldots&\ldots&\ldots&\ldots&\ldots&\ldots&\ldots&\ldots\\
Mg I & 3829.36 & 2.71 &$-$0.23 &\ldots& 154.7 & 292.4 & 178.3 & 174.9 &\ldots&\ldots& 215.2 & 70.5 & 166.5 &\ldots&\ldots&\ldots&\ldots&\ldots&\ldots&\ldots&\ldots& 122.7 &\ldots&\ldots&\ldots\\
Mg I & 5183.6 & 2.72 &$-$0.17 &\ldots& 178.4 & 366.4 &\ldots& 225.9 &\ldots& 257.5 & 272.5 &\ldots& 195.8 &\ldots& 239.8 &\ldots& 290.9 & 161.7 & 122.5 & 219.0 & 128.4 &\ldots& 248.4 &\ldots& 219.6 \\
Mg I & 4057.5 & 4.35 &$-$0.9 &\ldots& 33.4 & 123.0 &\ldots& 34.0 & 70.1 &\ldots&\ldots&\ldots&\ldots& 23.0 & 42.4 & 83.9 &\ldots& 14.6 &\ldots&\ldots& 10.0 & 18.3 &\ldots&\ldots&\ldots \\
\enddata
\tablecomments{\label{ew_tab_stub} Numbered columns refer to stars as follows:
1: BM235, 
2: HE 0033-2141, 
3: HE 0050$-$0918, 
4: HE 1210$-$2729, 
5: HE 2234$-$4757, 
6: BM028, 
7: BM209, 
8: BM308, 
9: HE 0012$-$5643, 
10: HE 0017-3646, 
11: HE 0048$-$1109, 
12: HE 0324$-$0122, 
13: BM056, 
14: BM121, 
15: HE~0007$-$1752,  
16: HE 0039$-$0216, 
17: HE 0429$-$4620,  
18: HE 1401$-$0010, 
19: HE 2138-0314, 
20: HE 2315$-$4306,  
21: HE 2319$-$5228, 
22: HE~2322$-$6125.
This Table is published in its entirety in the machine readable format. A portion is shown here for guidance regarding its form and content. }
\end{deluxetable*}

To determine stellar parameters and chemical abundances of 41 elements, 
we employ a 1D plane-parallel model atmosphere with $\alpha$-enhancement \citep{Castelli04}, and the 2017 version of the MOOG analysis code\footnote{https://github.com/alexji/moog17scat} \citep{Sneden73} with Rayleigh scattering included \citep{Sobeck11}. 
These tools have been integrated into an updated version of a custom-made analysis package\footnote{https://github.com/andycasey/smhr} first described in \citet{Casey14}. 

\begin{figure}[!ht]

  \includegraphics[clip=false,width=8.4cm 
   ]{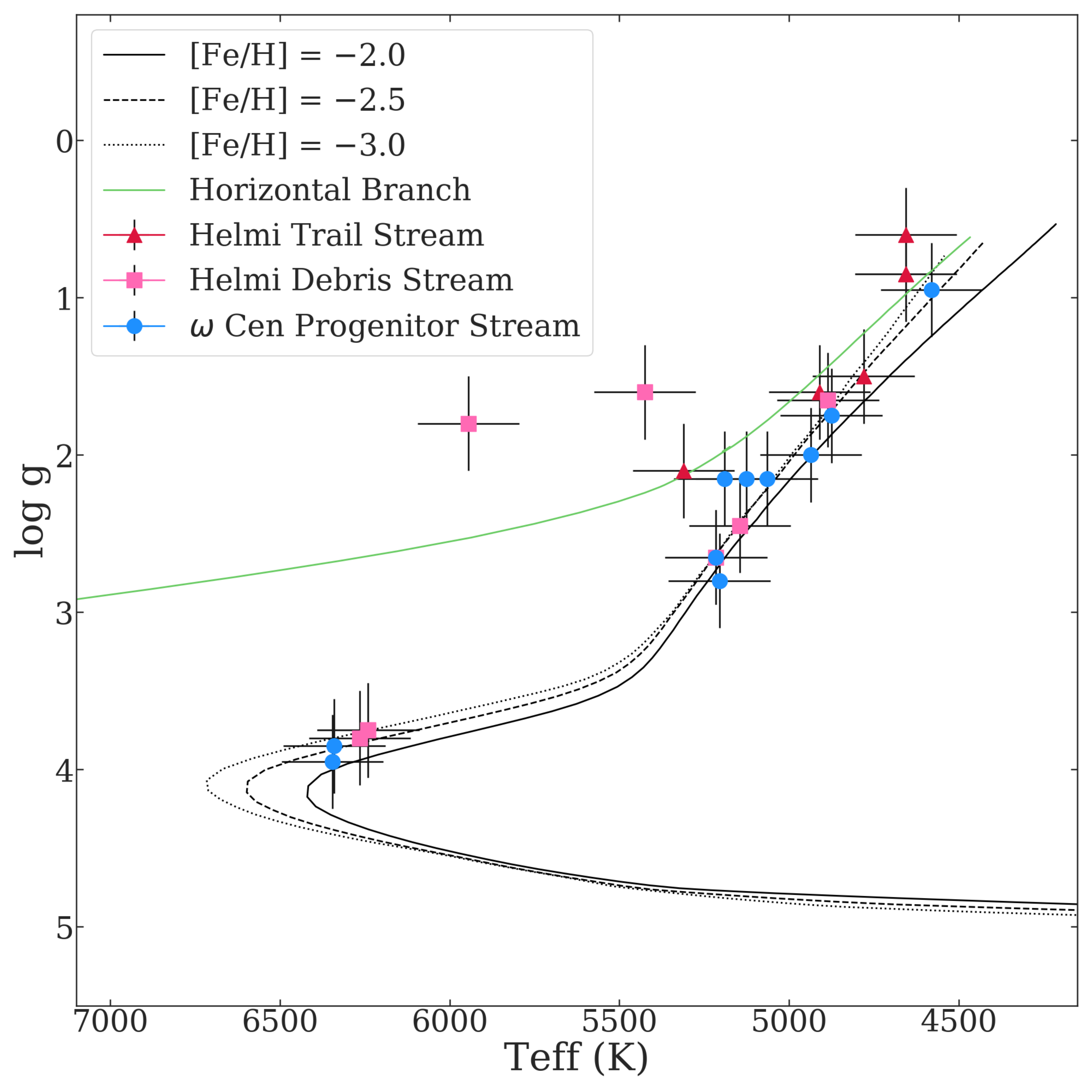} 
   \figcaption{\label{iso} Stellar parameters for all 21 stars, overplotted with 12\,Gyr isochrones of different metallicities. The green line shows a horizontal branch isochrone of 12\,Gyr \citep{Marigo17}.
   }

\end{figure}

We use Fe I and Fe II line measurements to determine the stellar parameters spectroscopically. We follow the procedure of \citet{Frebel13}. In an iterative process, we first derive metallicity, effective temperature (T$_{\rm eff}$), surface gravity ($\log$ {\it g}) and microturbulence (v$_{micr}$). We then apply T$_{\rm eff}$ corrections as described in \citeauthor{Frebel13}\ to rederive the stellar parameters. Table~\ref{stellpar} summarizes these results. Figure~\ref{iso} displays our adopted LTE stellar parameters overlaid with 12\,Gyr Y$^{2}$ isochrones with $\mbox{[Fe/H]} = -2.0$, $\mbox{[Fe/H]} = -2.5$, and $\mbox{[Fe/H]} = -3.0$ \citep{yy01}. The horizontal branch tracks are adopted from the PARSEC models \citep{Marigo17}\footnote{\url{http://stev.oapd.inaf.it/cgi-bin/cmd}}.

Our stellar parameters agree very well with the most metal-poor isochrone. Most stars are red giants.
Four are subgiants near the main sequence turnoff, and three stars (BM028, HE~0017$-$3646, and HE~0050$-$0918) are on the horizontal branch.  Two of the highly r-process-enhanced stars have been independently found by \citet{ezzeddine20}. \mbox{HE~0048$-$1109} and \mbox{HE~0007$-$1752}  (as \mbox{2MASS~J00512646$-$1053170} and \mbox{2MASS~J00101758$−$1735387}, respectively) are enhanced at the \mbox{$r$-II} level\footnote{See definition in Section~\ref{rproc_streams}.}, and our derived stellar parameters and abundances are in excellent agreement with their values.

\tabcolsep=0.11cm

\begin{deluxetable}{lrrrrrr}
\tablecaption{\label{stellpar} Stellar Parameters }
\tablewidth{0pc}
\tablehead{
\colhead{Star} &
\colhead{$\textrm{T}_\textrm{eff}$} &
\colhead{\logg} & \colhead{$\textrm{$v$}_\textrm{micr}$}&  \colhead{[Fe/H]} &\colhead{[C/Fe]$_{\rm corr}$} &\colhead{[N/Fe]$_{uncorr}$} \\
\colhead{}&\colhead{[K]} &
\colhead{[dex]} & \colhead{[\kms]}&  \colhead{[dex]}&  \colhead{[dex]} &  \colhead{[dex]}}
\startdata
 \multicolumn{7}{c}{Helmi trail stream}            \\\hline
BM235           & 4780 &    1.50 &  2.05 &  $-$1.63 & $-$0.04&0.30\\
HE~0033$-$2141	& 4910 &	1.60 &	1.70 &	$-$2.91 & 0.43&\nodata\\
HE~0050$-$0918	& 5310 &	2.10 &	2.00 &	$-$1.37 & $-$0.25&0.60\\
HE~1210$-$2729	& 4655 &	0.60 &	2.25 &	$-$3.01 & 0.36&1.38 \\
HE~2234$-$4757	& 4655 &	0.85 &	2.35 &	$-$2.83 & 0.77&0.93\\
\hline
 \multicolumn{7}{c}{Helmi debris stream}              \\\hline
BM028  &   5945 &   1.80 &   2.95 &$-$1.83& 0.38&0.70 \\
BM209  &   5215 &   2.65 &   1.45 &$-$2.27& 0.21&\nodata \\
BM308  &   4885 &   1.65 &   1.85 &$-$2.10& 0.31&\nodata \\
HE~0012$-$5643  &   6240 &   3.75 &  1.20 &  $-$3.03 & 1.22&\nodata \\
HE~0017$-$3646 &    5425 &   1.60 &  2.15 &  $-$2.45 & 0.08&\nodata \\
HE~0048$-$1109  &   6265 &   3.80 &   1.45&   $-$2.35& 0.61&\nodata \\
HE~0324$-$0122  &   5145 &   2.45 &  1.65 &   $-$2.41& 0.37&0.70 \\
\hline
 \multicolumn{7}{c}{$\omega$~Cen progenitor stream}              \\\hline
BM056      &       4875  &    1.75  &    1.85  & $-$2.02 & 0.34&0.40 \\ 
BM121      &       4580  &    0.95  &    2.55  & $-$2.48 & 0.37& 0.60 \\
HE~0007$-$1752 &   5215  &    2.65  &    1.45  & $-$2.36 & $-$0.29&\nodata \\
HE~0039$-$0216 &   6345  &    3.95  &    1.45  & $-$2.56 & 0.94&\nodata  \\
HE~0429$-$4620 &   5205  &    2.80  &    1.55  & $-$2.40 & 0.05& 0.70  \\
HE~1401$-$0010 &   6340  &    3.85  &    1.40  & $-$2.54 & 0.93&0.71  \\
HE~2138$-$0314 &   5190  &    2.15  &    1.75  & $-$3.08 & 0.95& 0.95  \\
HE~2315$-$4306 &   5125  &    2.15  &    1.75  & $-$2.39 & 0.31&\nodata  \\
HE~2319$-$5228 &   4935  &    2.00  &    1.85  & $-$3.18 & 1.63& 2.87 \\
HE~2322$-$6125 &   5065  &    2.15  &    1.65  & $-$2.61 & 0.30&\nodata  \\  
\enddata
\end{deluxetable}

\subsection{Abundance and stellar parameters uncertainties}
For each element, we estimate abundance uncertainties based on the spread in individual line abundances, and the data and fit quality. For elements with only one line available, we adopt a nominal uncertainty of 0.15\,dex. Overall, we adopt a nominal minimum uncertainty of 0.10\,dex for elements when calculated uncertainties indicate a value of $<$0.10\,dex.

Following \citet{Frebel13}, we assume that systematic uncertainty arises from uncertainty in the stellar parameters. We explore these effects for two representative sample stars, with a focus on elements Fe, Ba, La, Ce, Eu, Gd, and Dy to test for the robustness of the $r$-process patterns. We do so by changing one stellar parameter at a time by its typical uncertainty, which is 100\,K for T$_{\rm eff}$, 0.3\,dex for $\log g$, and 0.3\,\kms\ for $v_{micr}$. Those systemic uncertainties were chosen following \citet{Frebel10b} and adjusted for the better data quality available in this study.
The recorded changes in $\log\epsilon(X)$ for various elements are listed in Table~\ref{deltaparam} for the warmer star HE~0048$-$1109 with T$_{\rm eff} = 6265$\,K and $\mbox{[Fe/H]} =-2.35$, and the cooler star BM235 with T$_{\rm eff} = 4780$\,K and $\mbox{[Fe/H]} =-1.65$. We also add in quadrature our random uncertainties obtained from the abundance measurements (see Tables~\ref{hts} and \ref{hds}) to the uncertainties arising from the stellar parameter uncertainties to calculate the total sensitivity of our abundances.

Finally, we note that while other systematic uncertainties exist, arising from e.g., $\log gf$ uncertainties, choice of model atmospheres and radiative transfer codes, we do not further explore their influence on our final uncertainties as their impact is comparatively small compared to abundance uncertainties arising from the stellar parameters uncertainties.

\begin{deluxetable}{lrrrrr}
\tablecaption{\label{deltaparam} Abundance Uncertainties }
\tablewidth{0pc}
\tablehead{
\colhead{Element} & \colhead{Random} &
\colhead{$\Delta T_{eff}$} &
\colhead{$\Delta$ \logg} & 
\colhead{$\Delta$ $\textrm{$v$}_\textrm{micr}$}& \colhead{Root Mean}\\
\colhead{}& \colhead{error}&
\colhead{[$+$ 100 K]} &
\colhead{[$+$0.3 dex]} & 
\colhead{[$+$0.3 \kms]}&
\colhead{Square} 
} 
\startdata
 \multicolumn{6}{c}{HE~0048$-$1109}            \\\hline
Fe I	&0.12& 0.12 &	$-$0.02 &	$-$0.05& 0.18\\	
Ba II	&0.20& 0.02 &	0.04 &	$-$0.12&0.24 \\
La II	&0.10& 0.12 &	0.15 &	0.03& 0.22\\
Ce II	&0.30& 0.07 &    0.07 &	0.00 &0.32 \\
Eu II   &0.10& 0.05  &  0.05 &  $-$0.06& 0.14 \\
Gd II	&0.10& 0.11 &	0.09 &	$-$0.01 &0.18 \\
Dy II   &0.20& 0.11  &    0.08 & $-$0.02&0.24 \\
\hline
 \multicolumn{6}{c}{BM235}              \\\hline
Fe I	&0.15& 0.20 & $-$0.01 & $-$0.13& 0.28\\	
Ba II	&0.10& 0.10 &	0.15 &	0.03& 0.20 \\
La II	&0.10& 0.03 &	0.11 &	0.01& 0.16 \\
Ce II	&0.10& 0.06 &    0.12 &	$-$0.01& 0.17 \\
Eu II  &0.10&  0.08 &    0.15 &     0.03& 0.19 \\
Gd II	&0.15& 0.05 &	0.09 &	$-$0.01& 0.18 \\
Dy II   &0.15& 0.07 &    0.11 &  $-$0.03& 0.20\\
\enddata
\end{deluxetable}

For our discussion, we combine both our sample and the \citet{Roederer10} sample for a total of 17 stars. Two of our stars were also analyzed by \citeauthor{Roederer10}. These authors also determine spectroscopic stellar parameters but without any corrections or adjustments. This results in lower temperatures, albeit the effect is gradual, with cooler stars being much more affected than the warmer stars. This is reflected in the two examples discussed here. Overall, when factoring this systematic effect in, there is very good agreement between the stellar parameter determination of the two studies.

There is no significant discrepancy between our [Fe/H] and T$_{\rm eff}$ and abundance measurements for the warm subgiant BM028 (CD\,$-$36 1052 in \citeauthor{Roederer10}).  We derive 
[Fe/H] = $-1.83 \pm 0.10$, 
T$_{\rm eff} =5945 \pm 100$\,K, 
$\log g = 1.80 \pm 0.3$ and 
$v_{micr} = 2.95 \pm 0.3$\,\kms, 

whereas \citet{Roederer10} derived 
[Fe/H] = $-1.79 \pm 0.10$, 
T$_{\rm eff}=6070 \pm 200$\,K, 
$\log g = 2.30 \pm 0.3$ and 
$v_{micr} =3.20 \pm 0.3$\,\kms. 

For the red giant BM209 (HD~128279 in \citeauthor{Roederer10}), we do find some discrepancies in the stellar parameters, in line with the systematic differences in the analysis, as described above. We derive 
[Fe/H] = $-2.27 \pm 0.09$, 
T$_{\rm eff} =5215 \pm 100$\,K, 
$\log g = 2.65 \pm 0.3$ and 
$v_{micr}$ = $ 1.45 \pm 0.3 $\,\kms, 

whereas \citet{Roederer10} derived 
[Fe/H] = $-2.51 \pm 0.10$, 
T$_{\rm eff}=5050 \pm 200$\,K, 
$\log g = 2.35 \pm 0.3$ and 
$v_{micr} =1.50 \pm 0.3$\,\kms. 
The expected, lower T$_{\rm eff}$ by \citet{Roederer10} causes a lower surface gravity and [Fe/H].

\vspace{0.5cm}

\section{Chemical Abundances of Stellar Stream Stars}

As described in Section~\ref{sec3}, we obtain chemical abundance measurements for up to 41 elements using a combination of equivalent width analysis and spectrum synthesis for the 22 stars. The final abundances are summarized by stream in Tables~\ref{hts}, \ref{hds} and \ref{omg}. In the following, we present additional discussion on our abundance measurements of individual elements.

\startlongtable
\begin{deluxetable*}{lrrrrrrrrrrrrrrrrrr}

\tabletypesize{\tiny}
\tablewidth{0pc}
\tablecaption{\label{hts} Abundances of Helmi et al. trail stream, and one star in the Helmi et al.  Debris Stream}
\tablehead{
\colhead{} &
\colhead{} & \colhead{} & \colhead{}& \colhead{}& \colhead{}&\colhead{}&
\colhead{} & \colhead{} & \colhead{}&  \colhead{}& \colhead{} &\colhead{}&
\colhead{} & \colhead{} & \colhead{}& \colhead{}& \colhead{}}
\startdata
& 
\multicolumn{5}{c}{BM235}&\colhead{}& 
\multicolumn{5}{c}{HE~0033$-$2141}&\colhead{}&
\multicolumn{5}{c}{HE~0050$-$0918}&\colhead{}\\
\cline{2-6} \cline{8-12} \cline{14-18} \\
\colhead{Species} &\colhead{$\log\epsilon (\mbox{X})$} & \colhead{[X/H]} & \colhead{[X/Fe]}&  \colhead{$N$}& \colhead{$\sigma$} & &\colhead{$\log\epsilon (\mbox{X})$} & \colhead{[X/H]} & \colhead{[X/Fe]}&  \colhead{$N$}& \colhead{$\sigma$}&&\colhead{$\log\epsilon (\mbox{X})$} & \colhead{[X/H]} & \colhead{[X/Fe]}&  \colhead{$N$}& \colhead{$\sigma$}\\
\hline \\
CH	&	6.37	&	$-$2.06	&	$-$0.04	        &	2	&	0.10	&&		5.60	&	$-$2.83	&	0.43	&	3	&	0.10	&&	6.74	&	$-$1.69	&	$-$0.27	&	4	&	0.14\\
CN	&	6.60	&	$-$1.23	&	0.40		    &	1	&	0.30	&&	\nodata		&		\nodata	&		\nodata &	\nodata	&	\nodata	&&	7.06	&	$-$0.77	&	0.60		&	1	&	0.20\\
NH	&	6.40	&	$-$1.43	&	0.20		    &	1	&	0.30	&&	\nodata		&	\nodata		&	\nodata	&	\nodata	&	\nodata	&&	7.05	&	$-$0.78	&	0.59		&	1	&	0.20\\
O I 	&	7.79	&	$-$0.90	&	0.73		&	1	&	0.15	&&	\nodata		&	\nodata		&	\nodata	&	\nodata	&	\nodata	&&	\nodata	&	\nodata		&	\nodata		&	\nodata	& \nodata	\\
Na I	&	4.66	&	$-$1.58	&	0.05		&	4	&	0.15	&&		3.71	&	$-$2.53	&	0.38	&	2	&	0.10	&&	5.25	&	$-$0.99	&	0.38		&	2	&	0.10\\
Mg I	&	6.29	&	$-$1.31	&	0.31		&	8	&	0.11	&&		5.16 	&	$-$2.44	&	0.47	&	10	&	0.16	&&	6.47	&	$-$1.13	&	0.25		&	7	&	0.15 \\
Al I	&	\nodata		&		\nodata	&		\nodata &	\nodata	&	\nodata	&&		2.57	&	$-$3.88 	& $-$0.97 & 	1 	& 	0.15	&&	\nodata	&	\nodata		&	\nodata		&	\nodata	& \nodata	\\
Si I	&	6.22	&	$-$1.29	&	0.34		&	6	&	0.17	&&		5.27 	&	$-$2.24	&	0.67	&	1	&	0.15	&&	6.52	&	$-$0.99		&	0.39		&	4	&	0.15 \\
Ca I	&	5.10	&	$-$1.24	&	0.39		&	23	&	0.19	&&		3.67	& 	$-$2.67 	& 	0.24 & 	19 	& 	0.11	&&	5.36	&	$-$0.98	&	0.40		&	22	&	0.15\\
Sc II	&	1.82	&	$-$1.33	&	0.30		&	9	&	0.17	&&		0.18 	& 	$-$2.97 	& $-$0.06 & 	7	&	0.14	&&	1.12	&	$-$1.11	&	0.26		&	12	&	0.14\\
Ti I	&	3.59	&	$-$1.36	&	0.27		&	28	&	0.16	&&		2.04 	& 	$-$2.91 	&	0.00	& 	16 	&	0.13	&&	3.86	&	$-$1.09	&	0.28		&	26	&	0.17\\
Ti II	&	3.71	&	$-$1.24	&	0.39		&	38	&	0.17	&&		2.01 	& 	$-$2.94 	&$-$0.03	& 	42 	&	0.15	&&	4.10	&	$-$0.85	&	0.52		&	40	&	0.18\\
V II	&	2.21	&	$-$1.72	&	$-$0.09	    &	3	&	0.10	&&		0.80 	& 	$-$3.13 	&$-$0.22	& 	1 	&	0.15	&&	2.58	&	$-$1.35	&	0.02		&	3	&	0.12\\
Cr I	&	3.96	&	$-$1.68	&	$-$0.05	    &	20	&	0.17	&&		2.26 	& 	$-$3.38 	&$-$0.47	& 	8 	&	0.16	&&	4.17	&	$-$1.47	&	$-$0.10	&	18	&	0.12\\
Cr II	&	4.14	&	$-$1.50	&	0.12		&	3	&	0.10	&&		2.92 	& 	$-$2.72 	&	 0.19	& 	2 	&	0.20	&&	4.53	&	$-$1.11	&	0.26		&	3	&	0.10\\
Mn I	&	3.35	&	$-$2.08	&	$-$0.45	    &	9	&	0.15	&&		2.02 	& 	$-$3.41 	&$-$0.50	& 	5 	&	0.10	&&	3.83	&	$-$1.60	&	$-$0.22	&	6	&	0.10\\
Fe I	&	5.87	&	$-$1.63	&	0.00		&	231	&	0.15	&&		4.59 	& 	$-$2.91 	&	0.00	& 	170 	&	0.12	&&	6.13	&	$-$1.37	&	0.00		&	220	&	0.12\\
Fe II	&	5.84	&	$-$1.66	&	$-$0.03	    &	28	&	0.12	&&		4.58 	& 	$-$2.92 	&$-$0.01	& 	17 	&	0.10	&&	6.13	&	$-$1.37	&	0.00		&	21	&	0.10\\
Co I	&	3.29	&	$-$1.70	&	$-$0.07	    &	5	&	0.16	&&		2.16 	& 	$-$2.83 	&	0.08	& 	7 	&	0.10	&&	3.57	&	$-$1.42	&	$-$0.05	&	7	&	0.16\\
Ni I	&	4.55	&	$-$1.67	&	$-$0.04	    &	27	&	0.15	&&		3.05 	& 	$-$3.17 	&$-$0.26	& 	8 	&	0.13	&&	4.86	&	$-$1.36	&	0.01		&	22	&	0.16\\
Cu I	&	2.16	&	$-$2.03	&	$-$0.40	    &	2	&	0.17	&&	\nodata		&		\nodata	&		\nodata &	\nodata	&	\nodata	&&	2.31 &	$-$1.88	&	$-$0.51 	&	1 	&	0.15\\
Zn I	&	2.87	&	$-$1.69	&	$-$0.06	    &	2	&	0.10	&&		1.85 	&	 $-$2.71	&	0.20 &	 1 	&	0.10	&&	3.16	&	$-$1.40	&	$-$0.02	&	2	&	0.10\\
Sr II	&	1.33	&	$-$1.54	&	0.09		&	2	&	0.15	&&	$-$0.72 	& 	$-$3.59 	&$-$0.68 	&	 2 	&	0.20	&&	1.73	&	$-$1.14	&	0.23		&	3	&	0.10\\
Y II	&	0.30	&	$-$1.91	&	$-$0.28	    &	2	&	0.10	&&	$-$1.77 	& 	$-$3.98 	&$-$1.07 	&	 2	&	0.20	&&	0.83	&	$-$1.38	&	$-$0.01	&	3	&	0.10\\
Zr II	&	1.32	&	$-$1.26	&	0.37		&	7	&	0.20	&&	$-$0.57	&	$-$3.15	&$-$0.24	&	2	&	0.20	&&	1.75 &	$-$0.83	&	0.54 		&	2 	&	0.10\\
Ba II	&	0.63	&	$-$1.55	&	0.08		&	5	&	0.10	&&	$-$1.90	&	$-$4.08	&$-$1.17	&	3	&	0.20	&&	0.92	&	$-$1.26	&	0.11		&	5	&	0.15\\
La II	&$-$0.23	&	$-$1.33	&	0.29		&	14	&	0.10	&&	\nodata		&		\nodata	&		\nodata &	\nodata	&	\nodata	&&	0.10 &	$-$1.00	&	0.37 		&	12 	&	0.10 \\
Ce II	&$-$0.02	&	$-$1.60	&	0.03		&	7	&	0.10	&&	\nodata		&		\nodata	&		\nodata &	\nodata	&	\nodata	&&	0.32 &	$-$1.26	&	0.12 		&	3 	&	0.15\\
Pr II	&$-$0.49	&	$-$1.21	&	0.42		&	3	&	0.10	&&	\nodata		&		\nodata	&		\nodata &	\nodata	&	\nodata	&&$-$0.15&	$-$0.87	&	0.50		&	7	&	0.15\\
Nd II	&	0.20	&	$-$1.22	&	0.41		&	39	&	0.20	&&	\nodata		&		\nodata	&		\nodata &	\nodata	&	\nodata	&&	0.63	&	$-$0.79	&	0.58		&	27	&	0.15\\
Sm II	&$-$0.18	&	$-$1.14	&	0.49		&	22	&	0.20	&&	\nodata		&		\nodata	&		\nodata &	\nodata	&	\nodata	&&	0.20	&	$-$0.76	&	0.61		&	23	&	0.10\\
Eu II	&$-$0.63	&	$-$1.15	&	0.48		&	4	&	0.10	&&	$<-$2.50 	& 	$<-$3.02 	&$<-0.11$& 		&		&&$-$0.14&	$-$0.66	&	0.71		&	4	&	0.15\\
Gd II	&	-0.05	&	$-$0.93	&	0.70		&	5	&	0.15	&&	\nodata		&		\nodata	&		\nodata &	\nodata	&	\nodata	&&	0.33	&	$-$0.74	&	0.63		&	4	&	0.15\\
Dy II	&	0.00	&	$-$0.81	&	0.82		&	2	&	0.15	&&	$<-$1.96	&	$<-$3.06	&$<-$0.15&		&		&&	0.38	&	$-$0.72	&	0.66		&	2	&	0.15\\
Ho II	&$-$0.47	&	$-$0.95	&	0.68		&	2	&	0.15	&&	\nodata		&		\nodata	&		\nodata &	\nodata	&	\nodata	&&	\nodata	&	\nodata		&	\nodata		&	\nodata	& \nodata	\\
Er II	&$-$0.18	&	$-$1.10	&	0.53		&	2	&	0.15	&&	\nodata		&		\nodata	&		\nodata &	\nodata	&	\nodata	&&	0.28	&	$-$0.64	&	0.74		&	1	&	0.20\\
Hf II	&$-$0.50	&	$-$1.35	&	0.28		&	1	&	0.20	&&	\nodata		&		\nodata	&		\nodata &	\nodata	&	\nodata	&&	\nodata	&	\nodata		&	\nodata		&	\nodata	& \nodata	\\
Os I	&	0.33	&	$-$1.07	&	0.56		&	1	&	0.15	&&	\nodata		&		\nodata	&		\nodata &	\nodata	&	\nodata	&&	0.65	&	$-$0.75	&	0.62		&	1	&	0.20\\
Pb I	&	0.55	&	$-$1.20	&	0.43		&	1	&	0.20	&&	\nodata		&		\nodata	&		\nodata &	\nodata	&	\nodata	&&	\nodata	&	\nodata		&	\nodata		&	\nodata	& \nodata	\\
Th II	&$-$1.19	&	$-$1.21	&	0.42		&	1	&	0.10	&&	\nodata		&		\nodata	&		\nodata &	\nodata	&	\nodata	&& $-$0.71&	$-$0.73	&	0.64		&	1	&	0.10\\\\
\\\\
\\
& \multicolumn{5}{c}{HE~1210$-$2729}&\colhead{}& 
\multicolumn{5}{c}{HE~2234$-$4757}&\colhead{}&
\multicolumn{5}{c}{BM028}&\colhead{}\\
\cline{2-6} \cline{8-12} \cline{14-18} \\
\colhead{Species} &\colhead{$\log\epsilon (\mbox{X})$} & \colhead{[X/H]} & \colhead{[X/Fe]}&  \colhead{$N$}& \colhead{$\sigma$} & &\colhead{$\log\epsilon (\mbox{X})$} & \colhead{[X/H]} & \colhead{[X/Fe]}&  \colhead{$N$}& \colhead{$\sigma$}&&\colhead{$\log\epsilon (\mbox{X})$} & \colhead{[X/H]} & \colhead{[X/Fe]}&  \colhead{$N$}& \colhead{$\sigma$}\\
\hline \\
CH	&	5.05 	& 	$-$3.38 	&	0.36	 	&	2	&	0.10	&&		5.70	&	$-$2.73	&	0.77	&	5	&	0.10	&&	6.80 &	$-$1.63 	&  	0.38    	&   	2  	&   0.20\\
CN	&	6.20	&	$-$1.63	&	1.38		&	1	&	0.20	&&		5.87	&	$-$1.96	&	0.86	&	1	&	0.20	&&	\nodata	&	\nodata		&	\nodata		&	\nodata	& \nodata	\\
NH	&	\nodata		&		\nodata	&		\nodata &	\nodata	&	\nodata	&&		6.00	&	$-$1.83	&	1.00	&		&		&&	6.69 &  	 $-$1.13 	&   	0.70		&	1	&   0.30\\
O I &	\nodata		&		\nodata	&		\nodata &	\nodata	&	\nodata	&&	\nodata		&		\nodata	&		\nodata &	\nodata	&	\nodata	&&	\nodata	&	\nodata		&	\nodata		&	\nodata	& \nodata	\\
Na I	&	3.75 	& 	$-$2.49	&	0.52		&	3	&	0.10	&&		3.57 &	 $-$2.67	&	0.16 &	 4 	& 	0.12	&&	5.45 &   	$-$0.79 	&   	1.04    	&   	2   	&   0.15\\
Mg I	&	5.30 	& 	$-$2.30 	& 	0.71 		& 	9	&	0.15	&&		5.12 &	$-$2.48	&	0.34	&	12	&	0.14	&&	6.09 &   	$-$1.51 	&   	0.32    	&   	7   	&   0.12  \\
Al I	&	3.00 	& 	$-$3.45 	& 	$-$0.44 	& 	1	& 	0.15	&&		3.12	&	$-$3.33	&$-$0.51	&	1	&	0.15	&&	4.00 &   	$-$2.45 	&   	$-$0.56 	&   	1   	&   0.15	\\
Si I	&	5.49 	& 	$-$2.02 	& 	1.00 		& 	3	& 	0.15	&&		5.08	&	$-$2.43	&	0.39	&	3	&	0.10	&&	6.06 &   	$-$1.45 	&   	0.38    	&   	3   	&   0.10	\\
Ca I	&	3.72 	& 	$-$2.62 	& 	0.39 		& 	25	& 	0.16	&&		3.65	&	$-$2.69	&	0.14	&	22	&	0.15	&&	4.85 &   	$-$1.49 	&   	0.34    	&   	24 	&   0.10 \\
Sc II	&	0.24 	& 	$-$2.91 	& 	0.10 		& 	8	& 	0.12	&&		0.18 	&	$-$2.97	&$-$0.14	&	 7 	&	0.15	&&	1.31 &   	$-$1.84 	&   	$-$0.00 	&   	7   	&   0.12\\
Ti I	&	2.31 	& 	$-$2.64 	& 	0.37 		& 	22	& 	0.12	&&		2.21	&	$-$2.74	&	0.09	&	17	&	0.17	&&	3.42 &   	$-$1.53 	&   	0.30    	&   	18 	&   0.15\\
Ti II	&	2.30 	& 	$-$2.65 	& 	0.37 		& 	52	& 	0.15	&&		2.24	&	$-$2.71	&	0.12	&	47	&	0.16	&&	3.57 &   	$-$1.38	&   	0.45    	&   	13 	&   0.15\\
V II	&	1.17 	& 	$-$2.76 	& 	0.25 		& 	3	& 	0.10	&&		1.09	&	$-$2.84	&$-$0.01	&	3	&	0.10	&&	2.29 &   	$-$1.64	&   	0.19    	&   	4   	&   0.15 \\
Cr I	&	2.52 	& 	$-$3.12 	&	$-$0.10 	&	17	& 	0.16	&&		2.60	&	$-$3.04	&$-$0.22	&	17	&	0.16	&&	3.83 &   	$-$1.81 	&   	0.02    	&   	14 	&   0.15\\
Cr II	&	2.91 	& 	$-$2.73 	&	0.28 		& 	2	& 	0.10	&&		3.01	&	$-$2.63	&	0.20	&	2	&	0.10	&&	3.92 &   	$-$1.72 	&   	0.11    	&   	3   	&   0.10\\
Mn I	&	2.18 	& 	$-$3.25 	&	$-$0.24 	& 	5	& 	0.10	&&		2.05	&	$-$3.38	&$-$0.56	&	4	&	0.18	&&	3.19 &   	$-$2.24 	&	$-$0.41    	&   	6   	&   0.15\\
Fe I	&	4.49	&	$-$3.01	&	0.00		&	196	&	0.10	&&		4.67	&	$-$2.83	&	0.00	&	228	&	0.14	&&	5.67 &   	$-$1.83 	&	0.00    	&   	228 	&   0.11\\
Fe II	&	4.45	&	$-$3.05	&	$-$0.04	&	23	&	0.09	&&		4.62	&	$-$2.88	&$-$0.05	&	27	&	0.12	&&	5.64 &   	$-$1.86 	&	$-$0.02    	&   	26 	&   0.10\\
Co I	&	2.13	&	$-$2.86	&	0.16		&	8	&	0.10	&&		2.05	&	$-$2.94	&$-$0.12	&	8	&	0.12	&&	3.11 &  	$-$1.88 	&	$-$0.05    	&   	3   	&   0.10\\
Ni I	&	3.47	&	$-$2.75	&	0.26		&	13	&	0.15	&&		3.40	&	$-$2.82	&	0.01	&	15	&	0.16	&&	4.49 &   	$-$1.73 	&   	0.10    	&   	8   	&   0.15\\
Zn I	&	1.94	&	$-$2.62	&	0.39		&	2	&	0.10	&&		1.99	&	$-$2.57	&	0.26	&	3	&	0.10	&&	2.81 &   	$-$1.75 	&   	0.08    	&   	2  	&   0.10\\
Sr II	&$-$0.46	&	$-$3.33	&	$-$0.32	&	2	&	0.10	&&		0.17	&	$-$2.70	&	0.12	&	2	&	0.15	&&	1.54 &   	$-$1.33 	&   	0.51    	&   	3  	&   0.15\\
Y II	&$-$1.34	&	$-$3.55	&	$-$0.53	&	3	&	0.15	&&	$-$0.61	&	$-$2.83	&	0.00	&	3	&	0.17	&&	0.10 &   	$-$2.11 	&	$-$0.28    	&   	3  	&   0.10\\
Zr II	&$-$0.29	&	$-$2.88	&	0.14		&	1	&	0.30	&& 		0.27	&	$-$2.31	&	0.52	&	2	&	0.15	&&	0.90 &   	$-$1.68 	&  	0.16    	&   	5  	&   0.10\\
Ba II	&$-$1.88	&	$-$4.05	&	$-$1.04	&	2	&	0.20	&&	$-$0.60	&	$-$2.78	&	0.05	&	5	&	0.19	&&	0.47 &   	$-$1.71 	&  	0.12    	&   	5  	&   0.20\\
La II	&$<-$2.27	&	$<-$3.37	&	$<-$0.35	&		&		&&	$-$1.36	&	$-$2.46	&	0.37	&	6	&	0.10	&&$-$0.68&   	$-$1.78 	&   	0.05    	&   	5  	&   0.10 \\
Ce II&	\nodata		&		\nodata	&		\nodata &	\nodata	&	\nodata	&&	$-$1.21	&	$-$2.79	&	0.05	&	3	&	0.15	&&$-$0.45&   	$-$2.03 	&	$-$0.20    	&   	1  	&   0.20\\
Pr II	&$-$2.41	&	$-$3.13	&	$-$0.11	&	2	&	0.20	&&	$-$1.34	&	$-$2.06	&	0.14	&	6	&	0.15	&&$-$0.94&   	$-$1.66 	&  	0.17    	&   	2  	&   0.15\\
Nd II	&$<-$1.82	&	$<-$3.24	&	$<-$0.23	&		&		&&	$-$0.91	&	$-$2.33	&	0.50	&	15	&	0.19	&&$-$0.23&   	$-$1.65 	&   	0.18    	&   	3  	&   0.15\\
Sm II	&$<-$1.65	&	$<-$2.61	&	$<$0.40	&		&		&&	$-$1.19	&	$-$2.15	&	0.68	&	6	&	0.16	&&$-$0.57&   	$-$1.53 	&   	0.30    	&   	1  	&   0.20\\
Eu II	&$-$2.95 	&	$-$3.47 	&	$-$0.46 	&	2	& 	0.10	&&	$-$1.59	&	$-$2.11	&	0.72	&	5	&	0.10	&&$-$0.99&   	$-$1.51 	&   	0.33    	&   	5  	&   0.10\\
Gd II	&	\nodata		&		\nodata	&		\nodata &	\nodata	&	\nodata	&&	$-$1.08	&	$-$2.15	&	0.67	&	2	&	0.10	&&$<$0.09&  	 $<-$0.98	&	$<$0.85    &   	    	&        \\
Tb II   &	\nodata		&		\nodata	&		\nodata &	\nodata	&	\nodata	&&	\nodata		&		\nodata	&		\nodata &	\nodata	&	\nodata	&&$-$1.00&   	$-$1.30 	&   	0.53    	&   	1   	&   0.30 \\
Dy II	&	\nodata		&		\nodata	&		\nodata &	\nodata	&	\nodata	&&	$-$0.86	&	$-$1.96	&	0.87	&	1	&	0.30	&&$-$0.61&   	$-$1.71 	&   	0.12    	&   	1   	&   0.30\\
Ho II	&	\nodata		&		\nodata	&		\nodata &	\nodata	&	\nodata	&& 	$-$1.59	&	$-$2.07	&	0.76	&	2	&	0.20	&&$-$1.05&   	$-$1.53 	&   	0.30    	&   	2   	&   0.10\\
Er II	&	\nodata		&		\nodata	&		\nodata &	\nodata	&	\nodata	&&	$-$1.15	&	$-$2.07	&	0.76	&	1	&	0.30	&&$-$0.69&   	$-$1.61 	&   	0.22    	&   	1   	&   0.30\\
Tm II   &	\nodata		&		\nodata	&		\nodata &	\nodata	&	\nodata	&&	\nodata		&		\nodata	&		\nodata &	\nodata	&	\nodata	&&$<-$1.00& 	 $<-$1.10 &	$<$0.73    &       	&        \\
Hf II	&	\nodata		&		\nodata	&		\nodata &	\nodata	&	\nodata	&&	\nodata		&		\nodata	&		\nodata &	\nodata	&	\nodata	&&	\nodata	&	\nodata		&	\nodata		&	\nodata	& \nodata	\\
Os I	&	\nodata		&		\nodata	&		\nodata &	\nodata	&	\nodata	&&	$-$0.93	&	$-$2.33	&	0.50	&	1	&	0.20	&&	\nodata	&	\nodata		&	\nodata		&	\nodata	& \nodata	\\
Ir I 	&	\nodata		&		\nodata	&		\nodata &	\nodata	&	\nodata	&&	$-$0.55	&	$-$1.93	&	0.90	&	1	&	0.20 	&&	\nodata	&	\nodata		&	\nodata		&	\nodata	& \nodata	\\
Pb I	&	\nodata		&		\nodata	&		\nodata &	\nodata	&	\nodata	&&	\nodata		&		\nodata	&		\nodata &	\nodata	&	\nodata	&&	\nodata	&	\nodata		&	\nodata		&	\nodata	& \nodata	\\
Th II	&	\nodata		&		\nodata	&		\nodata &	\nodata	&	\nodata	&&	\nodata		&		\nodata	&		\nodata &	\nodata	&	\nodata	&&	\nodata	&	\nodata		&	\nodata		&	\nodata	& \nodata	\\
\enddata
 \tablecomments{
[X/Fe] ratios are computed with [Fe\,I/H] abundances
of the respective stars. Solar abundances have been taken from \citet{Asplund09}. For abundances measured from only one line, we adopt a nominal uncertainty of 0.10\,dex. } 
\end{deluxetable*}
\pagebreak

\newpage

\startlongtable
\begin{deluxetable*}{lrrrrrrrrrrrrrrrrr}

\tabletypesize{\tiny}
\tablewidth{0pc}
\tablecaption{\label{hds} Abundances of Helmi et al. Debris Stream}
\tablehead{
\colhead{} &
\colhead{} & \colhead{} & \colhead{}& \colhead{}& \colhead{}& \colhead{}&
\colhead{} & \colhead{} & \colhead{}&  \colhead{}& \colhead{}& \colhead{}&
\colhead{} & \colhead{} & \colhead{}& \colhead{}& \colhead{}}
\startdata
\colhead{}&
\multicolumn{5}{c}{BM209}&\colhead{}& 
\multicolumn{5}{c}{BM308}&\colhead{}& 
\multicolumn{5}{c}{HE~0012$-$5643}\\
\cline{2-6} \cline{8-12} \cline{14-18} \\
\colhead{Species} &
\colhead{$\log\epsilon (\mbox{X})$} & \colhead{[X/H]} & \colhead{[X/Fe]}& \colhead{$N$}& \colhead{$\sigma$}&\colhead{}&
\colhead{$\log\epsilon (\mbox{X})$} & \colhead{[X/H]} & \colhead{[X/Fe]}&  \colhead{$N$}& \colhead{$\sigma$} &\colhead{}&
\colhead{$\log\epsilon (\mbox{X})$} & \colhead{[X/H]} & \colhead{[X/Fe]}& \colhead{$N$}& \colhead{$\sigma$}\\
\hline \\
CH      & 6.35  & $-$2.08 & 0.21 &2 &0.10  &&   6.29    &   $-$2.14 &   0.31    &   2   &   0.10    &&   6.62 & $-$1.81 & 1.22 & 2 & 0.12 \\
NH      &	\nodata		&		\nodata	&		\nodata &	\nodata	&	\nodata	&&  5.73    &   $-$2.10 &   0.00    &   1   &   0.30   &&	\nodata	&	\nodata		&	\nodata		&	\nodata	& \nodata	\\
CN      & 6.20  &$-$1.63&1.38&1&0.20 &&   5.83    &   $-$2.00 &   0.10    &   1   &   0.30    &&	\nodata	&	\nodata		&	\nodata		&	\nodata	& \nodata	\\
Na I    & 4.17  &$-$2.07&0.21&3&0.10 &&   4.18    &   $-$2.06 &   0.03    &   2   &   0.10    &&  2.69 & $-$3.55 &$-$0.52 &2 & 0.15\\
Mg I    & 5.58  &$-$2.02&0.25&4&0.15 &&  5.65    &   $-$1.95 &   0.15    &   7   &   0.10    &&  4.59  & $-$3.01 &  0.02 & 5 & 0.15\\
Al I    & 3.42  &$-$3.03&$-$0.75&1&0.15  &&	\nodata		&		\nodata	&		\nodata &	\nodata	&	\nodata	&& 2.53 & $-$3.92 & $-$0.89 & 1 & 0.15 \\
Si I    & 5.64  &$-$1.87&0.40&2&0.15  &&5.74    &   $-$1.77 &   0.32    &   2   &   0.10    &&  4.60 & $-$2.91 & 0.12 & 1& 0.15\\
Ca I    & 4.43 &$-$1.91&0.37&18&0.10 && 4.47    &   $-$1.87 &   0.23    &   23  &   0.10    &&  3.65 & $-$2.69 & 0.35 & 10 & 0.10\\
Sc II   & 0.79  &$-$2.36&$-$0.09&7&0.15 && 0.91    &   $-$2.24 &   $-$0.14 &   10  &   0.10    && 0.15 & $-$3.00 & 0.04 & 2 & 0.13\\
Ti I    & 2.95  &$-$2.00&0.27&14&0.15  &&2.90    &   $-$2.05 &   0.10    &   31  &   0.15    &&	\nodata	&	\nodata		&	\nodata		&	\nodata	& \nodata	\\
Ti II   & 2.95  &$-$2.00&0.27&16&0.15  &&3.11    &   $-$1.84 &   0.25    &   48  &   0.15    && 2.21 & $-$2.74 & 0.30 & 13 & 0.10\\
V II    & 1.63  &$-$2.30&$-$0.03&1&0.15 && 1.87    &   $-$2.06 &   0.03    &   3   &   0.10    &&	\nodata	&	\nodata		&	\nodata		&	\nodata	& \nodata	\\
Cr I    & 3.14   &$-$2.18&$-$0.23&10&0.10 &&  3.33    &   $-$2.31 &   $-$0.22 &   19  &   0.10    && 2.51 & $-$3.13 & $-$0.10 & 6 & 0.10 \\
Cr II   & 3.46  &$-$2.18&0.09&1&0.15 &&  3.72    &   $-$1.92 &   0.18    &   3   &   0.15    &&	\nodata	&	\nodata		&	\nodata		&	\nodata	& \nodata	\\
Mn I    & 2.63  &$-$2.80&$-$0.53&6&0.10 &&   2.76    &   $-$2.67 &   $-$0.57 &   14  &   0.10    && 1.78 & $-$3.65 & $-$0.61 & 2 & 0.10\\
Fe I    & 5.23&$-$2.27&0.00&191&0.13 &&  5.40    &   $-$2.10 &   0.00    &   210 &   0.10    &&  4.47 & $-$3.03 & 0.00 & 87 & 0.13 \\
Fe II   & 5.21&$-$2.29&$-$0.02&19&0.09 &&  5.39    &   $-$2.11 &   $-$0.01 &   28  &   0.10    && 4.48 & $-$3.02 & 0.01 & 8 & 0.13\\
Co I    & 2.88&$-$2.11&0.16&2&0.10 &&  2.64    &   $-$2.35 &   $-$0.25 &   9   &   0.15    &&	\nodata	&	\nodata		&	\nodata		&	\nodata	& \nodata	\\
Ni I    & 3.69&$-$2.26&0.01&18&0.15 && 3.95    &   $-$2.27 &   $-$0.17 &   25  &   0.15    &&   3.13 & $-$3.09 & $-$0.06 & 5 & 0.08 \\
Cu I    &	\nodata		&		\nodata	&		\nodata &	\nodata	&	\nodata	&&  1.45    &   $-$2.74 &   $-$0.64 &   1   &   0.15   &&	\nodata	&	\nodata		&	\nodata		&	\nodata	& \nodata	\\
Zn I    & 2.30&$-$2.26&0.01&2&0.15&&  2.30    &   $-$2.26 &   $-$0.16 &   2   &   0.10    &&	\nodata	&	\nodata		&	\nodata		&	\nodata	& \nodata	\\
Sr II   & 0.20&$-$2.67&$-$0.40&2&0.10 && 0.63    &   $-$2.24 &   $-$0.14 &   2   &   0.10    &&   $-$0.32 & $-$3.19 & $-$0.16 & 2 & 0.10\\
Y II    &$-$0.82&$-$3.03&$-$0.76&1&0.20  && $-$0.47 &   $-$2.68 &   $-$0.58 &   3   &   0.15    && $<$0.29  &$<-$1.92 &$<$1.12  &   & \\
Zr II   &  $<$0.71&$<-$1.87&$<$0.41&\nodata	&	\nodata		&& 0.53    &   $-$2.05 &   0.05    &   4   &   0.15    &&  $<$0.80 &$<-$1.78 & $<$1.26 &   &     \\
Ba II   &$-$0.53&$-$2.71&$-$0.44&3&0.11 &&  $-$0.17 &   $-$2.35 &   $-$0.26 &   5   &   0.10    &&$<-$0.50 &$<-$2.45 &$<$0.55 &   &      \\
La II   &$-$1.26& $-$2.36&$-$0.08&2&0.15  &&  $-$0.95 &   $-$2.05 &   $-$0.04 &   8   &   0.10    &&	\nodata	&	\nodata		&	\nodata		&	\nodata	& \nodata	\\
Ce II   &$<-$0.49&$<-$2.07&$<$0.21&&  && $-$0.62 &   $-$2.20 &   $-$0.11 &   4   &   0.10   &&	\nodata	&	\nodata		&	\nodata		&	\nodata	& \nodata	\\
Pr II   &  $-$1.12&$-$1.84&0.44&2&0.10 && $-$1.03 &   $-$1.75 &   0.35    &   3   &   0.10    &&	\nodata	&	\nodata		&	\nodata		&	\nodata	& \nodata	\\
Nd II   &  $-$0.49&$-$1.91&0.36&1&0.10 && $-$0.48 &   $-$1.90 &   0.20    &   18  &   0.11    &&	\nodata	&	\nodata		&	\nodata		&	\nodata	& \nodata	\\
Sm II   &$<-$0.94&$<-$1.90&$<$0.37 && &&  $-$0.80 &   $-$1.76 &   0.34    &   15  &   0.15    && $<$0.50 &$<-$0.46 &$<$2.57 \\
Eu II   &$-$1.53&$-$2.05&0.22&3&0.10  && $-$1.26 &   $-$1.78 &   0.32    &   4   &   0.10    &&  $<-$0.86 &$<-$1.38 &$<$1.65 \\
Gd II  &	\nodata		&		\nodata	&		\nodata &	\nodata	&	\nodata	&& $-$0.60 &   $-$1.67 &   0.42    &   4   &   0.30   &&	\nodata	&	\nodata		&	\nodata		&	\nodata	& \nodata	\\
Tb II   &	\nodata		&		\nodata	&		\nodata &	\nodata	&	\nodata	&&$-$1.24 &   $-$1.54 &   0.55    &   1   &   0.30    &&	\nodata	&	\nodata		&	\nodata		&	\nodata	& \nodata	\\
Dy II &	\nodata		&		\nodata	&		\nodata &	\nodata	&	\nodata	&&$-$0.48 &   $-$1.58 &   0.51    &   2   &   0.30   &&	\nodata	&	\nodata		&	\nodata		&	\nodata	& \nodata	\\
Ho II  &	\nodata		&		\nodata	&		\nodata &	\nodata	&	\nodata	&&$-$1.12 &   $-$1.60 &   0.50    &   2   &   0.15    &&	\nodata	&	\nodata		&	\nodata		&	\nodata	& \nodata	\\
Er II   &	\nodata		&		\nodata	&		\nodata &	\nodata	&	\nodata	&&$-$0.79 &   $-$1.71 &   0.38    &   1   &   0.15    &&	\nodata	&	\nodata		&	\nodata		&	\nodata	& \nodata	\\
Tm II   &	\nodata		&		\nodata	&		\nodata &	\nodata	&	\nodata	&&	\nodata		&		\nodata	&		\nodata &	\nodata	&	\nodata	&&	\nodata	&	\nodata		&	\nodata		&	\nodata	& \nodata	\\
Os I    & $<$0.60&$<-$0.80&$<$1.47 &\nodata	&	\nodata		&&$-$0.40  &   $-$1.80 &   0.30    &   1   &   0.30    &&	\nodata	&	\nodata		&	\nodata		&	\nodata	& \nodata	\\
\hline \\
\colhead{}&
\multicolumn{5}{c}{HE~0017$-$3646}&\colhead{}& 
\multicolumn{5}{c}{HE~0048$-$1109}&\colhead{}& 
\multicolumn{5}{c}{HE~0324$-$0122}\\
\cline{2-6} \cline{8-12} \cline{14-18} \\
\colhead{Species} &
\colhead{$\log\epsilon (\mbox{X})$} & \colhead{[X/H]} & \colhead{[X/Fe]}& \colhead{$N$}& \colhead{$\sigma$}&\colhead{}&
\colhead{$\log\epsilon (\mbox{X})$} & \colhead{[X/H]} & \colhead{[X/Fe]}&  \colhead{$N$}& \colhead{$\sigma$} &\colhead{}&
\colhead{$\log\epsilon (\mbox{X})$} & \colhead{[X/H]} & \colhead{[X/Fe]}& \colhead{$N$}& \colhead{$\sigma$}\\
\hline \\
CH  	&      5.64		& 	$-$2.79	&	0.08	&	2	&	0.15  &&  6.69    &   $-$1.74 &   0.61    &   2   &   0.10&&    6.39 & $-$2.04 & 0.37 & 3 & 0.10\\\\ 
NH  &	\nodata		&		\nodata	&		\nodata &	\nodata	&	\nodata	&&	\nodata		&		\nodata	&		\nodata &	\nodata	&	\nodata	&&    6.02 & $-$1.81 & 0.60 & 1 & 0.20\\
CN  &	\nodata		&		\nodata	&		\nodata &	\nodata	&	\nodata	&&	\nodata		&		\nodata	&		\nodata &	\nodata	&	\nodata	&&   6.22 & $-$1.61 & 0.80 & 1 & 0.20\\
Na I  & 	4.25 		&	$-$1.99  	& 0.46 	& 	5	&	0.15&&  3.93    &   $-$2.31 &   0.04    &   2   &   0.10 &&  4.38 & $-$1.86 &  0.55 & 2 & 0.10 \\
Mg I 	&     	5.57 		&	$-$2.03 	&  0.42 	& 	8	&	 0.15  &&  5.55    &   $-$2.05 &   0.30    &   8   &   0.12&& 5.56 &$-$2.04  &  0.36  & 8 & 0.15 \\
Al I 	&      3.20 		&	$-$3.25 	& $-$0.80 	& 	2	& 	0.10 &&  3.10    &   $-$3.25 &   $-$0.90 &   1   &   0.15&& 3.37 &$-$3.08  &$-$0.68 & 1 & 0.15\\
Si I  	&      5.39 		&	$-$2.12 	&  0.33 	& 	1	& 	0.15  &&  5.53    &   $-$1.98 &   0.37    &   1   &   0.15&&   5.49 &$-$2.02 & 0.39  & 1 & 0.15\\
Ca I 	&      4.24 		&	$-$2.10 	& 0.35 	&	21	& 	0.15   &&  4.38    &   $-$1.96 &   0.39    &   21  &   0.10&&  4.32 &$-$2.02 &  0.39 & 15 & 0.10\\
Sc II 	&     	0.61 		&	$-$2.54 	&$-$0.09	& 	4 	& 	0.10  &&  0.96    &   $-$2.19 &   0.16    &   5   &   0.10&&  0.78 &$-$2.37 &  0.04 & 5 & 0.10\\
Ti I 	&     	2.75 		&	$-$2.20  	& 0.25 	&	13	& 	0.15 &&  3.06    &   $-$1.89 &   0.46    &   9   &   0.10 &&2.77 &$-$2.18 &   0.22  &12 & 0.15\\
Ti II  	&    	2.69 		& 	$-$2.26 	&  0.19 	&	32	& 	0.12  &&  3.05    &   $-$1.90 &   0.45    &   34  &   0.15&& 2.84 &$-$2.11 &  0.30 & 29 & 0.12\\
V II 	&      1.64 		&	$-$2.29 	&  0.16 	&	4 	&	0.15 &&  2.14    &   $-$1.79 &   0.56    &   3   &   0.15&&1.70 &$-$2.23 &  0.17 & 4 & 0.08\\
Cr I 	&     	3.02 		&	$-$2.62 	& $-$0.17 &	11	&	 0.10 &&  3.23    &   $-$2.41 &   $-$0.06 &   6   &   0.15  &&3.11 &$-$2.53 &$-$0.12 & 9 & 0.10\\
Cr II 	&      3.29 		&	$-$2.35 	&  0.10 	&	1 	&	 0.15 &&  3.43    &   $-$2.21 &   0.14    &   2   &   0.10 && 3.66 &$-$1.98 &  0.43 & 1 & 0.15\\
Mn I 	&      2.44 		&	$-$2.99 	& $-$0.54 	&	4 	&	 0.08 &&  2.59    &   $-$2.84 &   $-$0.49 &   3   &   0.10  &&2.71 &$-$2.72 &$-$0.31 & 7 & 0.10\\
Fe I 	&    	5.05 		&	$-$2.45 	& 0.00 	&	185 	&	 0.13 &&  5.15    &   $-$2.35 &   0.00    &   198 &   0.12 && 5.09 &$-$2.41  & 0.00 & 187 & 0.12\\
Fe II 	&     	5.04		&	$-$2.46 	& $-$0.01 	&	24 	& 	0.13 &&  5.14    &   $-$2.36 &   $-$0.01 &   20  &   0.15 && 5.09 &$-$2.41 &$-$0.01 & 17 & 0.10\\
Co I 	&      2.65 		&	$-$2.34 	&  0.00 	&	5 	& 	0.11  &&  2.88    &   $-$2.11 &   0.24    &   5   &   0.10 && 2.66 &$-$2.33 &  0.08 & 6 & 0.06\\
Ni I 	&      3.74 		&	$-$2.48 	& $-$0.03 	&	7 	&	0.10 &&  3.87    &   $-$2.35 &   0.00    &   11  &   0.15&&   3.86 &$-$2.36 &  0.05 & 8 & 0.08\\
Zn I 	&      2.50 		&	$-$2.06  	& 0.39 	&	2 	&	 0.15 && & & & & &&  2.29 &$-$2.27 &  0.14 & 2 & 0.10\\
Sr II 	&      0.42 		&	$-$2.45 	&   0.00 	&	2 	&	 0.15 &&  1.11    &   $-$1.76 &   0.59    &   2   &   0.10&& 0.73 &$-$2.14 &  0.27 & 2 & 0.10\\
Y II  	&  $-$0.86		&	$-$3.07 	&$-$0.62 	&	2 	&	 0.10   && 0.32    &   $-$1.89 &   0.46    &   3   &   0.10 &&$-$0.16 &$-$2.37 &  0.03 & 2 & 0.10\\
Zr II 	&      0.33 		&	$-$2.25 	&   0.19 	&	1 	&	 0.30  & & 1.01    &   $-$1.57 &   0.78    &   2   &   0.10 && 0.66   &$-$1.92 &0.49 &3 &0.20 \\
Ba II 	&  $-$0.80		&	$-$2.98 	&$-$0.53 	&	5 	&	 0.16  &&   0.45    &   $-$1.73 &   0.62    &   5   &   0.20  && 0.23 &$-$1.95 &  0.46 & 5 & 0.15\\
La II 	&  $-$1.54		&	$-$2.58 	&$-$0.13 	&	1 	&	 0.20  && $-$0.28 &   $-$1.38 &   0.97    &   6   &   0.10&&     $-$0.60 &$-$1.70 &  0.70 & 8 & 0.10\\
Ce II	&  $<-$0.72	&	$<-$2.30 	&$<$0.15 	&	  	&         && $-$0.01 &   $-$1.59 &   0.76    &   1   &   0.30&&  $-$0.28 & $-$1.86 & 0.55 & 3 & 0.10 \\
Pr II 	&  $<-$0.95	&	$<-$1.67 	&$<$0.77 	&	  	&        && $-$0.30 &   $-$1.02 &   1.33    &   2   &   0.20&& $-$0.76 &$-$1.48  & 0.92 & 8  & 0.15\\
Nd II &  $-$1.00 	&	$-$2.42 	&   0.03 	&	1	& 	 0.20  &&     0.12    &   $-$1.30 &   1.05    &   2   &   0.15 &&  $-$0.11  &$-$1.53 &0.87 &9 &0.15 \\
Sm II&	\nodata		&		\nodata	&		\nodata &	\nodata	&	\nodata	&& $-$0.14 &   $-$1.10 &   1.25    &   2   &   0.10   && $-$0.48 &$-$1.44 &  0.96 & 6 & 0.10\\
Eu II & $-$1.75 	 	& 	$-$2.27 	&0.18    	&	2  	&	0.10 &&$-$0.54 &   $-$1.06 &   1.29    &   5   &   0.10 &&$-$0.77 &$-$1.29  & 1.12 & 4 & 0.10\\
Gd II&	\nodata		&		\nodata	&		\nodata &	\nodata	&	\nodata	&& 0.07    &   $-$1.00 &   1.35    &   1   &   0.10  && $-$0.32  &$-$1.39 &  1.01 & 2 & 0.10\\
Tb II &	\nodata		&		\nodata	&		\nodata &	\nodata	&	\nodata	&&	\nodata		&		\nodata	&		\nodata &	\nodata	&	\nodata	&& $-$0.90  &$-$1.20 &  1.20 & 2 & 0.20\\
Dy II&	\nodata		&		\nodata	&		\nodata &	\nodata	&	\nodata	&&  0.09    &   $-$1.01 &   1.34    &   2   &   0.20 &&$-$0.15  &$-$1.25 &  1.16 & 3 & 0.22\\
Ho II &	\nodata		&		\nodata	&		\nodata &	\nodata	&	\nodata	&&	\nodata		&		\nodata	&		\nodata &	\nodata	&	\nodata	&&  $-$0.82 &$-$1.30  & 1.11 & 1 & 0.20\\
Er II &	\nodata		&		\nodata	&		\nodata &	\nodata	&	\nodata	&&	\nodata		&		\nodata	&		\nodata &	\nodata	&	\nodata	&& $-$0.40  &$-$1.31 &  1.09 & 1 & 0.20\\
Tm II &	\nodata		&		\nodata	&		\nodata &	\nodata	&	\nodata	&&	\nodata		&		\nodata	&		\nodata &	\nodata	&	\nodata	&& $-$1.10  &$-$1.20 & $-$1.21 & 1 & 0.30\\
Yb II &	\nodata		&		\nodata	&		\nodata &	\nodata	&	\nodata	&&	\nodata		&		\nodata	&		\nodata &	\nodata	&	\nodata	&& $-$0.60  &$-$1.44 &  0.97 & 1 & 0.20\\
Os I &	\nodata		&		\nodata	&		\nodata &	\nodata	&	\nodata	&&	\nodata		&		\nodata	&		\nodata &	\nodata	&	\nodata	&&  0.30  &$-$1.10 &  1.31 & 1 & 0.30\\
Ir I &	\nodata		&		\nodata	&		\nodata &	\nodata	&	\nodata	&&	\nodata		&		\nodata	&		\nodata &	\nodata	&	\nodata	&& 0.55  &$-$0.82 &  1.58 & 1 & 0.30\\
Th II 	&$-$1.60		&	$-$1.62 	&0.83	&	1 	&0.30&&$-$0.55 &   $-$0.57 &   1.78    &   1   &   0.30&&$-$1.32 &$-$1.34 &1.07 &1 &0.20\\
\enddata
 \tablecomments{
[X/Fe] ratios are computed with [Fe\,I/H] abundances
of the respective stars. Solar abundances have been taken from
\citet{Asplund09}.  For abundances measured from only one line, we adopt a nominal uncertainty of 0.10\,dex.} 
\end{deluxetable*} 

\pagebreak

\newpage

\startlongtable
\begin{deluxetable*}{lrrrrrrrrrrrrrrrrr}

\tablecaption{\label{omg} Abundances of stars in the $\omega$~Cen progenitor stream}
\tablehead{
\colhead{} &
\colhead{} & \colhead{} & \colhead{}& \colhead{}& \colhead{}&\colhead{}&
\colhead{} & \colhead{} & \colhead{}&  \colhead{}& 
\colhead{} &\colhead{}&
\colhead{} & \colhead{} & \colhead{}& \colhead{}& \colhead{}}
\startdata
\colhead{}&
\multicolumn{5}{c}{BM056}&\colhead{}& 
\multicolumn{5}{c}{BM121}&\colhead{}& 
\multicolumn{5}{c}{HE~0007$-$1752}\\
\cline{2-6} \cline{8-12} \cline{14-18} \\
\colhead{Species} &
\colhead{$\log\epsilon (\mbox{X})$} & \colhead{[X/H]} & \colhead{[X/Fe]}& \colhead{$N$}& \colhead{$\sigma$}&\colhead{}&
\colhead{$\log\epsilon (\mbox{X})$} & \colhead{[X/H]} & \colhead{[X/Fe]}&  \colhead{$N$}& \colhead{$\sigma$} &\colhead{}&
\colhead{$\log\epsilon (\mbox{X})$} & \colhead{[X/H]} & \colhead{[X/Fe]}& \colhead{$N$}& \colhead{$\sigma$}\\
\hline \\
CH &6.47 & $-$1.96 & 0.34 & 4 &0.20&&5.57 &$-$2.86 & 0.37  & 3  &0.10 &&5.77&$-$2.66&$-$0.29&2&0.10\\
NH &6.24&$-$1.59 &0.43  &1  &0.20&&5.80 &$-$2.03 & 0.45  & 1  &0.10&&5.14&$-$2.69&$-$0.33&1&0.10\\
O I &7.49 & $-$1.20 & 0.82 & 1 & 0.30 &&  7.00 &$-$1.69 & 0.79  & 1  &0.15&&	\nodata	&	\nodata		&	\nodata		&	\nodata	& \nodata	\\
Na I &4.40 & $-$1.84 & 0.18 & 3 & 0.15 &&  3.88 &$-$2.36 &   0.13  & 2  &0.09 && 3.29 & $-$2.95 & $-$0.59 & 2 & 0.10\\
Mg I &5.98 & $-$1.62 & 0.40 & 9 & 0.15&& 5.37 &$-$2.23 &   0.25  & 7  &0.13 && 4.91 &$-$2.69&$-$0.33&5&0.10 \\
Al I  &	\nodata		&		\nodata	&		\nodata &	\nodata	&	\nodata	&&	\nodata		&		\nodata	&		\nodata &	\nodata	&	\nodata	&& 3.50   &$-$2.95&$-$0.58&  1&0.15\\
Si I  & 5.88 & $-$1.63 & 0.39    & 4 & 0.15&&  5.38 &$-$2.13 &   0.36  & 2  &0.15&& 5.26   &$-$2.25&0.11   &  1&0.15\\
Ca I & 4.68 & $-$1.66 & 0.35    & 23& 0.15&& 4.05 &$-$2.29 &   0.19  &11  &0.06&& 4.19   &$-$2.15&0.21   & 26&0.10\\
Sc II & 1.15 & $-$2.00 & 0.02 & 8 & 0.15&& 0.71 &$-$2.44 & 0.04  & 9  &0.12&& $-$0.20&$-$3.35&$-$0.98& 3 &0.10\\
Ti I &3.14 & $-$1.81 & 0.21 & 30 & 0.15&&  2.71 &$-$2.24 &   0.24  &17  &0.15&& 2.33&$-$2.62&$-$0.26&9&0.11\\
Ti II &3.31 & $-$1.64 & 0.38 & 43 & 0.15&&  2.88 &$-$2.07 &   0.41  &28  &0.12 && 2.42&$-$2.53&$-$0.17&41&0.15\\
V II &2.03 & $-$1.90 & 0.11 & 4 & 0.15&& 1.49 &$-$2.44 &  0.04  & 3  &0.13&&0.93&$-$3.00&$-$0.64&3&0.10\\
Cr I &3.45 & $-$2.19 &$-$0.17 & 17 & 0.10&& 3.11 &$-$2.53 &$-$0.05  &15  &0.08&&3.03&$-$2.61&$-$0.24&16&0.13\\
Cr II &3.73 & $-$1.91 & 0.11 & 3 & 0.10&& 3.38 &$-$2.26 &   0.22  & 2  &0.10&&3.55&$-$2.09&0.27&3&0.15\\
Mn I &2.91 & $-$2.52 &$-$0.50 & 7 & 0.10&& 2.75 &$-$2.68 &$-$0.20  & 5  &0.10&&2.42&$-$3.01&$-$0.65&6&0.10\\
Fe I &5.48 & $-$2.02 & 0.00 & 256 & 0.12&& 5.02 &$-$2.48 &   0.00  &158 &0.15 &&5.14&$-$2.36&0.00&291&0.13\\
Fe II&5.45 & $-$2.05 & $-$0.04 & 27 & 0.09&&  5.06 &$-$2.44 &   0.00  &21  &0.10 &&5.12&$-$2.38&$-$0.01&28&0.11\\
Co I &2.95 & $-$2.04 & $-$0.02 & 6 & 0.15&& 2.53 &$-$2.46 &   0.03  & 7  &0.15&&2.25&$-$2.74&$-$0.38&7&0.15\\
Ni I &4.13 & $-$2.09 &$-$0.07 & 23 & 0.15&&  3.76 &$-$2.46 &   0.02  &11  &0.10&&3.46&$-$2.76&$-$0.40&15&0.15\\
Cu I&	\nodata		&		\nodata	&		\nodata &	\nodata	&	\nodata	&&  1.17 &$-$3.02 &$-$0.53  & 1  & 0.15&&	\nodata	&	\nodata		&	\nodata		&	\nodata	& \nodata	\\
Zn I &2.54 & $-$2.02 & 0.00 & 2 & 0.10&& 2.53 &$-$2.03 &   0.45  & 3  & 0.10&&	\nodata	&	\nodata		&	\nodata		&	\nodata	& \nodata	\\
Sr II &1.15&$-$1.72&0.30&2&0.10&& 0.20 &$-$2.67 &$-$0.19  & 2  & 0.15&&0.70 & $-$2.17 & 0.20&3&0.20\\
Y II &0.04 & $-$2.17 &$-$0.15 & 2& 0.10&&  $-$0.84 &$-$3.05 &$-$0.56  & 3  & 0.10 && $-$0.02&$-$2.23&0.13&5&0.11\\
Zr II &0.96&$-$1.62 &0.40&3 &0.10&&  0.32 &$-$2.26 &   0.22  & 4  & 0.15 && 0.83&$-$1.75&0.61&3&0.10\\
Mo I & 0.29&$-$1.59&0.43&1&0.20&&	\nodata		&		\nodata	&		\nodata &	\nodata	&	\nodata	&&	\nodata	&	\nodata		&	\nodata		&	\nodata	& \nodata	\\
Ba II &$-$0.08 & $-$2.26 &$-$0.06 & 5 & 0.10&& $-$0.78 &$-$2.96 &$-$0.48  & 2  & 0.15&&  0.36 &$-$1.81 &0.55 & 5 &0.11\\
La II &$-$0.69&$-$1.79&0.23&9&0.15&& $-$1.55 &$-$2.65 &$-$0.17  & 4  & 0.15 &&$-$0.37&$-$1.47&0.90&9&0.10\\
Ce II &$-$0.25&$-$1.83&0.19&20&0.15&& $-$1.23 &$-$2.81 &$-$0.33  & 3  & 0.15&&$-$0.15&$-$1.73&0.63&15&0.10\\
Pr II &$-$1.01&$-$1.73&0.29&7&0.10&&  $-$1.63 &$-$2.35 & 0.13  & 2  & 0.20&&$-$0.61&$-$1.33&1.03&2&0.10\\
Nd II &$-$0.35 & $-$1.77 & 0.24 & 21 & 0.15&& $-$1.00 &$-$2.42 & 0.07  & 11 & 0.10&&0.00&$-$1.42&0.95&38&0.11\\
Sm II &$-$0.82 & $-$1.78 & 0.24 & 15 & 0.15&& $-$1.39 &$-$2.35 &  0.13  & 8  & 0.15&&$-$0.30&$-$1.26&1.10&9&0.11\\
Eu II &$-$1.17 & $-$1.69 & 0.33 & 5 & 0.10&& $-$1.69 &$-$2.21 & 0.27  & 3  & 0.15&&$-$0.60&$-$1.12&1.25&5&0.10\\
Gd II &$-$0.77&$-$1.84&0.18&4&0.10&& $-$1.27 &$-$2.34 &   0.14  & 1  & 0.20&&$-$0.12&$-$1.19&1.18&4&0.15\\
Tb II &$-$1.31&$-$1.61&0.40&1&0.30&& $-$1.86 &$-$2.16 &   0.32  & 1  & 0.20&&$-$0.85 & $-$1.15 & 1.21 & 2&0.10\\
Dy II &$-$0.58&$-$1.68&0.34&2&0.10&& $-$0.96 &$-$2.06 &   0.42  & 1  & 0.30&&  0.11 & $-$0.99 & 1.37 & 8&0.15\\
Ho II &$-$1.34&$-$1.82&0.20&2&0.15&&  $-$1.65 &$-$2.13 &   0.35  & 1  & 0.20&& $-$0.67 & $-$1.15 & 1.21 & 2&0.15\\
Er II &$-$0.74&$-$1.66&0.36&1&0.15&& $-$1.28 &$-$2.20 & 0.29  & 2  & 0.20&& $-$0.19 & $-$1.11 & 1.25 & 7&0.15\\
Tm II&	\nodata		&		\nodata	&		\nodata &	\nodata	&	\nodata	&& $-$2.29 &$-$2.39 &  0.09  & 1  & 0.20&&$-$0.96 & $-$1.06 & 1.30 & 1&0.15\\
Yb II &$-$0.87&$-$1.71&0.31&1&0.30&&	\nodata		&		\nodata	&		\nodata &	\nodata	&	\nodata	&&$-$0.24 & $-$1.08 & 1.29 & 1&0.15\\
Os I &$-$0.35&$-$1.75&0.27&2&0.11&& $-$0.84 &$-$2.24 & 0.24  & 1  & 0.20&&0.30&$-$1.10&1.26&1&0.15\\
Ir I &$-$0.04&$-$1.42&0.60&1&0.30&&	\nodata		&		\nodata	&		\nodata &	\nodata	&	\nodata	&&	\nodata	&	\nodata		&	\nodata		&	\nodata	& \nodata	\\
Pb I &0.50&$-$1.25&0.77&1&0.30&&	\nodata		&		\nodata	&		\nodata &	\nodata	&	\nodata	&	\nodata	&	\nodata		&	\nodata		&	\nodata	& \nodata	\\
Th II &$-$1.75&$-$1.97&0.05&1&0.20&&  $-$2.26 &$-$2.28 & 0.20  & 1  & 0.10&&$-$1.16 & $-$1.18 & 1.18 & 1&0.10\\\\
\hline
\\
\colhead{}&
\multicolumn{5}{c}{HE~0039$-$0216}&\colhead{}& 
\multicolumn{5}{c}{HE~0429$-$4620}&\colhead{}& 
\multicolumn{5}{c}{HE~1401$-$0010}\\
\cline{2-6} \cline{8-12} \cline{14-18} \\
\colhead{Species} &
\colhead{$\log\epsilon (\mbox{X})$} & \colhead{[X/H]} & \colhead{[X/Fe]}& \colhead{$N$}& \colhead{$\sigma$}&\colhead{}&
\colhead{$\log\epsilon (\mbox{X})$} & \colhead{[X/H]} & \colhead{[X/Fe]}&  \colhead{$N$}& \colhead{$\sigma$} &\colhead{}&
\colhead{$\log\epsilon (\mbox{X})$} & \colhead{[X/H]} & \colhead{[X/Fe]}& \colhead{$N$}& \colhead{$\sigma$}\\
\hline \\
CH   &6.81&$-$1.62&0.94&1&0.20&&6.08 & $-$2.35 &0.05 & 2 & 0.10&&6.82 &$-$1.61 &0.93 &2 &0.10 \\
NH &	\nodata		&		\nodata	&		\nodata &	\nodata	&	\nodata	&&6.00&$-$1.83&0.71&2&0.10  &&6.00&$-$1.83&0.71&2&0.10\\
Na I &  3.49 &$-$2.75 &$-$0.19 &2 &0.15 &&3.96 & $-$2.28 & 0.12 & 2 & 0.10 &&3.49 & $-$2.75 & $-$0.21 & 2 & 0.10\\
Mg I & 5.23 &$-$2.37&0.19&4&0.11&&5.51 & $-$2.09 & 0.31 & 6 & 0.15 &&5.25 & $-$2.35 & 0.18 & 5 & 0.15\\
Al I &	2.89&$-$3.56&$-$1.00&1&0.15 &&3.08 & $-$3.37 & $-$0.97 & 1 & 0.15&& 2.90 & $-$3.55& $-$1.01& 1&0.15 \\
Si I &	5.00&$-$2.51&0.05&1&0.15 &&5.89 & $-$1.62 & 0.78 & 2 & 0.19&&5.25 & $-$2.26 & 0.28 & 1 & 0.15\\
Ca I &	4.22&$-$2.12&0.45&9&0.10 &&4.24 & $-$2.10 & 0.30 & 13 & 0.15&&4.20 & $-$2.14 & 0.40 & 14 & 0.15\\
Sc II &	0.81&$-$2.34&0.22&2&0.10 &&0.90 & $-$2.25 & 0.15 & 3 & 0.10&&0.73 &$-$2.42 & 0.12 &1 &0.15\\
Ti I &	3.18&$-$1.71&0.79&3&0.12 &&2.72 & $-$2.23 & 0.17 & 4 & 0.09&&2.89 & $-$2.06 & 0.48 & 2 & 0.10\\
Ti II &	2.87&$-$2.08&0.49&16&0.12 &&2.91 & $-$2.04 & 0.36 & 25 & 0.15&&2.76 & $-$2.19 & 0.35 & 20 & 0.15 \\
V II &	\nodata		&		\nodata	&		\nodata &	\nodata	&	\nodata	&&1.87 & $-$2.06 & 0.34 & 3 & 0.15 &&2.10 &$-$1.83 &0.71 &1 &0.15 \\
Cr I &	2.81&$-$2.83&$-$0.27&4&0.10 &&3.00 & $-$2.64 &$-$0.24 & 3 & 0.15&&2.84 & $-$2.80 &$-$0.27 & 3 & 0.10 \\
Mn I &	2.27&$-$3.16&$-$0.60&2&0.10&&2.72 & $-$2.71 &$-$0.31 & 3 & 0.10&&2.27 & $-$3.16 &$-$0.62 & 1 & 0.15\\
Fe I &4.94&$-$2.56&0.00&83&0.13 &&5.10&$-$2.40&0.00&147&0.19&&4.96 & $-$2.54 & 0.00 & 82 & 0.20\\
Fe II &	4.93&$-$2.57& 0.00&11&0.11&&5.09&$-$2.41& $-$0.01&13&0.10&&4.93 & $-$2.57 & $-$0.03 & 9 & 0.17\\
Co I &	2.68&$-$2.31&0.25&3&0.11&&2.78&$-$2.21&0.20&2&0.15&&	\nodata	&	\nodata		&	\nodata		&	\nodata	& \nodata	\\
Ni I &	   3.86&$-$2.36&0.20&5&0.11 &&3.95&$-$2.27&0.13&7&0.13 &&3.80 &$-$2.42&0.11 &3&0.15 \\
Zn I &	$<$2.70&$<-$1.86&$<$0.70& & && 2.17&$-$2.38&0.02&1&0.15 &&	\nodata	&	\nodata		&	\nodata		&	\nodata	& \nodata	\\
Sr II &	0.14&$-$2.73&$-$0.17&2&0.10	&&0.58&$-$2.29&0.11&2&0.15&& 0.28& $-$2.59& $-$0.06& 2& 0.10\\
Y II & 	$-$0.08 &$-$2.29&0.27&1&0.15 &&$-$0.46 &$-$2.67 &$-$0.27 &1  &0.20 &&$-$0.23  &$-$2.44  &0.10 &2 &0.15\\
Zr II & $<$0.90 &$<-$1.68&$<$0.88&\nodata	&	\nodata		&& 0.61 &$-$1.97&0.40&2&0.10 &&$<$1.18 &$<-$1.41 &$<$1.13&\nodata	&	\nodata	\\
Ba II & $-$0.68&$-$2.86&$-$0.30&1&0.30	&&$-$0.41&$-$2.59&$-$0.19&5&0.15 &&$-$0.79 &$-$2.97 &$-$0.43&1 &0.10\\
La II &	\nodata		&		\nodata	&		\nodata &	\nodata	&	\nodata	&&$-$1.19 & $-$2.29 & 0.38 & 3 & 0.30&&	\nodata	&	\nodata		&	\nodata		&	\nodata	& \nodata	\\
Ce II &	\nodata		&		\nodata	&		\nodata &	\nodata	&	\nodata	&&$<-$0.50 &$<-$2.08& $<$0.32 && &&	\nodata	&	\nodata		&	\nodata		&	\nodata	& \nodata	\\
Pr II &	\nodata		&		\nodata	&		\nodata &	\nodata	&	\nodata	&& $<-$0.90 &$<-$1.62&$<$0.82 &&	\nodata	&	\nodata		&	\nodata		&	\nodata	& \nodata	\\
Nd II &	\nodata		&		\nodata	&		\nodata &	\nodata	&	\nodata	&&$-$0.57 &$-$1.99&0.41&1&0.20&&	\nodata	&	\nodata		&	\nodata		&	\nodata	& \nodata	\\
Sm II &	\nodata		&		\nodata	&		\nodata &	\nodata	&	\nodata	&& $-$0.85 &$-$1.81&0.59&1&0.20&&	\nodata	&	\nodata		&	\nodata		&	\nodata	& \nodata	\\
Eu II &	$<-$0.95&$<-$1.47&$<$1.09&\nodata	&	\nodata		&& $-$1.35 & $-$1.87 &0.53 & 3 & 0.15&&$<-$1.90 &$<-$2.42  &$<$0.12 &\nodata	&	\nodata		\\\\
\hline
\\
\colhead{}&
\multicolumn{5}{c}{HE~2138$-$0314}&\colhead{} &
\multicolumn{5}{c}{HE~2315$-$4306}&\colhead{} &
\multicolumn{5}{c}{HE~2319$-$5228}\\
\cline{2-6} \cline{8-12} \cline{14-18} \\
\colhead{Species} &
\colhead{$\log\epsilon (\mbox{X})$} & \colhead{[X/H]} & \colhead{[X/Fe]}& \colhead{$N$}& \colhead{$\sigma$}&\colhead{}&
\colhead{$\log\epsilon (\mbox{X})$} & \colhead{[X/H]} & \colhead{[X/Fe]}&  \colhead{$N$}& \colhead{$\sigma$} &\colhead{}&
\colhead{$\log\epsilon (\mbox{X})$} & \colhead{[X/H]} & \colhead{[X/Fe]}& \colhead{$N$}& \colhead{$\sigma$}\\
\hline \\
CH   &   6.28&$-$2.15&   0.95& 2 &0.10 &&   6.33 & $-$2.10 &   0.31 & 2  & 0.10&&   6.88 &$-$1.55 &   1.63 &2  &0.05\\
NH   &5.46 &$-$2.37&0.71&1&0.15 &&5.48&$-$2.35&0.05&1&0.15&&  7.51 & $-$0.32 & 2.87 & 1 &0.20\\
Na I &   3.24&$-$3.00&   0.09& 3 &0.07 &&   4.14 & $-$2.10 &   0.29 & 2  & 0.10&&   4.91 &$-$1.33 &   1.85 &4  &0.14\\
Mg I &   4.94&$-$2.66&   0.42& 7 &0.18 &&   5.61 & $-$1.99 &   0.40 & 2  & 0.15&&   5.15 &$-$2.45 &   0.73 &5  &0.12\\
Al I &   2.58&$-$3.87&$-$0.78& 1 &0.15 &&   3.68 & $-$2.77 &  $-$0.38 & 1  & 0.15&&3.35    &$-$3.10  &0.08    & 1  & 0.15    \\
Si I &   5.71&$-$1.80&   1.29& 1 &0.15 &&   5.70 & $-$1.81 &   0.58 & 1  & 0.15&&   5.36 &$-$2.15 &   1.04 &1  &0.15\\
Ca I &   3.56&$-$2.78&   0.30& 14&0.17 &&   4.31 & $-$1.86 &   0.53 & 19 & 0.15&&   3.98 &$-$2.36 &   0.82 &19 &0.10\\
Sc II&  0.02&$-$3.13&$-$0.05&  3&0.10 &&   0.64 & $-$2.51 &$-$0.11 & 4  & 0.10&&  0.16 &$-$2.99 &   0.20 &4  &0.10\\
Ti I &   2.11&$-$2.84&   0.25&  9&0.19 &&   2.72 & $-$2.23 &   0.17 & 11 & 0.10&&   2.10 &$-$2.85 &   0.33 &7  &0.11\\
Ti II&   2.05&$-$2.90&   0.18& 26&0.19 &&   2.75 & $-$2.20 &   0.19 & 36 & 0.15&&   2.42 &$-$2.53 &   0.65 &23 &0.12\\
V II &   0.99&$-$2.94&   0.15&1  &0.15 &&   1.69 & $-$2.24 &  0.16 & 2  & 0.15&&   1.16 &$-$2.77 &   0.41 & 2 &0.15\\
Cr I &   2.31&$-$3.33&$-$0.25&  5&0.17 &&   3.07 & $-$2.57 &$-$0.18 & 14 & 0.10&&   2.25 &$-$3.39 &$-$0.21 &5  &0.11\\
Cr II&       &       &       &   &     &&   3.46 & $-$2.17 &   0.22 &  2 & 0.10&&	\nodata	&	\nodata		&	\nodata		&	\nodata	& \nodata	\\
Mn I &   1.75&$-$3.68&$-$0.60&  3&0.10 &&   2.64 & $-$2.79 &$-$0.39 &  5 & 0.15&&   1.50 &$-$3.93 &$-$0.75 &1  &0.15\\
Fe I &   4.42&$-$3.08&   0.00&110&0.13 &&   5.11 & $-$2.39 &   0.00 & 140& 0.12&&   4.32 &$-$3.18 &   0.00 &92 &0.15\\
Fe II &  4.40&$-$3.10&$-$0.01& 16&0.19 &&   5.06 & $-$2.44 &$-$0.05 &  11& 0.10&&   4.32 &$-$3.18 &$-$0.00 &10 &0.15\\
Co I &   2.18&$-$2.81&   0.28&  5&0.11 &&   2.74 & $-$2.25 &   0.14 &   7& 0.11&&   1.68 &$-$3.31 &$-$0.13 &2  &0.11\\
Ni I &   3.27&$-$2.95&   0.14&  6&0.15 &&   3.80 & $-$2.42 &$-$0.03 &   7& 0.13&&   3.15 &$-$3.07 &   0.11 &5  &0.11\\
Zn I &   1.78&$-$2.78&   0.31&  2&0.10 &&   1.83 & $-$2.73 &$-$0.33 &   1& 0.15&&   1.78 &$-$2.78 &   0.40 &1  &0.15 \\
Sr II&$-$0.24&$-$3.11&$-$0.02&  2&0.10 &&   0.21 & $-$2.66 &$-$0.27 &   2& 0.10&&$-$2.29 &$-$5.16 &$-$1.97 &1  &0.15\\
Y II &$-$1.04&$-$3.25&$-$0.17&  1&0.30 &&$-$0.63 & $-$2.84 &$-$0.44 &   1& 0.15&&	\nodata	&	\nodata		&	\nodata		&	\nodata	& \nodata	\\
Zr II& 0.20&$-$2.38&   0.71&  1&0.15 &&   0.30 & $-$2.28 & 0.11 &   2& 0.10&&	\nodata	&	\nodata		&	\nodata		&	\nodata	& \nodata	\\
Ba II&$-$1.83&$-$4.01&$-$0.93&  2&0.10 &&$-$0.84 & $-$3.02 &$-$0.62 &   3& 0.10&&$-$2.74 &$-$4.92 &$-$1.74 &1 &0.20\\
La II &	\nodata		&		\nodata	&		\nodata &	\nodata	&	\nodata	&&$<-$0.34&$<-$1.53&$<$0.87&\nodata	&	\nodata		&&$<-$0.30&$<-$1.40&$<$1.82&\nodata	&	\nodata		\\
Pr II &	\nodata		&		\nodata	&		\nodata &	\nodata	&	\nodata	&&	\nodata		&		\nodata	&		\nodata &	\nodata	&	\nodata	&&$<-$0.06&$<$1.64&$<$1.54&\nodata	&	\nodata		\\
Sm II &	\nodata		&		\nodata	&		\nodata &	\nodata	&	\nodata	&&$<-$0.20&$<-$1.16&$<$1.23&\nodata	&	\nodata		&&$<-$0.75&$<-$2.17&$<$1.01&&\\
Eu II&$<-$2.20&$<-$2.72&$<$0.36&\nodata	&	\nodata		&&$-$1.98 & $-$2.50 &$-$0.11 &   3& 0.10&&$<-$2.70 &$<-$3.20 &$<-$0.04 &\nodata	&	\nodata		\\
Ho II &	\nodata		&		\nodata	&		\nodata &	\nodata	&	\nodata	&&$<-$1.12&$<-$1.60&$<$0.79&\nodata	&	\nodata		&&	\nodata	&	\nodata		&	\nodata		&	\nodata	& \nodata	\\
\hline\\
\colhead{}&
\multicolumn{5}{c}{HE~2322$-$6125}\\
\cline{2-6}  \\
\colhead{Species} &
\colhead{$\log\epsilon (\mbox{X})$} & \colhead{[X/H]} & \colhead{[X/Fe]}& \colhead{$N$}& \colhead{$\sigma$}\\
\cline{0-6}\\
CH & 6.11 &$-$2.32&0.30&2&0.15\\
Na I  &    4.25 & $-$2.19 & 0.42 & 2 & 0.10 \\
Mg I  &    5.39 &$-$2.21 & 0.40 &3& 0.15   \\
Al I  &    3.23 &$-$3.22 & $-$0.61 &1& 0.15   \\
Si I  & 5.70&$-$1.81&0.80&1&0.15   \\
Ca I  &  4.14&$-$2.20&0.42&20&0.15   \\
Sc II &  0.45&$-$2.70&$-$0.09& 4 &0.10  \\
Ti I  &   2.58&$-$2.37&0.24&10&0.15  \\
Ti II &    2.62&$-$2.33&0.28&35&0.15   \\
V II  &   1.40&$-$2.53&0.08&2&0.15  \\
Cr I  &    2.80&$-$2.84&$-$0.23&11&0.15 \\
Cr II &    3.38 &$-$2.26 &   0.22  & 2  &0.10   \\
Mn I  &   2.36&$-$3.07&$-$0.46&4&0.15  \\
Fe I  &    4.89&$-$2.61&0.00&147&0.15   \\
Fe II &    4.89&$-$2.61&0.00&14&0.12   \\
Co I  &    2.75 &$-$2.24 &0.38 &5 &0.15  \\
Ni I  &   3.73&$-$2.49      &0.12&7    &0.15  \\
Cu I  &  1.17 &$-$3.02 &$-$0.53  & 1  & 0.15\\
Zn I  &    2.32&$-$2.24&0.37&3&0.10\\
Sr II &  0.20 & $-$2.67& $-$0.06& 2& 0.15 \\
Y II  &$-$0.46 &$-$2.67&$-$0.06&1&0.15\\
Zr II &   0.09 &$-$2.49&0.12&1&0.15 \\
Ba II & $-$0.72& $-$2.90& $-$0.29& 2& 0.10 \\
Sm II & $<$0.00& $<-$1.01& $<$1.01&  \\
Eu II & $-$1.73 &$-$2.25&0.37&2&0.15  \\
\enddata
 \tablecomments{
[X/Fe] ratios are computed with [Fe\,I/H] abundances
of the respective stars. Solar abundances have been taken from
\citet{Asplund09}. For abundances measured from only one line, we adopt a nominal uncertainty of 0.10\,dex.}
\end{deluxetable*}

\subsection{Carbon abundances, and CEMP star fractions}

We measure carbon abundances by performing spectrum synthesis on the CH \textit{G} bandhead at 4313\,{\AA} and the CH feature at 4323\,{\AA}. We then apply a carbon correction, as described in \citet{Placco14}, to account for surface depletion of carbon during the normal course of stellar evolution. Results are shown in Figure~\ref{light_ele}. Abundances range from [C/Fe]$_{\rm corr}$=$-$0.72 to [C/Fe]$_{\rm corr}$ = +1.90. Specifically, six stars in our data show significant carbon enhancement, and can thus be classified as carbon-enhanced metal-poor ($\mbox{[C/Fe]} > 0.7$, CEMP) stars, as defined by \citet{Aoki07}. An additional two stars have an enhancement $\sim$0.6, which we classify as borderline CEMP, and been included in our discussion of CEMP stars in the streams. 

We compare our findings to \citet{beers17}, who originally identified 7 stars as carbon-enhanced ([C/Fe]$ > +0.7$). While \citeauthor{beers17}\ find five CEMP stars in the $\omega$~Cen progenitor stream, we identified only four because we excluded the double-lined spectroscopic binary star HE~1120$-$0153. Both \citet{beers17} and this study find HE~0012$-$5643 to be the most carbon enhanced star in the Helmi debris stream, with [C/Fe] $= +1.2$. Note that in the following discussion, we do not include any stars with available upper limits of higher than $\mbox{[C/Fe]} = 1.0$; this applies specifically to two of the Helmi debris stars from \citet{Roederer10} with very high upper limits.

A large fraction of halo stars are carbon enhanced.  Approximately 30\% of the stars at [Fe/H] $= -2.0$ have [C/Fe] $\geq +0.5$, and  this fraction rises to 37\% at [Fe/H] $=-2.5$ (e.g., \citealt{Placco14}). For [C/Fe] $\geq +0.7$ the percentages become 20\% and 24\%, respectively. We identify 1/5 (20\%) of the Helmi trail stars, 3/15 (20\%) of the Helmi debris stars, and 4/10 (40\%) of the $\omega$~Cen progenitor stars as CEMP stars. 
When restricting our samples to [Fe/H] $< -2.0$, 
we find 1/3 (33\%) for both [C/Fe] $\geq +0.7$ and [C/Fe] $\geq +0.5$ for the Helmi trail stream, 

1/10 (10\%) for [C/Fe] $\geq +0.7$ and 3/10 (33\%) for [C/Fe] $\geq +0.5$ for the Helmi debris sample, and

4/9 (44\%) for both [C/Fe] $\geq +0.7$ and [C/Fe] $\geq +0.5$ for the $\omega$~Cen progenitor stream.

Further reducing the upper metallicity limit to [Fe/H] $<-2.5$, the fractions become 1/3 (33\%) for the Helmi trail stream,

1/2 (50\%) for the Helmi debris sample, and 

4/5 (80\%) for $\omega$~Cen progenitor stream, for either of the carbon abundance cutoffs.

Overall, these fractions roughly agree with what has been found among halo stars, even though the number statistics are relatively poor. Nevertheless, we also find the typical striking increase of higher CEMP fractions at lower metallicities, at least for the Helmi debris and $\omega$~Cen progenitor streams. The Helmi trail sample ($N=5$) is too small for any meaningful assessment. A large fraction of CEMP stars
has been found in Sculptor \citep{chiti18}. This suggests that it is not unlikely that the streams originated from a dwarf galaxy as the stream progenitors must have experienced a carbon enhancement similar to that of other larger dwarf galaxies. 
This provides a significant constraint on the nature of the progenitor. We thus conclude that the halo formed from streams that emerged from similar progenitor galaxies.

\subsection{Nitrogen abundances} 
\label{sec:nitrogen}
Data quality is generally poor around the NH feature at 3360\,{\AA}. Nevertheless, seven stars display mild N enhancements which allows for the feature to be measured. Three stars have [N/Fe] $\sim$ $+$0.45, one star has [N/Fe] $\sim$ $+$0.61, two stars have [N/Fe] $\sim$ $+$0.70, and one very strongly enhanced star in the $\omega$~Cen progenitor stream has [N/Fe] $=+$2.82. 

Figure~\ref{light_ele} (bottom left panel) shows [N/Fe] vs [C/Fe] in comparison with stars in the globular cluster $\omega$~Cen \citep{Marino12}. Generally, all stream stars broadly overlap with the halo and cluster stars with ranges of $\sim2$\,dex for both elements. The exception is the CEMP red giant HE~2319$-$5228, although some halo stars have reasonably similar abundances. HE~2319$-$5228 exhibits [N/Fe] $=+$2.82 (uncorrected) and [N/C] $=+$0.85. Its surface layers are not yet fully mixed with those experiencing CN-cycle burning ($T_{\rm eff}$ = 5000\,K), so any C and N abundance corrections should be relatively minor. According to \citet{Johnson07}, this characteristic also classifies HE~2319$-$5228 as a nitrogen enhanced metal-poor (NEMP) star, adding to the so far 80 other known NEMP stars \citep{simpson19}. 
Interestingly, HE~2319$-$5228 exhibits mild radial velocity variations that suggest binarity. The Gaia DR2 radial velocity is 286.2$\pm1$\,\kms \citep{gaia18}, while our value is 293.7$\pm$2\,\kms. We also note that HE~2319$-$5228 (with $\mbox{[Fe/H]}=-3.18$), along with HE~2138$-$0314 ($\mbox{[Fe/H]}=-3.08$) are the two lowest metallicity stars in the $\omega$~Cen progenitor stream sample. Finally, HE~2138$-$0314 exhibits [N/Fe] $=+$0.71 (uncorrected) and [N/C] $\sim$0.2. Overall, the fairly common mild N enhancement suggests that massive rotating progenitors stars may have provided C and N to the erstwhile progenitor host systems \citep{ekstroem08}.

\subsection{Light Elements ($Z\le30$) abundances}
\label{sec:light}

\begin{figure*}[!ht]
 \begin{flushleft}
    \includegraphics[width=18cm]{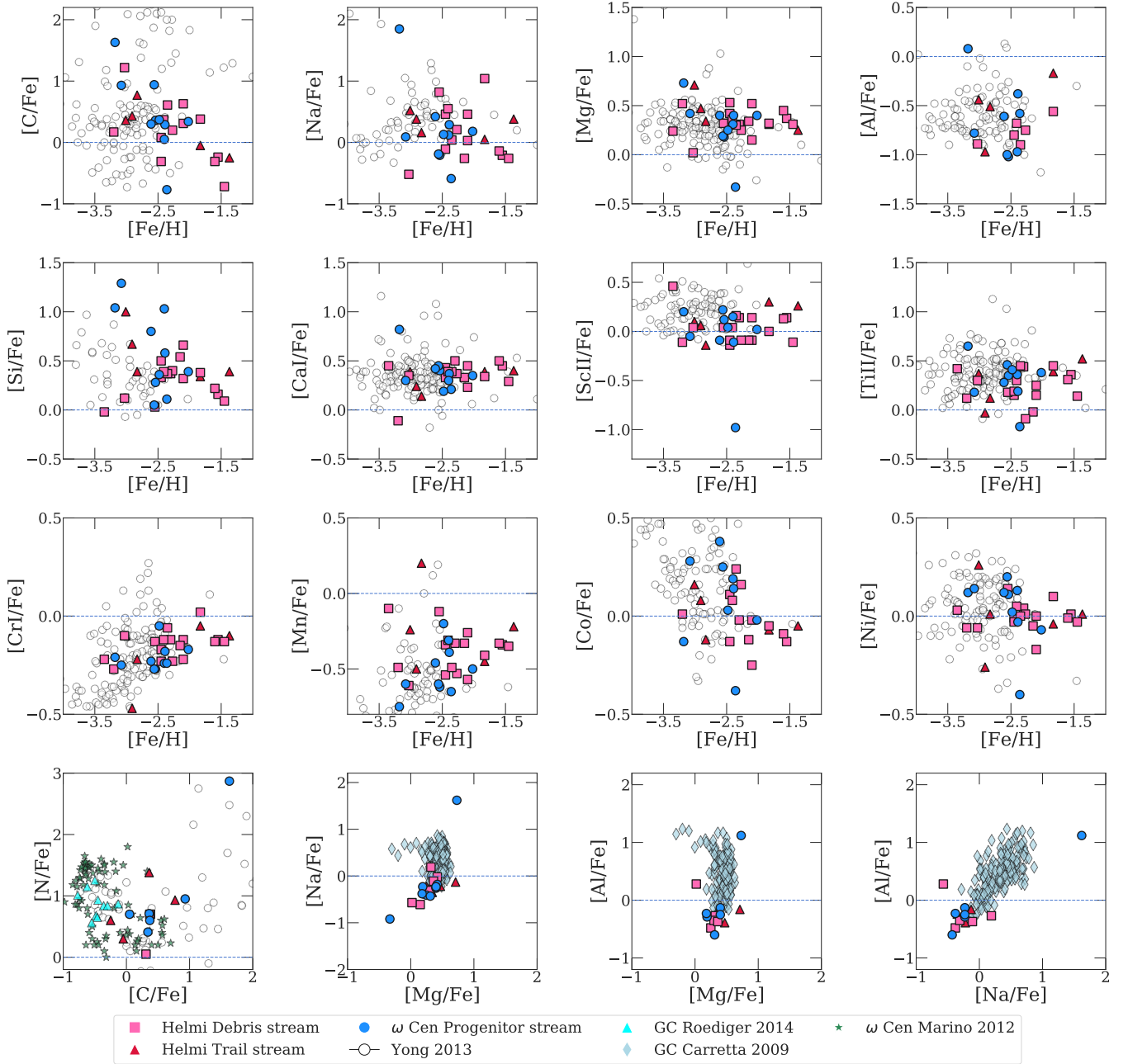}
    \caption{\label{light_ele} 
    Abundance ratios in our stream stars.  The symbol colors and shapes are indicated in the legend.  The [C/Fe] ratios have been corrected as described in \citet{Placco14}.  There is good agreement with the abundances in the non-CEMP stars from \citet{Yong13a}, which are shown for comparison as small black open circles. The bottom row shows several abundance ratio pairs that are commonly correlated or anti-correlated in globular cluster stars.  The globular cluster abundances, except for $\omega$~Cen, are adopted from \citet{carretta09} for all elements, except for [C/Fe] and [N/Fe], which are adopted from \citet{Roediger14}.  C and N abundance of $\omega$~Cen stars are from \citet{Marino12}. We note that  \citet{carretta09}'s and our values for Na I and Al I are corrected for NLTE (Na I; \citealt{Lind11}, Al I; \citealt{nordlander17}). }
 \end{flushleft}
\end{figure*}

Using equivalent width measurements or spectrum synthesis when appropriate (see \citealt{gull18} for more details), we determined chemical abundances of up to 16 light elements from sodium to zinc that are typically measured in metal-poor halo stars. This includes various $\alpha$- and iron-peak elements. Figure~\ref{light_ele} shows our abundance results combined with the Helmi debris stars of \citet{Roederer10}. We plot light elements [X/Fe] against [Fe/H] against a 
halo star sample of \citet{Yong13a}, for comparison. All abundances are in LTE, and we list LTE values in our abundance tables. For most element ratios, stars in all three streams generally do not show significant differences from each other, and they also do not stand out from the majority of halo stars with similar metallicities. 

In addition, in the bottom row, we present relationships among [C/Fe], [N/Fe], [Na/Fe], [Mg/Fe], and [Al/Fe] for our sample stars, as well as those of globular cluster and $\omega$~Cen stars. Na and Al abundances have been corrected for NLTE for comparison with results from \citet{carretta09}. For the correction of our values, we use the tools provided by \citealt{Lind11} for Na I and \citealt{nordlander17} for Al I, while \citet{carretta09} utilized corrections from \citet{gratton99}. We note here that the \citet{Johnson10} $\omega$~Cen stars do not have Na~\textsc{i} NLTE values available so we refrain from plotting them in the Figure (bottom right panel); however, their LTE values agree well with the \citet{carretta09} NLTE ones, modulo any NLTE effects.

In general, the light-element abundance ratios in the stars in the streams agree with the halo field stars and the so-called first-generation globular cluster stars.
In other words, few stars in our sample show enhanced [Na/Fe] or [Al/Fe] ratios or depleted [Mg/Fe] ratios
(e.g., \citealt{carretta09mgal}), relative to the field stars.
The stars in our sample are chemically distinct from the so-called second-generation stars in present-day metal-poor globular clusters, 
so it is unlikely that they were formed in a globular cluster-like environment.

The CEMP and NEMP star \mbox{HE~2319$-$5228} in the $\omega$~Cen stream is also evident in Figure~\ref{light_ele}, where its enhanced [Na/Fe], [Mg/Fe], and [Al/Fe] ratios also stand out from the rest of the sample.

This signature is found in roughly half of CEMP stars,
and its incidence increases at the lowest metallicities
\citep{Norris13}.
This signature also resembles that found in several of the most iron-poor stars \citep{christlieb02, frebel05}, and may thus indicate an origin scenario from a different mechanism than most CEMP stars in the field.

We note that two stars, HE~1210$-$2729 in the Helmi trail stream and HE~2322$-$6125 in the $\omega$~Cen progenitor stream, show very similar and unusual enhancements in Na, Mg, and Si: [Na/Fe] = $+$0.59 and $+$0.61, [Mg/Fe] = $+$0.71 and $+$0.64, and [Si/Fe]= $+$1.00 and $+$0.80, respectively. Interestingly, HE~1210$-$2729 ([Fe/H]= $-$3.01) is the most metal-poor star in the Helmi trail stream. While it does not show significant carbon enhancement, there is some nitrogen enhancement.
HE~2322$-$6125 ([Fe/H]= $-$2.61) is the second most metal-poor star in the $\omega$~Cen progenitor stream but does not display the C and N enhancements. Since these abundance patterns loosely resemble those of the most iron-poor stars, e.g., HE~1327$-$2326 \citep{frebel05}, they may be indicative of the earliest chemical enrichment events that their respective host systems experienced. 

One star in the $\omega$~Cen progenitor stream, HE~0007$-$1752, appears to be unusual, both with regard to other stream and halo stars. It has unprecedentedly low [Sc/Fe], with $\mbox{[Sc/Fe]}=-0.98$, nearly 1\,dex below all other stars. This measurement is based on three lines, with a standard deviation of 0.1\,dex, so it should be reliable. Interestingly, several other abundance ratios in this star are also unusually low:\ 
$\mbox{[Mg/Fe]}=-0.17$, based on 5 lines, $\sigma=0.17$;
$\mbox{[Ti~\textsc{ii}/Fe]}=-0.17$, based on 41 lines, $\sigma=0.15$;
$\mbox{[Co~\textsc{i}/Fe]}=-0.38$, based on 7 lines, $\sigma=0.15$; and
$\mbox{[Ni~\textsc{i}/Fe]}=-0.40$, based on 15 lines, $\sigma=0.15$.
Furthermore, we highlight that \textit{all other} light element abundance ratios, including [V/Fe], are among the lowest of our sample and other halo stars. Curiously, this star also shows one of the highest levels of $r$-process enhancement ($\mbox{[Eu/Fe]}= +1.25$; r-II star). 
It is unclear what the origin of this peculiar pattern is.
There is precedent for [Sc/Fe], [Ti/Fe], and [V/Fe] varying together \citep{Sneden16,cowan20,ou20}, and
there is also precedent for unusually low [Sc/Fe] ratios \citep{casey15,ji20_car}.

Overall, these stars somewhat resemble the Fe-rich metal-poor stars \citep{Yong13a, Jacobson15, Mcwilliam18, sakari19} although specific nucleosynthesis and chemical evolution modeling may be needed to illuminate the conditions that occurred in their natal gas clouds.

\subsection{Neutron-capture element abundances}

We obtained chemical abundances for each star for up to 22 neutron-capture elements from strontium to thorium.
We mostly use spectrum synthesis to account for blending and hyperfine structure of absorption features. For Ce, Nd, Gd, and Er, however, abundances were derived from equivalent widths. For Ba and Eu, we use $r$-process isotope ratios as given in \citet{Sneden08}. Our final neutron-capture abundance measurements of all stars are summarized in Table~\ref{rproc}. Total uncertainties in neutron-capture element abundances range from 0.1 to 0.3\,dex, depending on the level of blending and how well these blends could be accounted for. We take as uncertainties the standard error, as derived for small samples \citep{Keeping62}. In all cases, we adopt 0.1\,dex as the minimum nominal uncertainty. For elements with only one available line, we adopt an uncertainty between 0.1 and 0.3\,dex, depending on the quality of the measurement.

\begin{deluxetable*}{lrrrrrll}
\tablewidth{0pt}
\tabletypesize{\small}
\tablecaption{\label{rproc} Ba, Eu and C abundances, and $r$-process classifications }
\tablehead{
\colhead{Star} & 
\colhead{[Fe/H]} &  
\colhead{[Eu/Fe]}  &  
\colhead{[Ba/Fe]}  & 
\colhead{[Ba/Eu]}  & 
\colhead{$\mbox{[C/Fe]}_{\rm corr}$} & 
\colhead{Classification} &
\colhead{Comments}
}
\startdata
\multicolumn{8}{c}{Helmi trail stream}  \\\hline
HE~2234$-$4757	&	$-$2.83 &$ $0.72&$-$0.05&$-$0.77 &$+$0.77& r-I & \\	
HE~0050$-$0918	&	$-$1.37 &$ $0.71&$ $0.11&$-$0.60 &$-$0.25& r-I& 11.2\,Gyr  \\
BM235           &   $-$1.63 &$ $0.48&$ $0.08&$-$0.40 &$-$0.04& r-I; r+s& 10.8\,Gyr  \\
HE~1210$-$2729	&	$-$3.01 &$-$0.46&$-$1.04&$-$0.58 &$+$0.36& r-0      & \\  
HE~0033$-$2141	&	$-$2.91 &$<-$0.11&$-$1.17&$>-$1.06&$+$0.43&\nodata   & \\
\hline
\multicolumn{8}{c}{Helmi debris stream}  \\\hline
HE~0048$-$1109  & $-$2.35 &$ $1.29 &$ $0.62 &$-$0.67 &$ $0.61  & r-II & AB, $-$15.0\,Gyr \\
HE~0324$-$0122  & $-$2.41 &$ $1.12 &$ $0.46 &$-$0.66 &$ $0.37  & r-II & 10.3\,Gyr \\
BD\,+30 2611    & $-$1.50 &   0.65 &   0.06 &$-$0.59 & $-$0.72 & r-I  & R10, 13.6\,Gyr\\
BD\,+29 2356    & $-$1.59 &   0.41 &   0.09 &$-$0.32 & $-$0.24 & r-I; r+s & R10, AB, 2.3\,Gyr\\
CS~29513-032    & $-$2.08 &   0.39 & 0.76 &$ $0.37 &      0.62 & r-I; r+s & R10\\
HD~175305       & $-$1.73 &   0.35 &   0.06 &$-$0.29 & $-$0.31 & r-I; r+s & R10, AB, 0.5\,Gyr\\
HD~119516       & $-$2.26 &   0.34 &$-$0.02 &$-$0.36 &$<-$0.42 & r-I; r+s & R10\\
BM028           & $-$1.83 &$ $0.33 &$ $0.12 &$-$0.21 & $ $0.38 & r-0; r+s & (CD\,$-$36 1052; R10) \\
BM308           & $-$2.10 &$ $0.32 &$-$0.26 &$-$0.58 &$ $0.31  & r-0      &  \\
BM209           & $-$2.27 &$ $0.22 &$-$0.44 &$-$0.66 &$ $0.21  & r-0      & (HD~128279; R10) \\
HE~0017$-$3646  & $-$2.45 &$ $0.18 &$-$0.53 &$-$0.71 &$ $0.08  & r-0      & AB, $-$22.5\,Gyr  \\
BD\,+10 2495    & $-$2.31 &   0.13 &$-$0.01 &$-$0.14 & $-$0.31 & r-0; r+s & R10, AB, $-$2.8\,Gyr\\
HD~237846       & $-$3.29 &$-$0.30 &$-$0.79 &$-$0.49 &    0.17 & r-0 & R10\\
CS~29513-031    & $-$2.64 &$<$1.14 &$-$0.31 &$>-$1.45& $<$1.20 & \nodata  & R10  \\
CS~22876-040    & $-$2.34 &$<$1.15 &   0.12 &$>-$1.03& $<$0.59 & \nodata  & R10 \\ 
HE~0012$-$5643  & $-$3.03 &$<$1.51 &$<$0.45& \nodata&    1.22 & \nodata  & \\  
CS~22948-093    & $-$3.36 &$<$1.91 &   0.37 &$>-$1.54& $<$1.72 &\nodata& R10 \\
\hline
\multicolumn{8}{c}{$\omega$~Cen progenitor stream}  \\\hline
HE~0007$-$1752 & $-$2.36 &$ $1.25&$ $0.60& $-$0.65&$-$0.29& r-II & 10.8\,Gyr \\
HE~0429$-$4620 & $-$2.40 &   0.53&$-$0.19& $-$0.72 & 0.05& r-I & \\
BM056          & $-$2.02 &$ $0.33&$-$0.06& $-$0.39 & 0.34& r-I; r+s & 11.7\,Gyr  \\
HE~2322$-$6125 & $-$2.61 &$ $0.37&$-$0.29& $-$0.66 & 0.30& r-I&\\
BM121          & $-$2.48 &   0.27&$-$0.48& $-$0.75 & 0.37& r-0 &  11.2\,Gyr  \\
HE~2315$-$4306 & $-$2.39 &$-$0.11&$-$0.62& $-$0.51 & 0.31& r-0 &     \\
HE~2319$-$5228 & $-$3.18 &$<-$0.04&$-$1.74&$>-$1.70& 1.63&\nodata &  \\
HE~1401$-$0010 & $-$2.54 &$<$0.12&$-$0.43&$>-$0.55 & 0.93&\nodata & \\
HE~2138$-$0314 & $-$3.08 &$<$0.36&$-$0.93&$>-$1.29 & 0.95&\nodata & \\
HE~0039$-$0216 & $-$2.56 &$<$1.09&$-$0.30&$>-$1.39 & 0.94&\nodata &  \\
\enddata
\tablecomments{Stars in each stream are sorted by decreasing [Eu/Fe] ratios. For reference, the pure $r$-process ratio is $\mbox{[Ba/Eu]}=-0.71$ \citep{Sneden08}. We consider $-0.8<\mbox{[Ba/Eu]}<-0.45$ as the range covered by a pure main $r$-process signature \citep{frebel18}. Higher values reflect an additional $s$-process contribution, provided the neutron-capture pattern principally follows the pattern for combined $r+s$ enrichment, as shown in Figure~\ref{natureplot}. ``R10'' denotes stars from \citet{Roederer10}. ``AB'' denotes actinide boost stars. Stellar ages and other star names are given where available.}
\end{deluxetable*}

\begin{figure*}[!ht]
 \begin{center}
  \includegraphics[clip=false,width=17.7cm 
  ]{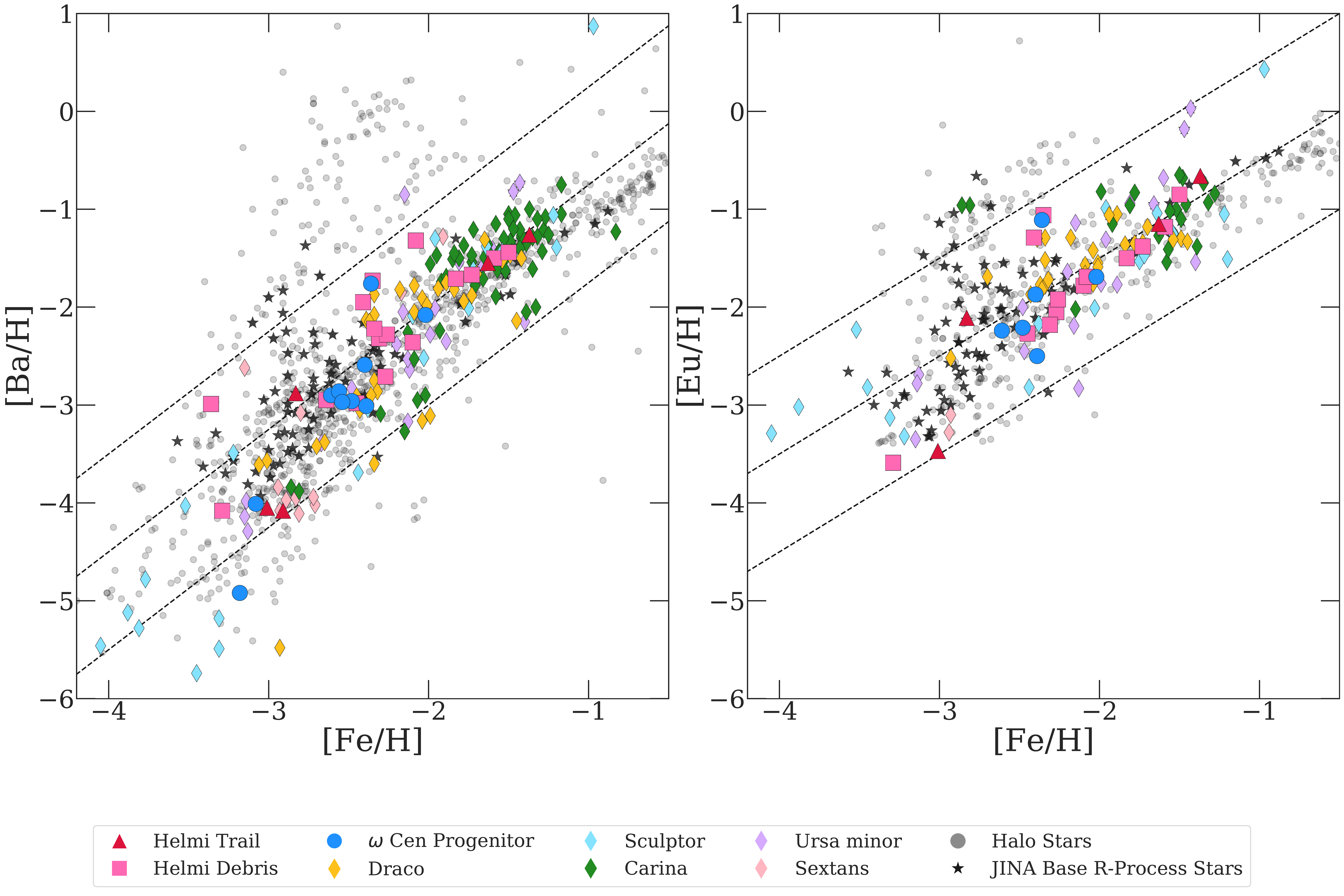} 
  \figcaption{\label{natureplot} [Ba/H] and [Eu/H] abundance ratios of our stream stars as a function of metallicity. The symbol colors and shapes are indicated in the legend.
  Literature data are adopted from
  \citet{Shetrone03},  \citet{Venn12}, and  \citet{Norris17} for Carina;
  \citet{Shetrone01},   \citet{Fulbright04}, and  \citet{Cohen09}   for Draco;
  \citet{Tafelmeyer10},  \citet{Kirby12},  \citet{Jablonka15}, and \citet{Simon15} for Sculptor;  \citet{Shetrone01} and  \citet{Tafelmeyer10}  for Sextans;  \citet{Shetrone01}   for Ursa Minor;
  and the JINAbase compilations (\citealt{abohalima18}, and references therein)
  for ordinary halo stars and $r$-process enhanced halo stars. Dashed lines with slopes of 1.25 for [Ba/H] and 1.0 for [Eu/H] are provided to guide the eye. See text for discussion.  }
 \end{center}
\end{figure*}

In Figure~\ref{natureplot}, we show the [Ba/H] and [Eu/H] ratios of the stream stars.
Recall that we added the Helmi debris stream stars analyzed by \citet{Roederer10} to our sample. For comparison, we also show abundances of other dwarf galaxy stars and halo stars as compiled in \citet{frebel10} and \citet{abohalima18}. We also show halo stars with $-0.8<\mbox{[Ba/Eu]}<-0.45$, which depicts an abundance signature dominated by the $r$-process. Generally, the stream stars follow the trend set by the $r$-process-dominated halo stars very closely. We defer to Section~\ref{sr_ba_eu} for a more complete discussion of these results.

For 13 stars (eight in the Helmi trail and debris streams, five in the $\omega$~Cen progenitor stream), neutron-capture element abundances generally agree well with the scaled solar $r$-process pattern. The other 10 stars display a pattern where a number of abundances of elements between Ba to Th do not entirely match the scaled main solar $r$-process pattern. Instead, elements Ba to Sm and Yb to Hf are slightly enhanced compared with elements Eu to Tm, suggesting an additional small contribution of neutron-capture elements from a different source. We attribute this excess to the $s$-process  which would make all these stars ``$r+s$'' stars (see Section~\ref{sproc} for further discussion). 
 
\begin{figure*}[!ht]
 \begin{center}
  \includegraphics[clip=false,width=17.7cm 
  ]{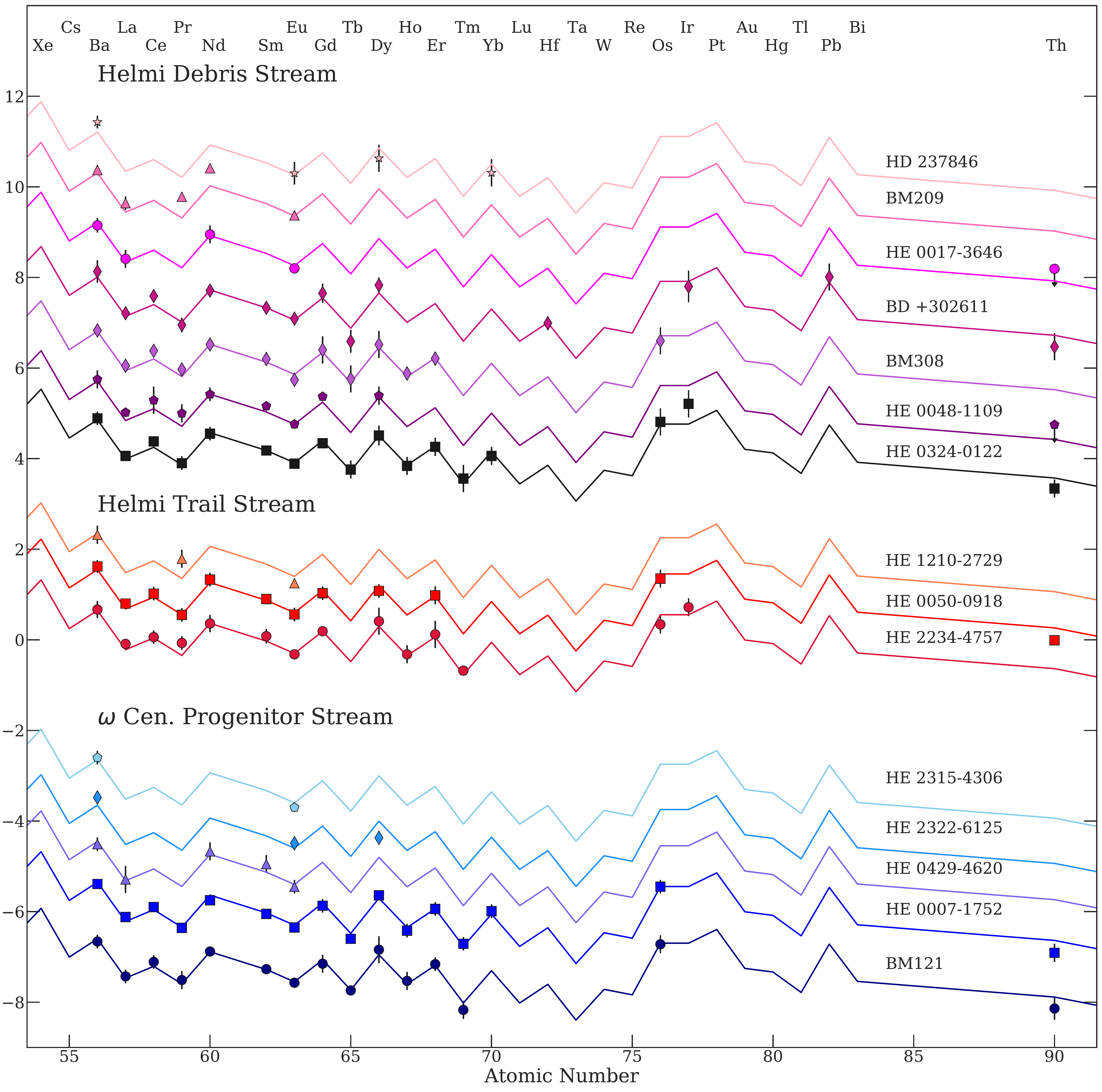} 
  \figcaption{\label{rpattern} Stellar abundances of neutron-capture elements overlaid with the solar $r$-process pattern \citep{Burris00} scaled to Eu. Patterns for each star are arbitrarily offset for easy inspection. The figure shows stars in all three streams that display a pure main $r$-process pattern,
  as defined by their [Ba/Eu] ratio agreeing well with that of the scaled solar $r$-process pattern \citep{Burris00}}
 \end{center}
\end{figure*}

We then define ``pure'' $r$-process stars to have [Ba/Eu] close to the solar $r$-process ratio of $\mbox{[Ba/Eu]}=-0.71$, i.e., $-0.8<\mbox{[Ba/Eu]}<-0.45$ to account for measurement uncertainties. The patterns of the ``pure'' $r$-process stars in all three streams are shown in Figure~\ref{rpattern}. 

\begin{figure*}[!ht]
 \begin{center}
  \includegraphics[clip=false,width=17.7cm 
  ]{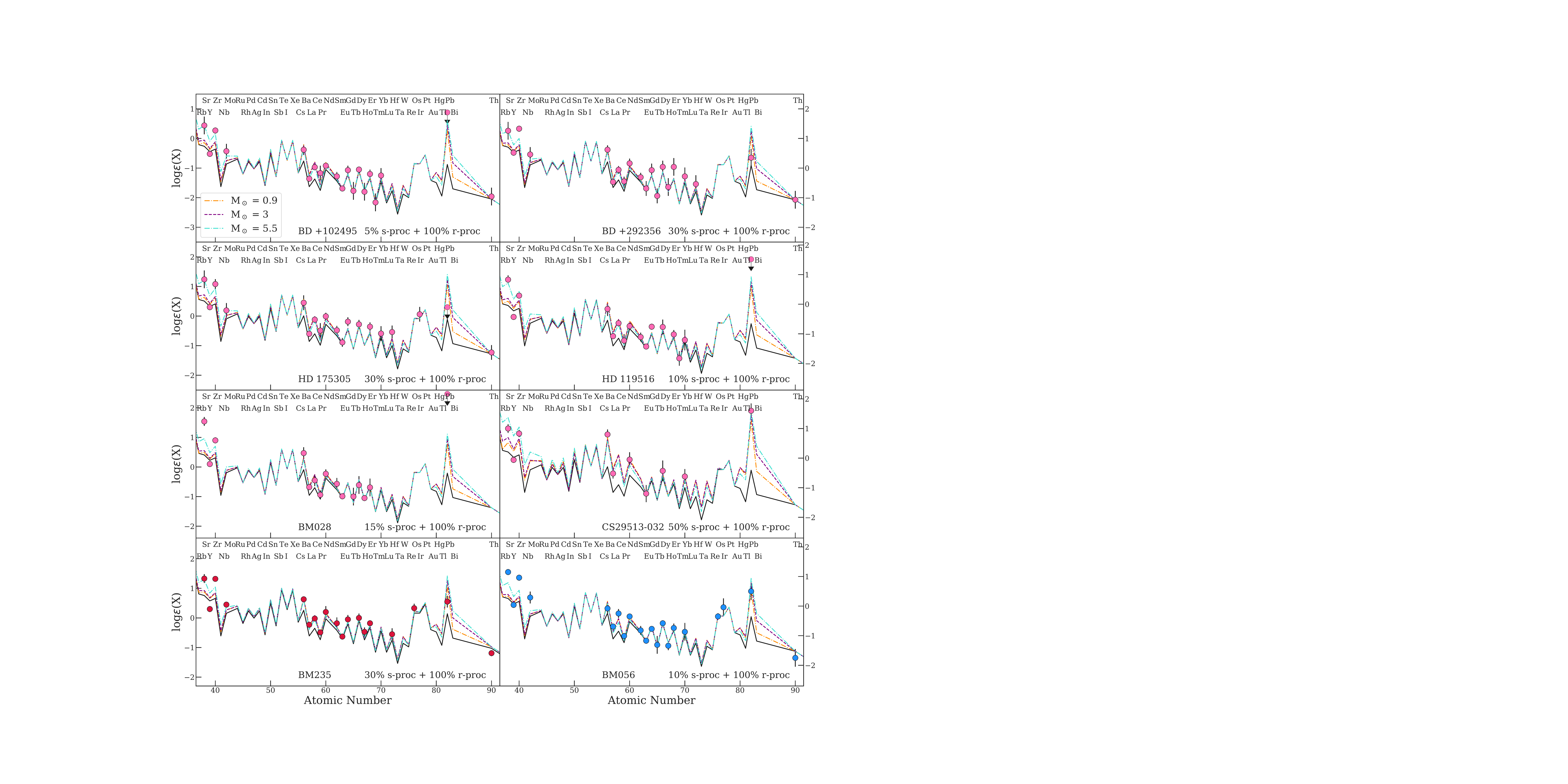} 
  \figcaption{\label{rspattern} Stellar abundances of neutron-capture elements of the eight ``$r+s$'' stars. The first six stars belong to the Helmi debris stream (pink dots) , BM235 to the Helmi trail stream (red dots) and BM056 to the $\omega$ Cen Progenitor Stream (blue dots). Overlaid is the solar $r$-process (solid) pattern (\citealt{Burris00}, but with a modified Y abundance, following \citealt{Arlandini99}) scaled to Eu and three $r+s$ yield (dashed) patterns. The $r+s$ yields consists of the $r$-process pattern scaled to Eu added to a appropriate fraction of the metal-poor $s$-process yield \citep{Lugaro12} resulting from 0.9, 3 and 5.5\,M$_{\odot}$ progenitor models (orange, purple, and cyan dashed lines, respectively), respectively. The amounts of each process contributing to the shown $r+s$ yields are listed as a percentage in each panel. See text for discussion.
  }
 \end{center}
\end{figure*}

We classify stars as $r+s$ stars if they display a somewhat higher [Ba/Eu] ratio and if the abundances of elements Ba to Sm principally follow the odd-even behavior set by the $r$-process pattern with abundances scaled to Eu (considering that Eu is predominantly produced in the $r$-process) but appear with somewhat offset. This translates to stars with $\mbox{[Ba/Eu]}\geq-0.45$.  In Figure~\ref{rspattern}, we display the eight stars that have these higher [Ba/Eu] ratios. We overlay the abundance patterns with a composite pattern consisting of the sum of the solar $r$-process pattern scaled to the stellar Eu value and a fraction of a low-metallicity (with $Z=0.0001$ and [Fe/H] = $-$2.3; \citealt{Lugaro12}) $s$-process pattern. We vary the $s$-process contribution to match the stellar abundance pattern after scaling it to the observed Eu abundance. Results are shown by the dashed lines in Figure~\ref{rspattern}. 
We compared $s$-process yields of models of progenitor AGB stars with different masses and other parameters \citep{Lugaro12}. Using the fractional yields of several models with different masses, we construct $r+s$ yields that principally are able to match the data quite well. In Figure~\ref{rspattern}, we show three $r+s$ models based on the 0.9, 3, and 5.5\,M$_{\odot}$ models, 
with the respective scaled $s$-process percentages (the same for all models) listed in each panel. 

We arrive at this percentage the following way: 

Upon scaling the relevant model $s$-process yield to the Eu abundances of the initial scaled solar composition model \citep{Lugaro12}, we then subtracted the initial abundances from the $s$-process pattern to arrive at a pure $s$-process signature scaled to Eu of zero. This approximates our assumption that Eu observed in our stream stars arises solely from the $r$-process as this is the dominant process and Eu is more easily produced in the $r$-process. 

A fraction of this signature is then added to the $r$-process pattern that has been scaled to the observed Eu abundances of the star to achieve a combined model yield that reproduces the observed neutron-capture element signature between Ba and Pb. For seven stars, the $s$-process fraction that leads to matching the data is between 5 and 30\%. The eighth star, CS~29513$-$032, requires 50\% -- not surprisingly since it has the strongest contribution as discussed above.

These combined $r+s$ yields, based on the 0.9 and 3\,M$_{\odot}$ models, are very similar overall, and they fit the abundances in the element region between the second and third $r$-process peaks very well, especially the abundances that do not match the scaled solar $r$-process pattern, Ba to Sm (when scaled to Eu). The largest difference is in Pb, but it does not exceed 0.35\,dex, with the 0.9\,M$_{\odot}$ model resulting in lower $r+s$ yields. 
The higher predicted Pb abundances do not match most of our observed values as well as the lower mass models do. The derived LTE Pb abundances, however, could be underestimated by several tenths of a dex because of NLTE effects \citep{mashonkina12,roederer20}, which would help reconcile the differences.

The Sr, Y, and Zr abundances, however, are not well matched by any model.
The 5.5\,M$_{\odot}$ model generally provides the closest match for Sr and Zr, and it fits the Ba to Eu region quite well.

We note that Sr, Y, Zr and Pb are the elements most affected by the input assumptions about the extent of partial mixing that drives $s$-process nucleosynthesis. Under different partial mixing assumptions, Sr, Y, and Zr decrease (relative to, e.g., Eu) although they remain above values produced with the lower mass yields. However, the relative Pb abundance of the $r+s$ yield decreases significantly when no partial mixing is introduced, and the Pb abundance falls significantly below results involving, e.g., the 0.9\,M$_{\odot}$ model. 
We tested these issues in only a limited way since other low-$Z$ high mass models were calculated without a partial mixing zone, which is inferior compared to results calculated with an implemented partial mixing zone (A.\ Karakas, priv.\ comm.). Hence, further investigations of the effects of partial mixing at various masses would be very insightful, and they might eventually help to identify the dominant mass of the progenitor stars that were responsible for the $s$-process enrichment in early star forming regions. 

In summary, the deviations of the stellar data points from the pure scaled $r$-process pattern are well fit by combined $r+s$ yields and thus may reflect small contributions by an $s$-process to an underlying $r$-process pattern. For one star, \mbox{CS~29513-032}, an $r+s$ pattern had already been pointed out by \citealt{Roederer10}). 
This star is well fit by our combined $r+s$ model yield, as seen in Figure~\ref{rspattern}. However, details remain to be worked out.  The Pb measurements seem to be better fit by $r+s$ yields based on lower mass models, while the Sr, Y, and Zr abundances appear to be well matched by involving contributions the $s$-process yields from higher mass models. More generally, this issue may simply reflect the case that likely a distribution of AGB progenitor stellar masses and other parameters were responsible for the observed abundance patterns in our sample of stars. A full AGB population synthesis modeling of the birth gas cloud would thus be helpful for further investigating this topic. Exploring contributions from the weak component of the \rproc, which were invoked by \citeauthor{Roederer10}\ to explain the deviations from the scaled solar \rproc\ residuals in the Helmi stream stars, may also offer a fruitful direction for future investigations.

\begin{figure*}[ht!]
 \begin{center}
  \includegraphics[clip=false,width=17.7cm]{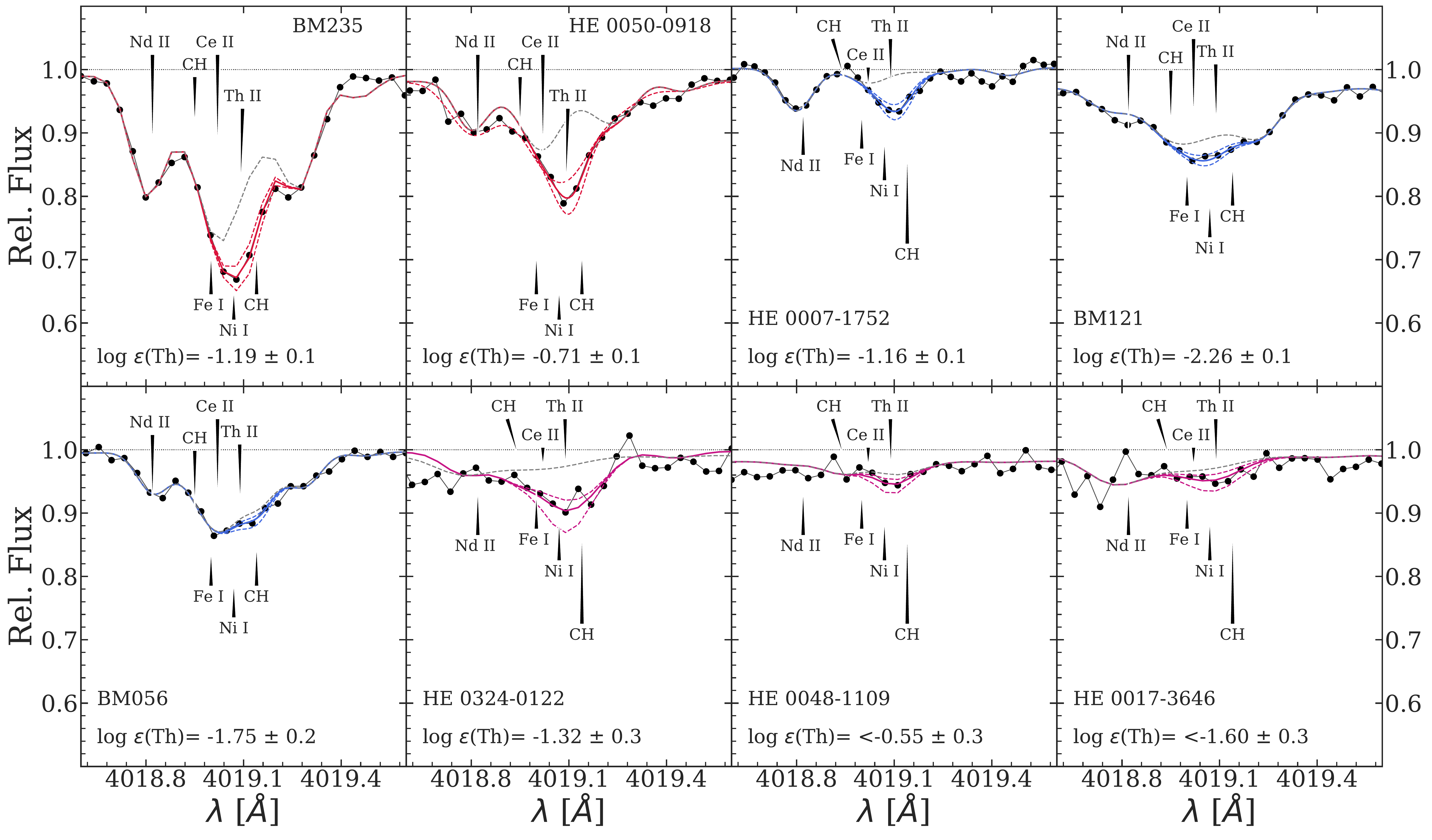} 
  \figcaption{\label{thorium} Spectra around the Th\,\textsc{ii} line at 4019\,$\AA$. The observed spectra are shown in black. Synthetic spectra are denoted as solid colored lines, and synthetic spectra are shown as dashed colored lines. The gray dashed lines represent the synthetic spectra with no thorium. }
 \end{center}
\end{figure*}

We detected thorium in eight stars. Two of the stars are in the Helmi trail stream, and three are in each of the Helmi debris and $\omega$~Cen progenitor streams. Figure~\ref{thorium} shows spectra of the Th\,\textsc{ii} line at 4019\,{\AA} in all eight stars, together with synthetic spectra with different abundances. Additionally, \citet{Roederer10} derived Th abundances for four stars belonging to the Helmi debris stream, bringing the total to 12 stars with Th abundance measurements. 
The thorium abundances are further discussed in Section~\ref{cosmo}.

\section{Discussion} \label{rproc_streams}
\subsection{R-process enhanced stream progenitors}

In Figures~\ref{rpattern} and \ref{rspattern}, we show the neutron-capture element abundance patterns of elements between Ba and Th of 23 of 24 stars in our three streams as they are all agreeable with that of the universal, main $r$-process pattern. The 24th star is too weak-lined for assessment. However, there is a large range of absolute abundance levels and some variation in the ``pureness'' of the  $r$-process signature.

We follow established conventions and classify the level of $r$-process enhancement following the nomenclature suggested in \citet{frebel18}. The classification of each star is listed in Table~\ref{rproc}. 
Three stars are strongly $r$-process enhanced ($r$-II), with [Eu/Fe] $>+$1.0 and [Ba/Eu] $<$ 0.0. Three stars are moderately $r$-process enhanced ($r$-I stars) with $+$0.3 $<$ [Eu/Fe] $\leq+$1.0 and [Ba/Eu] $<$ 0.0. 
Five stars clearly show an $r$-process pattern but technically do not fall in the previously established categories of $r$-process \textit{enhanced} stars due to their low [Eu/Fe] ratio, $< +0.3$. For convenience, we will refer to those stars as ``$r$-0'' stars. A sixth star only has a Eu/Ba ratio available, but it is consistent with that of the main $r$-process pattern.

For eight additional stars, Ba is detected, but Eu is not.
Finally, for the 24th star, we measure upper limits for both Eu and Ba only, since the star is warm and also carbon enhanced. This precluded assessment of the existence of an $r$-process pattern; however, given that all other stars show it, we assume that this star would also display an $r$-process pattern, albeit at undetectable levels.
Overall, all stars cover a significant range of enrichment levels of $-0.5 < \mbox{[Eu/Fe]} < +1.3$, or a factor of 60.

This clearly shows that our three streams chemically resemble dwarf galaxies which are known to contain $r$-process-enhanced stars.
These include Reticulum\,II which has seven of nine $r$-process enhanced stars \citep{Ji16a}, and Tucana\,III, which has four out of five stars identified as $r$-process enhanced \citep{hansen17, marshall19}, and many of the classical dwarf spheroidal galaxies (e.g., Draco, \citealt{Cohen09}; Carina, \citealt{Venn12}, \citealt{Shetrone03}; Sculptor, \citealt{Simon15}, \citealt{Jablonka15}; Sextans, \citealt{Shetrone01}, \citealt{Tafelmeyer10}; Gaia Sausage/Enceladus, \citealt{aguado21}).
This behavior aligns with expectations that all systems that experienced significant chemical evolution and accretion processes  included a variety of nucleosynthesis sources, including one or more events that produced heavy $r$-process elements.

\subsection{$s$-process enrichment in the progenitor streams} \label{sproc}

The existence of eight $r+s$ stars suggests that stars in the respective progenitor systems formed from gas that was enriched by stellar winds rich in $s$-process material emanating from asymptotic giant branch (AGB) stars \citep{Karakas10} as part of the system's own chemical evolution. Furthermore, this may have occurred wide-spread but somewhat inhomogeneously as not all stars show this additional $s$-process component and, when they do, it varies. We note that the behavior found here (except perhaps for \mbox{CS~29513-032}) is different from the abundance signature of the star studied in \citet{gull18} which is the result of a mass transfer event across a binary system. This scenario is based on an erstwhile AGB star that produced and then transferred $s$-process material onto the now-observed secondary star. The mass transfer scenario is unlikely in the majority of our stream stars given, e.g., their overall moderate Ba and C abundances, and the lack of radial velocity variations for all but one star.

Seven of the $r+s$ stars have an underlying $r$-process pattern commensurate with an $r$-I enrichment level, as evidenced by their [Eu/Fe] ratios (and Eu being largely unaffected by any $s$-process contamination). The remaining star has an underlying pattern at the $r$-0 level. We thus conclude that these stars are mildly enhanced $r+s$ stars, moderately enriched in both $r$ and $s$-process elements, unlike mass-transfer $r+s$ stars.

\begin{figure*}[!t]
 \begin{center}
  \includegraphics[clip=false,width=17.7cm]{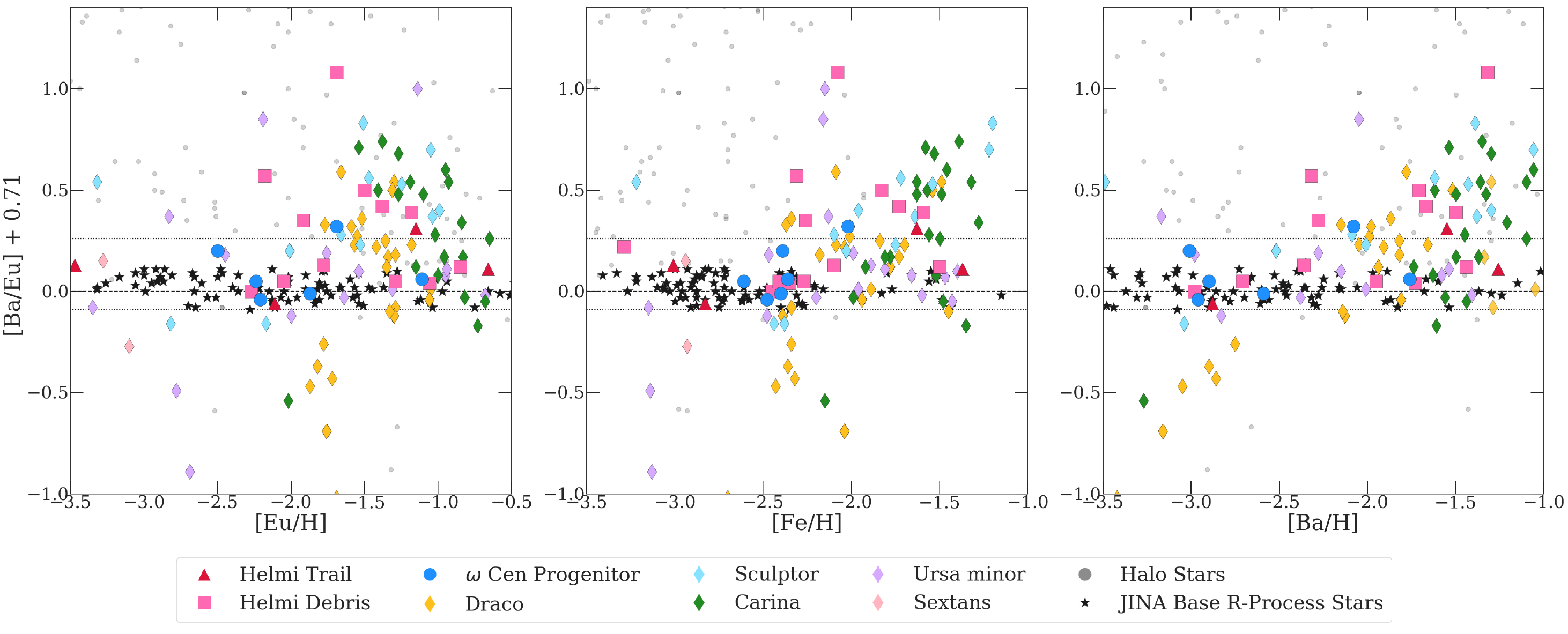} 
  \figcaption{\label{delta} Difference between observed Ba abundances and the scaled solar $r$-process pattern (scaled to Eu), $\log \epsilon {\rm (Ba/Eu)}_{\star}$-$ \log \epsilon {\rm (Ba/Eu)}^r_{\odot}$, for our stream stars and other dwarf galaxy stars (colors and sources are the same as in Figure~\ref{natureplot}). At lower metallicities, stars match the pattern more often than at higher [Fe/H], owing to the increasing contribution by the $s$-process (or potentially other processes). Stars in Carina, Sculptor, and Draco appear to be significantly affected this way while Ursa Minor shows the opposite behavior.}
 \end{center}
\end{figure*}

In Figure~\ref{delta}, we further explore these excess $s$-process contributions, with respect to the pure $r$-process ratio ($-$0.71) as a function of [Eu/H], [Fe/H], and [Ba/H]. There are preferentially more stars with Ba excess found at [Eu/H] $>-$2, [Fe/H] $>-$2, and $\mbox{[Ba/H]} >-2.5$. This would be expected because those stars likely formed at later times after chemical evolution had already resulted in significant enrichment (and metal mixing) of the system. At lower metallicities, any $s$-process contributions appear to be minimal. This shows that $r$-process enrichment dominates early in the evolution, and $r$- and $s$-processes operate independent of each other. $s$-process enrichment then appears at higher metallicity, progressively washing out any $r$-process signatures.  Along those lines, the Helmi debris stars show significant $s$-process excesses with increasing [Eu/H], [Fe/H] and [Ba/H]. Unfortunately, the Helmi trail and $\omega$~Cen progenitor streams have only one $r+s$ star each so meaningful conclusions cannot be reached.

\subsection{Abundance trends of Ba and Eu} \label{sr_ba_eu}

We now return to Figure~\ref{natureplot} to consider the more general abundance trends of Ba and Eu. Regarding [Ba/H] vs.\ [Fe/H], the stream stars well follow the overall trend set by the pure main $r$-process halo stars, with the $r+s$ stars naturally being slightly higher (recall that, regardless, nearly all $r+s$ stars are within only 0.5\,dex of the pure $r$-process Ba abundance). The correlation of [Ba/H] with [Fe/H] for the pure main $r$-process stars appears linear although not without scatter. A slope of $\sim$1.25 matches the main branch well, as is indicated by the inner dashed line that we add for illustrative purposes. The other two dashed lines are $\pm$1\,dex offset from the main branch, for reference.

A similar behavior is found for [Eu/H] vs.\ [Fe/H]. There is an extremely tight main branch, that is matched well by a shallower slope of 1. 
Naturally, the Eu abundances do not contain any significant $s$-process contribution, so this behavior traces only $r$-process chemical evolution. The difference in the ratios of the [Ba/H] and [Eu/H] slopes (with [Fe/H]) thus signifies the contributions by the $s$-process 
leading to Ba levels increasing faster with time. 

For completeness, we note that that the [Sr/H] versus [Fe/H] evolution as traced by our and other dwarf galaxy stars will be discussed in a separate paper (Frebel et al.\ 2020, in prep).

\subsection{The Nature of the $\omega$~Cen Stream Progenitor System}

Kinematic evidence suggests that the unusual globular cluster $\omega$~Cen and this stream are related.
If so, we suggest that the stream did not originate from $\omega$~Cen itself, but rather that they both originated from a common dwarf galaxy progenitor.
That progenitor must have been quite massive. The present-day stellar mass of the $\omega$~Cen globular cluster is $\approx 4 \times 10^{6}$\,M$_{\odot}$ \citep{dsouza13}, so the progenitor's stellar mass must have been comparable to or perhaps much greater than that of the most massive classical dwarf spheroidal galaxies, like Leo~I or Fornax ($5.5 \times 10^{6}$ and $20 \times 10^{6}$\,$M_{\odot}$, respectively; \citealt{mcconnachie12}). Within this context, we present evidence indicating that $\omega$~Cen and the stream are chemically distinct from one another.

First, we detect two stars in the stream with $\mbox{[Fe/H]}<-3.0$. 
Three other stars have $\mbox{[Fe/H]} < -2.5$, which is the lower bound of [Fe/H] ratios found in $\omega$~Cen today \citep{johnson20}.
These stars also are more metal-poor than the metallicity floor for globular clusters, $\mbox{[Fe/H]} \approx -2.5$ (e.g., \citealt{beasley19}), and the recently-discovered globular clusters or their putative tidal-stream remnants with
$-2.9 \leq \mbox{[Fe/H]} \leq -2.5$ \citep{roederer2019,larsen20,wan20}.
A massive dwarf galaxy progenitor would have likely hosted stars with a range of metallicities spanning several dex, including a metal-poor tail extending $\sim$~1--2\,dex below the peak of the metallicity distribution (e.g., \citealt{kirby11b}).
The metallicity of the dominant population in the $\omega$~Cen globular cluster,
\mbox{[Fe/H]}~$\approx -1.7$ \citep{Johnson10}, is itself perhaps lower than the mean metallicity of the progenitor system (cf.\ M54 in Sagittarius; \citealt{carretta10m54,hyde15}). We note here 
that the fact that all stars in our sample are more metal-poor than the expected peak of the progenitor galaxy metallicity distribution function is unsurprising. Our sample is somewhat biased toward more metal-poor stars, because stars in our sample were originally selected by \citet{Frebel06b}, who had loosely targeted bright metal-poor stars.
Nevertheless, it would be unprecedented to find stars with $\mbox{[Fe/H]} < -3.0$ in $\omega$~Cen, but such stars are found in other dwarf galaxies (e.g., \citealt{frebel10a}).
The extremely metal-poor stars found in the $\omega$~Cen progenitor stream thus favor an association with the progenitor dwarf galaxy, but not $\omega$~Cen itself.

Secondly, we find one star in this stream (\mbox{HE~0007$-$1752},
$\mbox{[Fe/H]} = -2.36 \pm 0.13$; Section~\ref{sec:light}) that has low [$\alpha$/Fe] ratios and unusual ratios among the Fe-group elements. 
Low [$\alpha$/Fe] ratios are typically found in more massive dwarf galaxies that have experienced Fe enrichment by Type~Ia supernovae (e.g., \citealt{Tolstoy04,Venn12}). 
This combination of low [$\alpha$/Fe] and unusual ratios among the Fe-group elements is not found, however, among any of the $> 10^{3}$ $\omega$~Cen stars studied to date.

Finally, the progenitor experienced $r$-process enrichment, but these 10~stars show that the level of Eu (Ba) enhancement varies by factors of $> 20$ ($1400$) relative to Fe or H.
We identify 1 $r$-II star, 2 $r$-I stars, 2 $r$-0 stars, 1 $r+s$ star with an underlying $r$-I signature, and 4~stars with indeterminate levels because only meaningless upper limits are available on the Eu abundance.
The existence of consistent $r$-process abundance patterns but with such a wide range of enhancement levels suggests that one or more $r$-process events inhomogeneously enriched the birth gas cloud.
This is not a common feature of globular clusters generally \citep{roederer11a}, or $\omega$~Cen specifically (e.g., \citealt{Johnson10}).
A wide range of heavy-element abundance ratios is, however, a common characteristic of low-metallicity (\mbox{[Fe/H]}~$< -2$) stars in dwarf galaxies (e.g., \citealt{cohen10,Tsujimoto15a}).

Regarding the heavy element nucleosynthetic history in the $\omega$~Cen stream progenitor system, our results indicate that even the most metal-poor stars  formed after one or more neutron-capture processes enriched the star-forming gas.
Barium is always detected in these stars, and in most cases the [Ba/Fe] ratios are sub-solar.
This result supports the conclusion of \citet{roederer13} that at least one neutron-capture process was active in the earliest stellar generations in all environments, including the $\omega$~Cen progenitor system.
Whenever we can also detect Eu in the $\omega$~Cen stream stars, the [Ba/Eu] ratios implicate the $r$-process as the origin of the heaviest elements in these stars.

More metal-rich stars in $\omega$~Cen formed later, after a substantial amount of $s$-process material from AGB stars was present \citep{smith00,Johnson10,marino11,pancino11b}.
The one $r+s$ star in the $\omega$~Cen stream sample, BM056
exhibits signatures of $s$-process enhancement, but at a level below that found in $\omega$~Cen.
Its \mbox{[Ba/Fe]} ratio, $-0.06 \pm 0.10$, is much lower than that of the more metal-rich stars in $\omega$~Cen (\mbox{[Ba/Fe]}~$\approx +0.5$ to $+1.0$), and its \mbox{[Pb/Fe]} ratio, $+0.77 \pm 0.30$, is comparable to that found in $\omega$~Cen (\mbox{[Pb/Fe]}~$\approx +0.35$ to $+0.65$, after correcting the results of \citealt{dorazi11} to our $\log gf$ scale).
This star, with \mbox{[Fe/H]} $= -2.02 \pm 0.12$, is the most metal-rich star in our $\omega$~Cen progenitor stream sample, and its metallicity is comparable to the metallicity range where the first traces of $s$-process material are found in $\omega$~Cen.

In summary, our results show that the $r$-process appears to have dominated the production of heavy elements in stars with \mbox{[Fe/H]}~$< -2$ in the progenitor. 
Only in more metal-rich stars found in the cluster itself did large amounts of $s$-process material accumulate.

\subsection{Are the Helmi trail and debris streams chemically distinct?}\label{distinct}

Given the somewhat uncertain association of the Helmi trail and debris streams, it is worthwhile to turn to the chemical abundance trends and signatures to investigate whether any clues could be obtained in support of the answer as to whether the streams can be chemically associated with each other, and hence with one progenitor.

The light elements do not show any significant differences between the two streams, or compared to the halo (see Figure~\ref{light_ele}). There are a few outliers, but on average there is no systematic difference with respect to halo stars. The star \mbox{HE~0012$-$5643} could be one such outlier that shows low [$\alpha$/Fe] ratios. This star in the Helmi debris stream has [Mg/Fe]$= +0.02$ and [Si/Fe]$= +0.12$, which are lower than average, but it also has [Ti/Fe]$= +0.30$ and [Ca/Fe] = $+0.35$, which are average. The [$\alpha$/Fe] and [Fe/H] ratios reveal information about the transition of the production of Fe from dominated by core-collapse supernovae to Type~Ia supernovae. A low ratio could thus indicate local late(r)-time Fe enrichment within the progenitor. The same overall agreeable behavior is found among heavy elements (see Figure~\ref{natureplot}); all abundances fall well within the range of the Milky Way halo stars.

While we examined the abundance trends for differences, it should be stated here again the one major traits that both streams in fact have in common: their ubiquitous $r$-process enrichment. This would perhaps be the strongest piece of evidence that both streams are chemically associated and thus share a progenitor. However, the coincidental fact (all targets were chosen based on the \citealt{beers17} kinematic study) that $\omega$~Cen also shows this behavior somewhat weakens the case, as it is apparent that $r$-process enrichment is not a unique signature as such (cf.\ \citealt{roederer13}; \citealt{aguado21}). Finally, we note that the trail stream does not contain any actinide boost stars, perhaps owing to the smaller sample size, and it displays significantly less scatter in [Th/Eu] than the Helmi debris stream stars (see Section~\ref{cosmo}).

Nevertheless, a potentially different actinide content may not necessarily preclude a common progenitor, as the trail stream could have been stripped first, with any actinide-enhanced nucleosynthesis happening on later time scales. 
Alternatively, any larger dwarf galaxy would likely be too large for locally produced $r$-process material to be fully mixed within the probably many pockets of star formation. This is reflected in the light element abundances where a few stars show outlier abundances in several element ratios. It is thus clear that additional data on Th, especially in stars in the trail stream, would help to provide further constraints on local actinide production.

In conclusion, we cannot find any significant differences between the chemical abundance trends set by the Helmi debris and trail streams that would suggest these streams to have originated from two separate progenitors. On the other hand, we uncover no chemical evidence that definitively links them, either.  The majority of halo stars display near-identical trends, and halo stars clearly did not all originate in one progenitor. This is demonstrated by the fact that the $\omega$~Cen progenitor shows essentially the same overall abundance trends and patterns. There thus remains no unique answer based on chemical abundances to the question of whether the Helmi trail and debris streams are associated. However, their metal-rich stars might differ after all, since galaxies of different mass evolve similarly at early times, but differently at late times.

\subsection{Actinide variations in the streams and implications for the r-process site} \label{cosmo}

We initially attempted to derive cosmo-chronometric ages of stars with Th detections.
Thorium is solely produced by the $r$-process.

We assume Eu to also be exclusively made in the $r$-process, even in the case of the $r+s$ stars, because any $s$-process contribution to their Eu abundances would principally be very small and the overall $s$-process component of any $r+s$ stars is small to be begin with. Stellar ages of metal-poor stars enhanced in $r$-process elements can principally be obtained by comparing, e.g., their Th/Eu abundance ratio with predictions for the initial production ratio, as made in a putative $r$-process event. Assuming that the stellar signatures follow the scaled main solar $r$-process, one can use the equation 
$\Delta$t = 46.78$\times$[$\log$(Th/Eu)$_{\rm initial}-\log(\rm{Th/Eu})_{\rm now}$] from \citet{Cayrel01} to derive such ages, where log(Th/Eu)$_{\rm initial}$ refers to the Th/Eu ratio created in the original nucleosynthesis event.

\begin{figure}[ht]
  \includegraphics[clip=false,width=8.4cm]{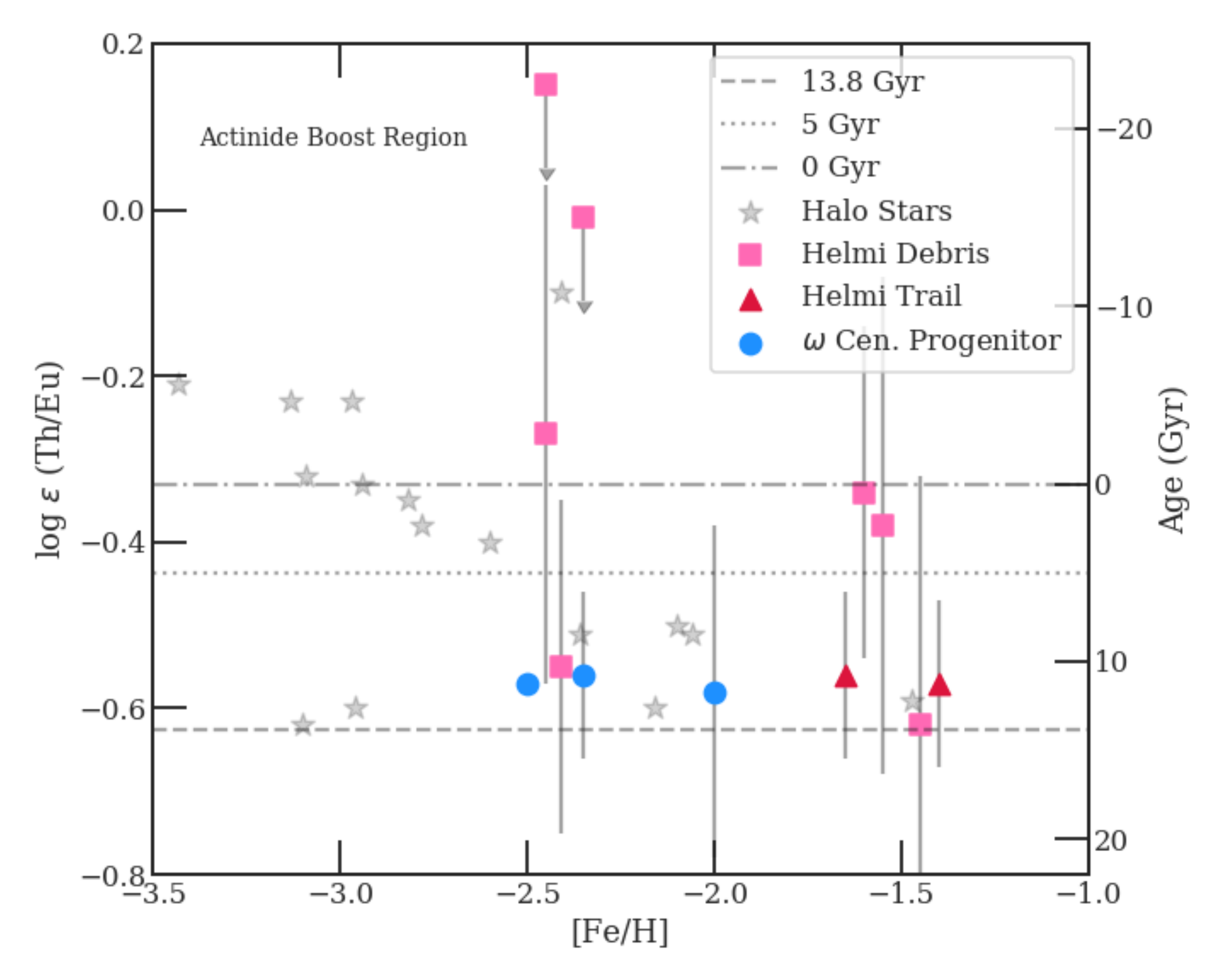} \includegraphics[clip=false,width=7.8cm]{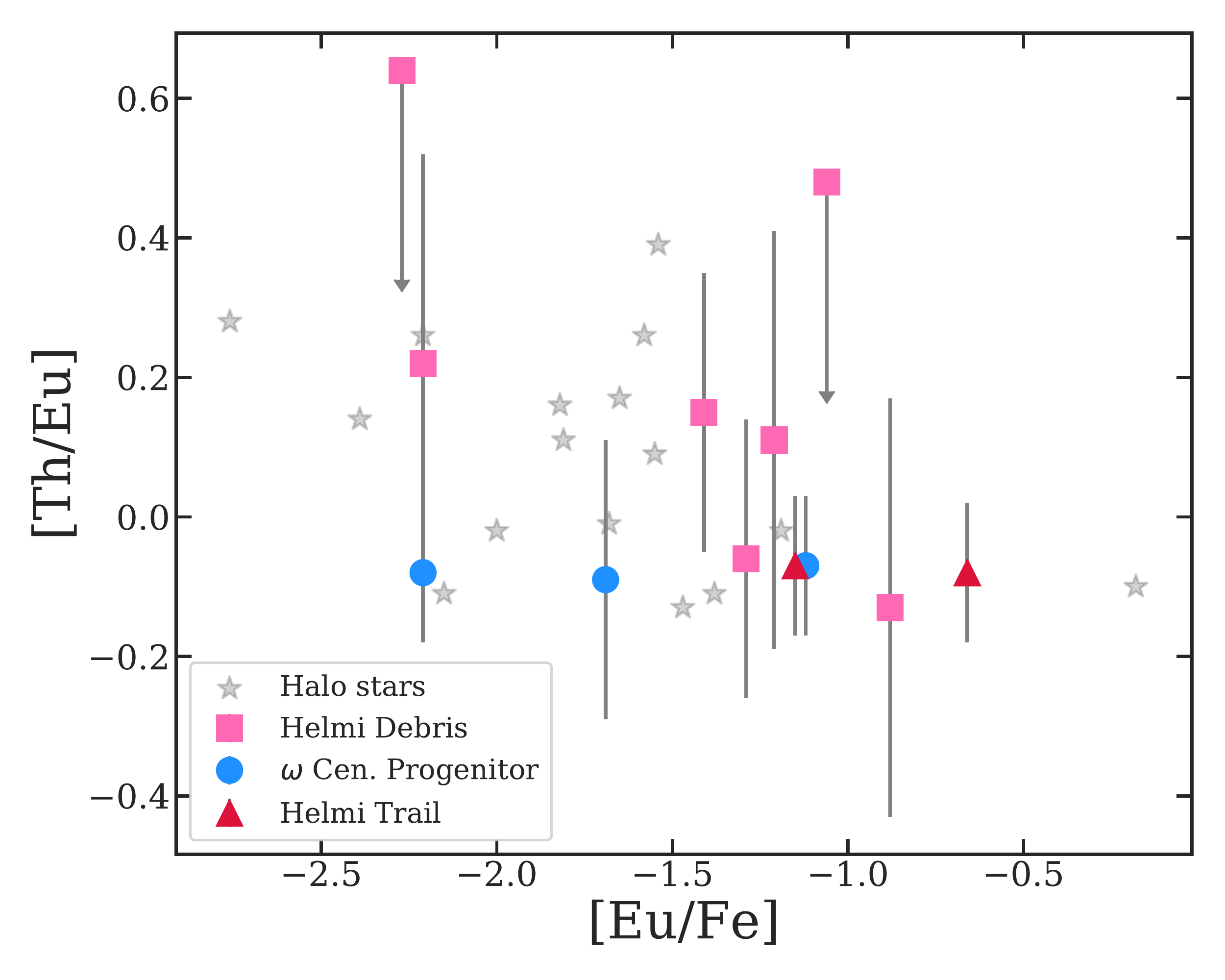} 
  \figcaption{\label{ages} Top: Th/Eu abundances as a function of [Fe/H] for the different streams. Corresponding ages are also indicated, calculated based on the \citet{Schatz02} production ratios. Bottom: [Th/Eu] as a function of [Eu/Fe]. We  include $r$-I and $r$-II halo stars with available Th measurements for comparison, using data selected with JINAbase (\citealt{Roederer14b}, \citealt{Honda04}), \citealt{MAS10}, \citealt{SIQ14}, \citealt{JOH02a}, \citealt{COW02}, \citealt{Westin00}, \citealt{Ivans06}, \citealt{Sneden03}, \citealt{Hayek09}, \citealt{Aoki07b}).}
  \end{figure}

As can be seen in Figure~\ref{ages}, there is an extremely large range of 0.4\,dex in (detected) Th/Eu values present among the Helmi debris stars. This implies that conditions for $r$-process nucleosynthesis must have been somewhat variable in order to produce such a spread. It also showcases that, at face value, age measurements span a range of 16\,Gyr across this system. Using initial production ratios from \citet{Schatz02} based on $r$-process waiting-point calculations, we derive individual ages of 13.6, 10.3, 2.3, 0.5, $-2.8$, $>$\,$-$15.0, and $>$\,$-$22.5\,Gyr (see also Table~\ref{rproc}).  Systematic uncertainties also need to be considered. If different production ratios are chosen (which corresponds to a change in Th/Eu ratios of only about 0.05\,dex), ages will change accordingly. For example, the \citet{hill17} production ratios are based on a high-entropy wind model of \citet{Farouqi10}, and typically result in much larger ages by several Gyr. 

Clearly, many of these ages are non-sensical but principally not unheard of. One phenomenon that leads to too young (and negative) ages is that of the ``actinide boost'' \citep{Schatz02}. Those stars show an excess of Th with respect to the scaled solar $r$-process pattern and accounting for billion year long decay times. Indeed, at least three of the Helmi debris stream stars (HD~175305, \mbox{BD+29 2356}, \mbox{BD+10 2495}) 
appear to be actinide boost stars (with HE~0048$-$1109 and HE~0017$-$3646 remaining candidates).
This corresponds to 18\% at face value, and 23\% among just the identified $r$-process stars. This is not unlike the actinide fraction among halo stars with about 25-30\% \citep{mashonkina2014}, all of which remains poorly understood \citep{holmbeck2019}. 
Contrary to the halo actinide boost stars, the stream stars investigated here are known to have a common progenitor which enables much better insight into the conditions of actinide production. For example, the Th/Eu spread may simply reflect multiple $r$-process events that occurred in a given host system. The $r$-process pattern would not be affected by this but the decaying Th might be if the events were sufficiently spread apart in time for some decay to occur between events. The actinide boost stars have metallicities of $\mbox{[Fe/H]} =-2.3$ and $\sim-1.6$. This might leave room for $r$-process events to have occurred a significantly different times.
Uranium abundances, albeit difficult to obtain because of with the shorter half-life of 4.7\,Gyr for $^{238}$U, would be able to confirm or refute such a scenario. Detailed star formation and chemical evolution modeling would also be very helpful to further explore these ideas. 

Another explanation for higher Th/Eu ratio might be that accreted $r$-process rich gas and/or that $r$-process nucleosynthesis operates differently under different conditions and sites. Details are being actively studied \citep{holmbeck19} to constrain local conditions that drive certain $r$-process yields. If conditions for $r$-process nucleosynthesis are eventually quantifiable, including knowing the number of $r$-process events, then matching production ratios could be used for age determinations. In the absence of that all ages are unfortunately not reliable due to unknown initial conditions. One can turn it around, though. Assuming that all stars in the Helmi debris stream have nearly the same age, to within 2\,Gyr or $<$0.05\,dex, Figure~\ref{ages} shows the 0.4\,dex spreads at a given [Fe/H] and also that across [Eu/Fe]. These  may only be achieved by varying actinide production and/or with inhomogeneous mixing occurring in the stream progenitor.

While no actinide boost stars are identified in the samples of the Helmi trail and the $\omega$~Cen progenitor streams, they do have two and three stars with measured Th and Eu, respectively. They paint a more well-behaved picture. Owing to near identical Th/Eu ratios, the two stars in the Helmi trail stream have ages of $10.8 \pm 4.7$ and $11.2 \pm 4.7$\,Gyr \citep{Schatz02}, with an average of $10.8 \pm 4.7$\,Gyr.  Clearly, (or perhaps owing to the smaller number of stars), the nucleosynthetic history of $r$-process enrichment appears to be much simpler for this stream and its progenitor, with indications for only one prior $r$-process event having taken place. At face value, this roughly agrees with the dynamical simulation results of an age of 11-13\,Gyr postulated by \citet{koppelman19}. 
The case is similar for the $\omega$~Cen progenitor stream. Given, once again, their very similar Th/Eu ratios, we calculate stellar ages for the three stars of $10.8\pm 4.7$, $11.2\pm 4.7$, and $11.7\pm 9.3$\,Gyr, with an average of 11.3\,Gyr, and in agreement with expectations for the age of its putative progenitor of at least 10\,Gyr (which is the estimated infall time; \citealt{bekki03}). With no actinide boost stars present in these two streams, this is once again suggestive of conditions that may have been more homogeneous or somehow different from those in the Helmi debris stream progenitor.

In summary, obtaining Th abundances of more than one star belonging to the same stream provides a principally helpful way to explore a system's age and nucleosynthetic history. Having available Th measurements for all stars in our sample would increase statistics to derive better, more firm observational conclusions to confront theoretical models and scenarios for actinide production in dwarf galaxies (e.g., \citealt{holmbeck19}).

\section{Conclusion and Summary}\label{concl}

We present detailed chemical abundance measurements for 22 stars in three different stellar streams, the Helmi debris, Helmi trail, and $\omega$~Cen progenitor streams. We combined our Helmi debris sample with the results for ten stars by \citet{Roederer10}, for a total of 32 stars.

23 stars in the three streams display heavy neutron-capture element signatures associated with the $r$-process. The three streams thus provide a missing link by connecting the sites of early $r$-process enrichment in (massive) dwarf galaxies with the Milky Way's population of $r$-process stars. Broadly, the hypothesis that $r$-process halo stars originated in dwarf galaxies that were eventually accreted can thus be examined with any chemical similarities and differences; in our case with three streams. 

Contrary to stars located in any dwarf galaxies, high $S/N$ spectra can be obtained for these bright objects that (now) residing in the halo. This offers the chance for studying dwarf galaxy chemical signatures and their overall chemical evolution in much greater detail than was is possible with relatively few, difficult to access, faint dwarf galaxy stars. This point is illustrated by the fact that the small $s$-process contributions that slightly change the $r$-process signature could be discerned for these stars because of the good data quality. This would likely not have been possible, or not to such extend, with any spectra of lower quality. 

The fortuity of having accessible these bright stars also expands on the numbers and mass range of accessible progenitor systems that can be studied compared to the existing (and accessible to high-resolution spectroscopy) satellite dwarfs. This advantage helps to better understand the role massive dwarf galaxies played in the build-up of the old halo, and should be taken advantage of more in the future, especially if additional streams are found. 

Based on our analysis of the 32 stars in the Helmi trail and debris stream and the $\omega$~Cen progenitor stream, we summarize additional key findings. 

\begin{enumerate} 

\item All stars with sufficient data quality (23 or 32) display neutron-capture elements that follow the scaled solar $r$-process pattern. Thorium is detected in eight of these stars. The remaining nine stars did not have sufficient data quality to enable assessment of the presence of $r$-process material, leaving open the possibility that all these stars show an $r$-process signature. 

\item For the two Helmi streams combined, we obtain fractions of 41\% of $r$-0 stars, 47\% of $r$-I stars, and 12\% of $r$-II stars. 

For $\omega$~Cen progenitor stream, we find 33\% of $r$-0 stars, 50\% of $r$-I stars, and 17\% of $r$-II stars. These are calculated only considering sample stars with a detected $r$-process signatures, and thus higher than the expected fraction for Milky Way halo stars. Taking the full sample, the numbers decrease somewhat to 32\%, 36\%, and 9\% for the two Helmi streams combined, and 20\%, 30\%, and 10\% for the $\omega$~Cen progenitor stream. These values should be regarded as lower limits given the unknown $r$-process enhancement status of the remaining sample stars.

\item In total, eight stars can be classified as $r+s$ stars. These stars contain a small component of $s$-process material, as evidenced by their neutron-capture element signatures not entirely matching that of the scaled solar $r$-process. These contributions likely originated from local (inhomogeneous) enrichment by earlier generation of AGB stars and offers a glimpse into the chemical enrichment by lower mass stars.

\item Six stars are CEMP stars (19\%) and one is a NEMP star. Eight stars have $\mbox{[C/Fe]}>0.6$, corresponding to a frequency of 22\%. We find an increase of carbon abundances with lower metallicities, similar to the trend found for halo stars \citep{Placco14}. The $\omega$~Cen CEMP fraction suggests the progenitor was more likely a dwarf galaxy than a globular cluster.

\item Cosmo-chronometric ages can principally be obtained from observed Th/Eu ratios in $r$-process enhanced stars even though uncertainties are large, typically around 5\,Gyr. A large spread of 0.4\,dex in [Th/Eu] values in the Helmi debris stream suggests a complex history of $r$-process enrichment and actinide production, including that by multiple $r$-process sources and sites. Stars in the other two streams show consistent [Th/Eu] abundances for which canonical age dating would suggest an old age. 

\end{enumerate}

Assuming that a system such as the Helmi streams progenitor(s) experienced one or more prolific $r$-process event early on in its history, stars with a resulting $r$-process signature would eventually be expected to be predominantly found in its outer region following additional star formation in the center. The large observed $r$-process fraction may thus stem from the fact that the observed streams were most likely part of a halo or the outer part of their progenitor. The same could hold for the $\omega$~Cen progenitor stream. Alternatively, these progenitors themselves may have experienced early accretion events of small(er) $r$-process galaxies, resulting in a potential acquisition of $r$-process stars.

This issue can be addressed with a cosmological simulation of galaxy formation to gain insight into the accretion histories of large galaxies. The Caterpillar simulation suite \citep{griffen16} consists of high-resolution dark matter-only cosmological simulations of Milky-way-mass halos. Analyzing 32 zoom-in simulations in combination with the abundance matching relation from \citet{Garrison_Kimmel17}, we find that galaxies with stellar masses of $2$-$10\times10^{8}$\,M$_{\odot}$ were typically accreted by the Milky Way-mass halo around $z = 1.1^{+1.0}_{-0.8}$, where the uncertainty describes 68\% scatter across the halos in the simulation. 
$z = 1.1^{+1.0}_{-0.8}$ corresponds to an age of around 8\,Gyr (with a range of 4 to 11\,Gyr), in line with dynamical simulation results for the acceretion time of the Helmi (5-8\,Gyr; \citealt{koppelman19}) and $\omega$~Cen (10\,Gyr; \citealt{bekki03}) progenitors.

We then estimated the number of ultra-faint dwarf (UFD) type galaxies that a galaxy with a stellar mass of $\sim$10$^{8-9}$\,M$_{\odot}$ might have accreted throughout its own formation history. We theoretically define an ultra-faint dwarf galaxy as a galaxy that was massive enough to form stars prior to reionization, but not massive enough to continue star formation after reionization. For their minimum total mass prior to reionization, we use $5\times10^{7}$\,M$_{\odot}$, motivated by the atomic cooling threshold \citep{bromm_yoshida11} and estimates of the number of surviving ultra-faint dwarfs orbiting the Milky Way (see \citealt{brauer19}). For the maximum total mass after reionization, we use $2\times10^{9}$\,M$_{\odot}$, the filtering mass estimated by radiation-hydrodynamic simulations of reionization \citep{ocvirk16}.

We also assume instantaneous reionization at $z = 8$, motivated by \citet{aubert18} who found that progenitor halos with virial mass M$_{vir}$($z = 0$) $< 10^{11}$\,M$_{\odot}$ reionized around the globally averaged 50\% reionization at $<z_{reion}>$ = 7.8. 
With these definitions, a galaxy with a stellar mass of $\sim$10$^{8-9}$\,M$_{\odot}$ typically accreted $20 \pm 7$ UFDs throughout its formation history, where the uncertainty describes 68\% scatter across the halos in the simulation. If about 10-15\% of those UFDs were $r$-process enriched like Reticulum\,II \citep{Ji16a}, then typically only 1-3 of the absorbed dwarfs would have brought in $r$-process stars.

While not unlikely, this suggests that more massive dwarf galaxies would typically need to experience multiple own $r$-process events to produce a population of $r$-process stars to account for the high fraction of observed $r$-process stars in the three streams. 
Looking ahead, with more stream stars being identified with, e.g., Gaia data, detailed chemical studies of additional streams will be helpful to further characterize the population of dwarf galaxies that assembled the Milky Way halo, and to learn about their respective environments and chemical evolution histories.

\acknowledgements
We thank C.\ Sneden for providing us with an up-to-date version of neutron-capture linelists.
M.G.\ acknowledges  support  from  the  MIT  UROP  program and support of the Schweizerischen Studienstiftung. 
AF acknowledges support from NSF grant AST-1716251, the Silverman (1968) Family Career Development Professorship, and thanks the Wissenschaftskolleg zu Berlin for their wonderful Fellow's program and generous hospitality. 
I.U.R.\ acknowledges support from
grants AST~1815403 and
PHY~1430152 (Physics Frontier Center/JINA-CEE)
awarded by the U.S.\ National Science Foundation (NSF).
A.P.J. is supported by a Carnegie Fellowship and the Thacher Research Award in Astronomy.

This work also benefited from support by the National Science Foundation under Grant No. PHY-1430152 (JINA Center for the Evolution of the Elements).

This work made extensive use of NASA's Astrophysics Data System Bibliographic Services and the python libraries 
\texttt{numpy} \citep{numpy}, 
\texttt{scipy} \citep{scipy}, 
\texttt{matplotlib} \citep{matplotlib},
and \texttt{astropy} \citep{astropy}.

\pagebreak
\end{document}